%% file: 0_Main.tex
\documentclass[conference]{IEEEtran}
\IEEEoverridecommandlockouts
\usepackage{cite}
\usepackage{amsmath,amssymb,amsfonts}
\usepackage{algorithmic}
\usepackage{graphicx}
\usepackage{textcomp}
\usepackage{amsthm}
\usepackage{xcolor}
\usepackage{bbding}
\usepackage{makecell}
\usepackage{algorithm}  
\usepackage{color}
\usepackage{multirow}
\usepackage{bm}
\usepackage{array}
\usepackage{booktabs}
\usepackage{tabularx,booktabs}
\usepackage{multicol}
\usepackage{cleveref}
\usepackage{balance}
\usepackage{lastpage}
\usepackage{tabulary}
\usepackage{subfigure}
\usepackage{etoolbox}
\usepackage{bbm}
\usepackage{enumitem}
\usepackage{mathrsfs}
\usepackage{multicol}
\usepackage{amsfonts,amssymb}
\usepackage{ulem}
\usepackage{fancyhdr}
\usepackage{cancel}
\usepackage{setspace}
\usepackage{stackengine}
\usepackage{threeparttable}

\DeclareMathOperator*{\argmin}{arg\,min}

\usepackage{amsthm}
\theoremstyle{definition}
\newtheorem{definition}{Definition}

\usepackage{nomencl}
\makenomenclature

\def\BibTeX{{\rm B\kern-.05em{\sc i\kern-.025em b}\kern-.08em
    T\kern-.1667em\lower.7ex\hbox{E}\kern-.125emX}}

\newcommand{\blue}[1]{\textcolor{black}{#1}}

\newcommand{\nomenclaturewww}[2]{\nomenclature{\textcolor{black}{#1}}{\textcolor{black}{#2}}}

\begin{document}

\bstctlcite{IEEEexample:BSTcontrol}

\title{%A Comprehensive Tutorial on Optimization Techniques for RIS-aided Future Wireless Communications: Model-based, Heuristic and Machine Learning Approaches\\
A Survey on Model-based, Heuristic, and Machine Learning Optimization Approaches in RIS-aided Wireless Networks\\
\thanks{
This work was supported in part by the Natural Sciences and Engineering Research Council of Canada (NSERC) Canadian Collaborative Research and Training Experience Program (CREATE) under Grant 497981; in part by the Canada Research Chairs Program; in part by the U.S National Science Foundation under Grants CNS-2128448 and ECCS-2335876; in part by the CHIST-ERA grant under the project CHIST-ERA-20-SICT-005; in part by the Engineering and Physical Sciences Research Council (EPSRC) under Project EP/W035588/1; and in part by the PHC Alliance Franco-British Joint Research Programme under Grant 822326028.

H. Zhou and M. Erol-Kantarci are with the School of Electrical Engineering and Computer Science, University of Ottawa, Ottawa, ON K1N 6N5, Canada. (emails:\{hzhou098, melike.erolkantarci\}@uottawa.ca).

Yuanwei Liu is with the School of Electronic Engineering and Computer Science, Queen Mary University of London, London E1 4NS, U.K. (email:yuanwei.liu@qmul.ac.uk).

H. Vincent Poor is with the Department of Electrical and Computer Engineering,
Princeton University, Princeton, NJ 08544 USA (e-mail: poor@princeton.edu).} 

}

\author{\IEEEauthorblockN{Hao Zhou, \IEEEmembership{Member, IEEE}, Melike Erol-Kantarci, \IEEEmembership{Senior Member, IEEE}, Yuanwei Liu, \IEEEmembership{Senior Member, IEEE},\\ and H. Vincent Poor, \IEEEmembership{Life Fellow, IEEE}} }

\maketitle

\thispagestyle{fancy}            %更改plain状态，首页格式设为fancy
\chead{This paper has been accepted by IEEE Communications Surveys and Tutorials. } 

\renewcommand{\headrulewidth}{1pt}      %把页眉线的宽度设为零，即去掉页眉线
\pagestyle{plain}

\begin{abstract}
Reconfigurable intelligent surfaces (RISs) have received considerable attention as a key enabler for envisioned 6G networks, for the purpose of improving the network capacity, coverage, efficiency, and security with low energy consumption and low hardware cost. However, integrating RISs into the existing infrastructure greatly increases the network management complexity, especially for controlling a significant number of RIS elements. To unleash the full potential of RISs, efficient optimization approaches are of great importance. 
This work provides a comprehensive survey on optimization techniques for RIS-aided wireless communications, including model-based, heuristic, and machine learning (ML) algorithms. 
In particular, we first summarize the problem formulations in the literature with diverse objectives and constraints, e.g., sum-rate maximization, power minimization, and imperfect channel state information constraints. 
Then, we introduce model-based algorithms that have been used in the literature, such as alternating optimization, the majorization-minimization method, and successive convex approximation. Next, heuristic optimization is discussed, which applies heuristic rules for obtaining low-complexity solutions. Moreover, we present state-of-the-art ML algorithms and applications towards RISs, i.e., supervised and unsupervised learning, reinforcement learning, federated learning, graph learning, transfer learning, and hierarchical learning-based approaches. Model-based, heuristic, and ML approaches are compared in terms of stability, robustness, optimality and so on, providing a systematic understanding of these techniques. 
Finally, we highlight RIS-aided applications towards 6G networks and identify future challenges.   
\end{abstract}


\begin{figure*}[!t]
\centering
\setlength{\abovecaptionskip}{0pt} 
\includegraphics[width=0.95\linewidth]{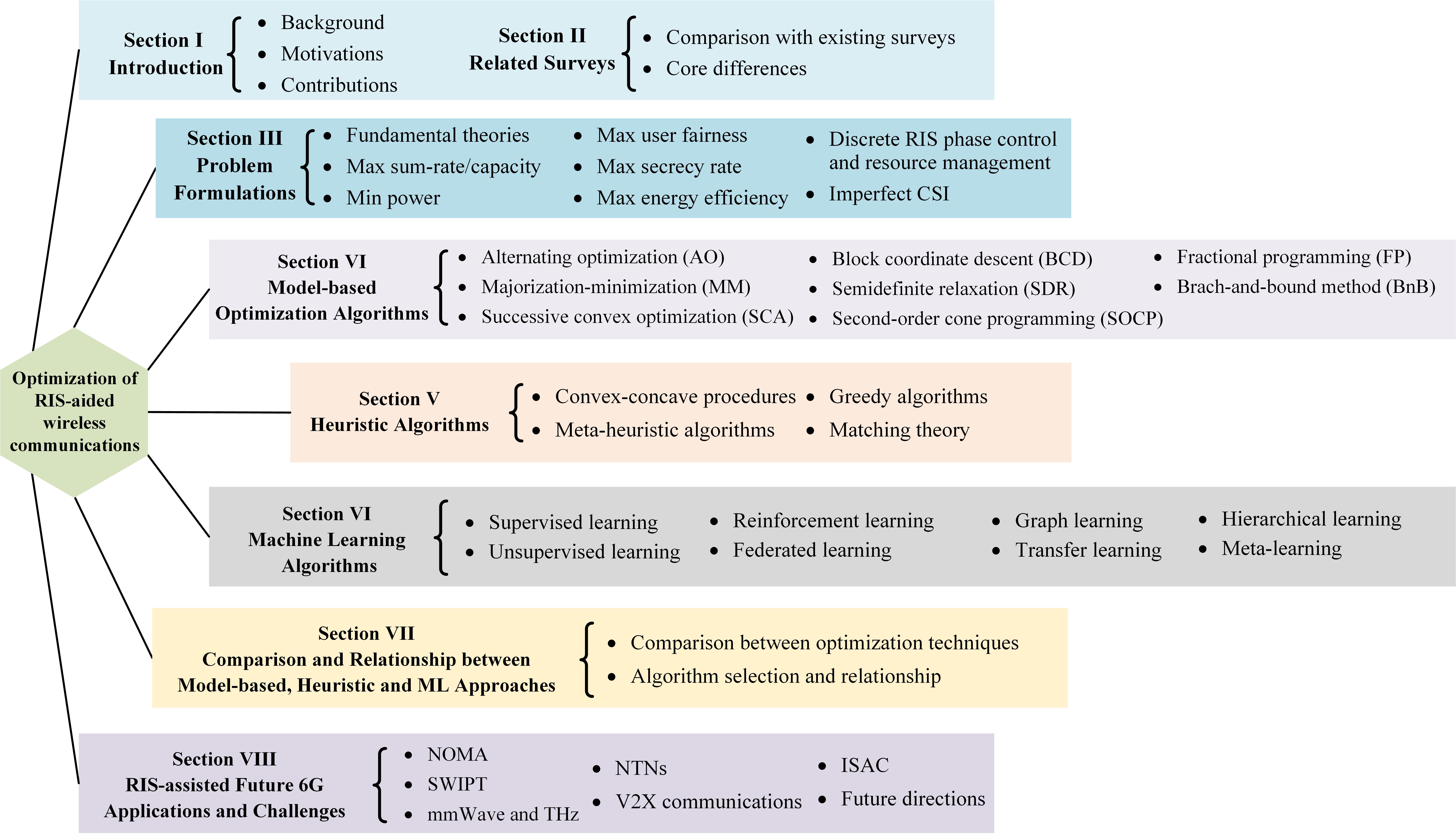}
\caption{\blue{\textbf{Organization and key topics covered in this work.}}}
\label{fig1}
\vspace{-5pt}
\end{figure*}

\mbox{}
\nomenclaturewww{$p$}{Base station transmit power}
\nomenclaturewww{$\bm{G}$}{The channel gain from BS antennas to RIS elements}
\nomenclaturewww{$h^{R}_{k}$}{The channel gain from RIS elements to user $k$}
\nomenclaturewww{$h^{D}_{k}$}{The channel gain from BS antennas to user $k$}
\nomenclaturewww{$N_{0}$}{The noise power}
\nomenclaturewww{$\theta_{n}$}{Phase shift of RIS element $n$}
\nomenclaturewww{$\bm{\Theta}$}{Diagonal matrix that includes all $\theta_{n}$}
\nomenclaturewww{$M$}{The total number of base station antennas}
\nomenclaturewww{$K$}{The total number of single-antenna users}
\nomenclaturewww{$N$}{The total number of RIS elements}
\nomenclaturewww{$P_{max}$}{Maximum transmission power of the base station}
\nomenclaturewww{$w_{k}$}{The weight of user $k$}
\nomenclaturewww{$\gamma_k$}{The SINR of user $k$}
\nomenclaturewww{$\sigma$}{The efficiency of the transmit power amplifier.}
\nomenclaturewww{$P_{UE}$}{Hardware static power consumed by one user}
\nomenclaturewww{$P_{BS}$}{Total hardware static power consumption in BS}
\nomenclaturewww{$P_{R}(\kappa)$}{Power consumption of one RIS reflecting element with resolution $\kappa$}
\nomenclaturewww{$\varrho$}{RIS phase shift resolution}
\nomenclaturewww{$R_L$,$R_E$}{Data rate of legitimate user and eavesdropper}
\nomenclaturewww{$\chi$}{A binary decision variable}
\nomenclaturewww{$Pr()$}{Probability}
\nomenclaturewww{$e_{ch}$}{Channel estimation error}
\nomenclaturewww{$x$}{The control variable in an optimization problem}
\nomenclaturewww{$\hat{x}$}{Optimal control variable}
\nomenclaturewww{$I$}{Total number of control variables}
\nomenclaturewww{$l$}{The index of iteration number}
\nomenclaturewww{$i$}{The index of variable number in a set}
\nomenclaturewww{$f(x)$,$F(x)$}{Objective/utility function}
\nomenclaturewww{$g(x)$}{Surrogate function}
\nomenclaturewww{$\mathscr{X}$}{A convex closed set}
\nomenclaturewww{$g(x|x^{l-1})$}{$g(x|x^{l-1})$ is an approximation function of $f(x)$ at the iteration $l$, and "$|$" in $g(x|x^{l-1})$ means that $x^{l-1}$ is on this function}
\nomenclaturewww{$\omega$,$\omega'$}{Neural network weight of main and target networks.}
\nomenclaturewww{$\omega^A$,$\omega^C$}{Neural network weight of actor and critic networks.}
\nomenclaturewww{$\mathcal{T}$}{Minibatch size}
\nomenclaturewww{$\alpha$}{Learning rate}
\nomenclaturewww{$\eta$}{Discount factor}
\nomenclaturewww{$s$,$s'$}{Current and the next state in a Markov decision process}
\nomenclaturewww{$a$}{ Action in a Markov decision process}
\nomenclaturewww{$r$}{ Reward in a Markov decision process}
\nomenclaturewww{$V(s)$}{State value at $s$ in Q-learning}
\nomenclaturewww{$Q(s,a)$}{State-action value in Q-learning}
\nomenclaturewww{$\mathscr{E}$}{Error function}
\nomenclaturewww{$\mathcal{U}$, $\mathcal{B}$}{ Sets of players in matching theory}
\nomenclaturewww{$\mathbb{E}$}{Expected value}
\nomenclaturewww{$u$,$u'$,$b$,$b'$}{Matching theory players}
\nomenclaturewww{$A,B,C,D$}{Constant real matrix }
\printnomenclature

\input{1_Introduction}

\input{2_RelatedWork}

\input{3_Background}

\input{5_Model_based}

\input{5_Heuristic_based}

\input{6_Machine_learning}

\input{7_Applications_and_Challenges}

\input{8_Conclusion}

%\red{The ref papers have been checked and revised, including "Proc. " for conference papers, abbreviations for journal names, capitalization in titles, and so on.}

\normalem
\bibliographystyle{IEEEtran}
\bibliography{Reference}

\begin{IEEEbiography}[{\includegraphics[width=1in,height=1.2in,clip,keepaspectratio]{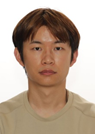}}]{Hao Zhou} received his Ph.D. degree from University of Ottawa, Canada, in 2023. He is an incoming Postdoctoral Research Fellow at Department of Computer Science, McGill University. Before that, he got his B.Eng. and M.Eng degrees from Huazhong University of Science and Technology in 2016, and Tianjin University in 2019, respectively, in China. His research interests include microgrid energy trading, 5G network slicing, mmWave, reconfigurable intelligent surfaces, and so on. He is devoted to developing machine learning algorithms for Smart Grid and 5G/6G networks. He won the best paper reward at the 2023 IEEE ICC conference, and the 2023 Outstanding Self-financed Abroad Chinese Students Award.     
\end{IEEEbiography}

\begin{IEEEbiography}[{\includegraphics[width=1in,height=1.2in,clip,keepaspectratio]{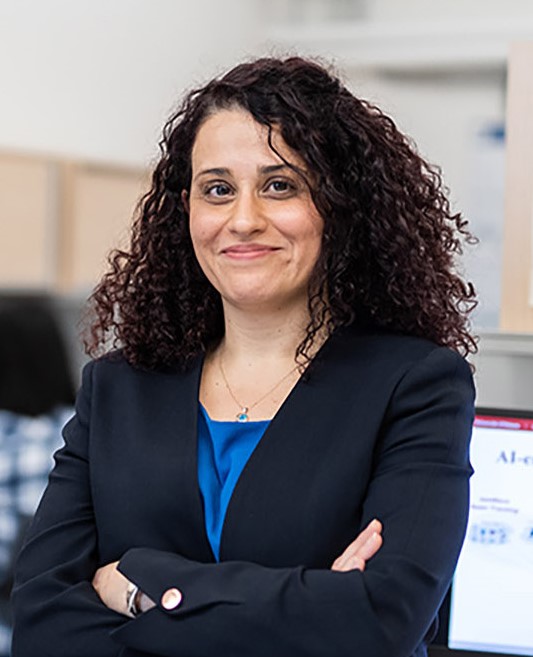}}]{Melike Erol-Kantarci} is Chief Cloud RAN AI$\backslash$ML Data Scientist at Ericsson and Canada Research Chair in AI-enabled Next-Generation Wireless Networks and Full Professor at the School of Electrical Engineering and Computer Science at the University of Ottawa. She is the founding director of the Networked Systems and Communications Research (NETCORE) laboratory. She has received numerous awards and recognitions. Dr. Erol-Kantarci is the co-editor of three books on smart grids, smart cities and intelligent transportation. She has over 200 peer-reviewed publications. She has delivered 70+ keynotes, plenary talks and tutorials around the globe. Dr. Erol-Kantarci is on the editorial board of the IEEE Transactions on Communications, IEEE Transactions on Cognitive Communications and Networking and IEEE Networking Letters.  She has acted as the general chair and technical program chair for many international conferences and workshops. Her main research interests are AI-enabled wireless networks, 5G and 6G wireless communications, smart grid and Internet of Things. Dr. Erol-Kantarci is an IEEE ComSoc Distinguished Lecturer, IEEE Senior member and ACM Senior Member.    
\end{IEEEbiography}

\begin{IEEEbiography}[{\includegraphics[width=1in,height=1.2in,clip,keepaspectratio]{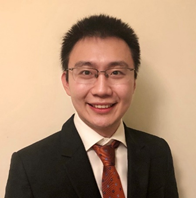}}]{Yuanwei Liu} (S'13-M'16-SM'19-F’24), \qquad \qquad \qquad (http://www.eecs.qmul.ac.uk/$\sim$yuanwei) received the PhD degree in electrical engineering from the Queen Mary University of London, U.K., in 2016. He was with the Department of Informatics, King’s College London, from 2016 to 2017, where he was a Post-Doctoral Research Fellow. He has been a Senior Lecturer (Associate Professor) with the School of Electronic Engineering and Computer Science, Queen Mary University of London, since Aug. 2021, where he was a Lecturer (Assistant Professor) from 2017 to 2021.  His research interests include non-orthogonal multiple access, reconfigurable intelligent surface, integrated sensing and communications, and machine learning. \\
Yuanwei Liu is a Fellow of the IEEE, a Web of Science Highly Cited Researcher, an IEEE Communication Society Distinguished Lecturer, an IEEE Vehicular Technology Society Distinguished Lecturer, the academic Chair for the Next Generation Multiple Access Emerging Technology Initiative, the rapporteur of ETSI Industry Specification Group on Reconfigurable Intelligent Surfaces on work item of “Multi-functional Reconfigurable Intelligent Surfaces (RIS): Modelling, Optimisation, and Operation”, and the UK representative for the URSI Commission C on “Radio communication Systems and Signal Processing”. He was listed as one of 35 Innovators Under 35 China in 2022 by MIT Technology Review. He received IEEE ComSoc Outstanding Young Researcher Award for EMEA in 2020. He received the 2020 IEEE Signal Processing and Computing for Communications (SPCC) Technical Committee Early Achievement Award, IEEE Communication Theory Technical Committee (CTTC) 2021 Early Achievement Award. He received IEEE ComSoc Outstanding Nominee for Best Young Professionals Award in 2021. He is the co-recipient of the Best Student Paper Award in IEEE VTC2022-Fall, the Best Paper Award in ISWCS 2022, the 2022 IEEE SPCC-TC Best Paper Award, and the 2024 IEEE ICCT Best Paper Award. He serves as the Co-Editor-in-Chief of IEEE ComSoc TC Newsletter, an Area Editor of IEEE Communications Letters, an Editor of IEEE Communications Surveys \& Tutorials, IEEE Transactions on Wireless Communications, IEEE Transactions on Vehicular Technology, IEEE Transactions on Network Science and Engineering, and IEEE Transactions on Communications (2018-2023). He serves as the (leading) Guest Editor for Proceedings of the IEEE on Next Generation Multiple Access, IEEE JSAC on Next Generation Multiple Access, IEEE JSTSP on Intelligent Signal Processing and Learning for Next Generation Multiple Access, and IEEE Network on Next Generation Multiple Access for 6G. He serves as the Publicity Co-Chair for IEEE VTC 2019-Fall, the Panel Co-Chair for IEEE WCNC 2024, Symposium Co-Chair for Cognitive Radio \& AI-Enabled Networks for IEEE GLOBECOM 2022 and Communication Theory for IEEE GLOBECOM 2023. He serves as the chair of Special Interest Group (SIG) in SPCC Technical Committee on Signal Processing Techniques for Next Generation Multiple Access, the vice-chair of SIG in SPCC Technical Committee on Near Field Communications for Next Generation Mobile Networks, and the vice-chair of SIG WTC on Reconfigurable Intelligent Surfaces for Smart Radio Environments.  
\end{IEEEbiography}

\begin{IEEEbiography}[{\includegraphics[width=1in,height=1.2in,clip,keepaspectratio]{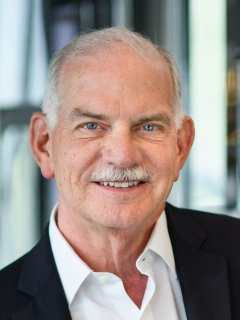}}]{H. Vincent Poor} (S’72, M’77, SM’82, F’87) received the Ph.D. degree in EECS from Princeton University in 1977.  From 1977 until 1990, he was on the faculty of the University of Illinois at Urbana-Champaign. Since 1990 he has been on the faculty at Princeton, where he is currently the Michael Henry Strater University Professor. During 2006 to 2016, he served as the dean of Princeton’s School of Engineering and Applied Science. He has also held visiting appointments at several other universities, including most recently at Berkeley and Cambridge. His research interests are in the areas of information theory, machine learning and network science, and their applications in wireless networks, energy systems and related fields. Among his publications in these areas is the recent book Machine Learning and Wireless Communications. (Cambridge University Press, 2022). Dr. Poor is a member of the National Academy of Engineering and the National Academy of Sciences and is a foreign member of the Chinese Academy of Sciences, the Royal Society, and other national and international academies. He received the IEEE Alexander Graham Bell Medal in 2017.    
\end{IEEEbiography}

\end{document}

%% file: 1_Introduction.tex
\section{Introduction}

While 5G has entered the commercialization phase, the research community has started the exploration of future 6G networks. Compared with previous generations, 6G networks are expected to present more stringent performance requirements, i.e., terabits per second (Tbps) data rates for virtual reality, and more than $10^7/km^2$ connection densities with significantly lower latencies than 5G networks \cite{zhang20196g}. One of the main obstacles to wireless network evolution is the uncontrollable radio environment with reflections, diffractions, and scattering.   
Recently, reconfigurable intelligent surfaces (RISs) have emerged as promising techniques to enhance wireless signal propagation \cite{ertu}.   
In particular, the core feature of RISs is to manipulate the signal propagation path by intelligently configuring numerous small elements. Each RIS element can independently tune the phase of the incident signal, creating a smart radio environment\cite{emil}. RISs not only are technically attractive but also require low energy consumption and hardware cost, making it promising to enhance the spectral efficiency for real-world deployments. 
Given these advantages, RISs can be combined with other emerging techniques, including multiple-input multiple-output (MIMO),
millimeter-wave (mmWave) communications, unmanned aerial vehicle (UAV) networks, non-orthogonal multiple access (NOMA), vehicle-to-everything (V2X) networks, and so on\cite{mu2021simultaneously,mu2021intelligent}. Many existing studies and implementations have demonstrated RIS's capability of improving network capacity, coverage, energy efficiency, and security. 

Despite their potential, integrating RISs into wireless networks will significantly increase the complexity of network management \cite{marco}. For example, each RIS element requires independent phase-shift configurations, leading to large solutions spaces for optimization algorithms. The RIS configuration is more complicated when other control variables are jointly involved, such as beamforming, spectrum allocation, NOMA decoding order, or UAV trajectory design. Therefore, advanced optimization techniques are of paramount importance to handle such complexity and take full advantage of RISs. 
Motivated by the importance of optimization techniques, this work provides a comprehensive overview of optimization techniques of RIS-aided wireless communications, including model-based, heuristic, and machine learning (ML) approaches. There are several surveys devoted to the theory, design, analyses, and applications of RISs \cite{Almo,gong,moha,mohadz,cunh,rawa}. However, this work is different from existing surveys and tutorials by systematically summarizing and analyzing the optimization techniques for RIS-aided wireless networks, providing detailed comparisons, as well as including more state-of-the-art ML techniques. Specifically, as shown in Fig. \ref{fig1}, we focus on the following aspects:

\begin{table*}[!t]
\caption{Comparison of this work with existing surveys }
\centering
\small
\setstretch{1.5}
\begin{threeparttable} 
\resizebox{1\textwidth}{!}{%
\begin{tabular}{|m{0.65cm}<{\centering}|m{0.3cm}<{\centering}|m{0.4cm}<{\centering}|m{0.5cm}<{\centering}|m{0.5cm}<{\centering}|m{0.5cm}<{\centering}|m{0.65cm}<{\centering}|m{0.25cm}<{\centering}|m{0.55cm}<{\centering}||m{0.45cm}<{\centering}|m{0.9cm}<{\centering}|m{0.75cm}<{\centering}|m{1.1cm}<{\centering}||m{1.3cm}<{\centering}|m{1.55cm}<{\centering}|m{1.7cm}<{\centering}|m{1.05cm}<{\centering}|m{1.05cm}<{\centering}|m{1.05cm}<{\centering}|m{1.4cm}<{\centering}|m{1.05cm}<{\centering}|}
\hline \rule{0pt}{8pt} 
\multirow{5}*{\makecell{Ref.}} & \multicolumn{20}{c|}{\large RIS control and optimization-related contributions\tnote{1}}\\
\cline{2-21} \rule{0pt}{15pt} 
  &\multicolumn{8}{c||}{ Model-based approaches}&\multicolumn{4}{c||}{ Heuristic algorithms} & \multicolumn{8}{c|}{ Machine learning-based methods} \\
\cline{2-21} 
 & AO & MM & SCA & BCD & SDR & SOCP & FP & BnB & CCP & Meta-heuristic & Greedy method & Matching theory& Supervised learning & Unsupervised learning & Reinforcement learning & Federated learning  & Graph learning & Transfer learning  & Hierarchical learning & Meta- learning\\
\hline
\cite{Almo} & \checkmark & \checkmark & \checkmark & \checkmark &  &  &  &   &   & &  &   & \checkmark  &   &   &   &  &   &  &\\
\hline
\cite{gong} & \checkmark & \checkmark & \checkmark  & \checkmark  & \checkmark &  &  &   &     &     &   &    & \checkmark   &    &  \checkmark  &   &  &   & & \\
\hline
\cite{moha} & \checkmark &  \checkmark &   &    & \checkmark &    &   &    &   &   & &   &   &   & \checkmark  &   &  &   & & \\
\hline
\cite{mohadz} & \checkmark & \checkmark & \checkmark & & \checkmark & \checkmark &  & &  &   &   &  &   &   & \checkmark  &   &  &   & &\\
\hline
\cite{cunh} & \checkmark & \checkmark & \checkmark & \checkmark & \checkmark &  &  \checkmark &   & \checkmark  &  \checkmark &   & &  &   & \checkmark  &  &  &   & &\\
\hline
\cite{rawa} & \checkmark  &   &   &  &  \checkmark &   &   &   &   &  &   &   &  &  &  &   &  &   & & \\ %7
\hline
\cite{kfai} &  &  &  &  &   &  &  & &  &  &  &  & \checkmark & \checkmark &  \checkmark & \checkmark  &  &   & & \\ %8
\hline
\cite{yliu} & \checkmark &  & \checkmark  &  & \checkmark &   &   & \checkmark &  &  & \checkmark &  \checkmark &  \checkmark &  \checkmark  & \checkmark  & \checkmark  &  &   & & \\  %9
\hline
\cite{zheng2022survey} & \checkmark &  \checkmark  & \checkmark  &   &   &   &   &   &  &   &   &  \checkmark & \checkmark  &  &   &   &  &   & & \\
\hline
This work &  \checkmark  &  \checkmark  &  \checkmark  &  \checkmark  &  \checkmark  &  \checkmark    & \checkmark   &  \checkmark  &  \checkmark   &  \checkmark   &   \checkmark  &  \checkmark  &  \checkmark  & \checkmark &  \checkmark   &  \checkmark   &  \checkmark  &  \checkmark   &  \checkmark & \checkmark\\
\hline
\end{tabular}}

 \begin{tablenotes}    
        \footnotesize       
        \item[1] There are many surveys and tutorials on RISs recently, but Table \ref{tab1} focuses on studies that include control and optimization sections.   
\end{tablenotes} 
      
\end{threeparttable}  
\label{tab1}
\vspace{5pt}
\end{table*}

1) Problem formulations: \blue{We first introduce the fundamental theories of RIS technology, and then provide an overview of the problem formulations for optimizing RIS-aided wireless networks,} including maximization of sum-rate/capacity, energy efficiency, user fairness, and secrecy rate, and minimization of power consumption. In addition, we consider discrete RIS phase shifts and resource management problems that include integer control variables, and imperfect channel state information (CSI) with different error model constraints.
    
2) Model-based methods: In this work, model-based methods refer to algorithms that rely on specific optimization models with full knowledge of the defined problem\footnote{Note that some machine learning algorithms are also model-based, but here we use “model-based” to best describe the common features of a type of optimization algorithms.}. Model-based algorithms usually have demanding requirements for the properties and forms of problem formulations, e.g., convexity, continuity, and differentiability. We include the following model-based algorithms for optimizing RIS-aided wireless networks: alternating optimization (AO), the majorization-minimization (MM) method, successive convex optimization (SCA), block coordinate descent (BCD), semidefinite relaxation (SDR), second-order cone programming (SOCP), fractional programming (FP) and branch-and-bound (BnB). 
    
3) Heuristic algorithms: These algorithms apply heuristic rules for problem-solving. They provide more efficient alternatives to conventional model-based methods by sacrificing optimality and accuracy for low complexity and fast solutions. Heuristic algorithms can be used to solve NP-hard problems or serve as baselines and supplements for other algorithms. In this survey, we review the convex-concave procedure (CCP) algorithm, meta-heuristic algorithms, greedy algorithms, and matching theory for optimizing RIS-aided wireless networks.

4) ML algorithms: ML algorithms are recognized as promising solutions for wireless network optimization\cite{eldar2022machine}. ML techniques do not need full knowledge of the defined problem, and they learn from data or interact with environments to find hidden patterns. We present state-of-the-art ML techniques for optimizing RIS-aided wireless networks, including supervised and unsupervised learning, reinforcement learning (RL), federated learning (FL), graph learning, transfer learning, hierarchical learning, and meta-learning. We provide in-depth analyses for algorithm features and applications towards RISs, i.e., the dataset acquisition of neural networks for RIS phase-shift optimization, and loss function definitions of unsupervised neural networks for data rate maximization. In addition, we compare model-based, heuristic, and ML approaches in terms of optimality, robustness, stability, and so on.

5) Applications and challenges towards 6G networks: We give an overview of RIS-assisted applications towards envisioned 6G networks, including NOMA, simultaneous wireless information and power transfer (SWIPT), mmWave and THz communications, nonterrestrial networks (NTNs), V2X communications, and integrated sensing and communication (ISAC). Moreover, we identify research challenges for the control and optimization of RISs.

In summary, the main contribution of this work is that we systematically survey the optimization techniques for RIS-aided wireless networks, ranging from problem formulations to the features and applications of various approaches. Our work aims to be a roadmap for researchers to optimize RIS-aided wireless networks. 
The rest of this work is organized as follows. Section \ref{sec-relat} reviews related work, while Section \ref{sec-bac} presents the problem formulations. Section \ref{sec-model}, \ref{sec-heu} and \ref{sec-ml} introduce model-based, heuristic, and ML optimization approaches, respectively, and we compare these three approaches in Section \ref{sec-compa}.
Section \ref{sec-futu} includes RIS-aided applications towards 6G networks and identifies future challenges. Finally, Section \ref{sec-con} concludes this survey.

%% file: 2_RelatedWork.tex
\section{Related Surveys}
\label{sec-relat}

There are many research directions relating to RISs, including channel modelling and estimation, signal processing, performance analysis, passive beamforming, and hardware designs. This work focuses on optimization techniques due to their paramount importance, and Table \ref{tab1} compares this work with existing surveys in terms of control and optimization-related contributions. 

Table \ref{tab1} shows that most existing works focus on model-based approaches, including AO, MM, SCA, and SDR. The main reason is that these techniques have been widely applied, e.g., using AO to decouple joint active and passive beamforming, and applying MM and SCA to approximate non-convex objectives. Then, heuristic algorithms are usually considered as low-complexity alternatives and supplements. For example, greedy algorithms are used for element-by-element RIS phase-shift control, and matching theory is applied for resource allocation. However, despite their importance, heuristic approaches are omitted in many existing surveys. Meanwhile, ML algorithms have been widely used for wireless network management, but existing surveys are limited in supervised learning and RL. In addition, some newly emerging techniques, such as graph learning and hierarchical learning, are not mentioned in existing surveys.       

More specifically, in many existing studies \cite{Almo,gong,moha,mohadz}, optimization techniques are very briefly discussed by introducing the algorithm titles that have been used in the literature, but the motivations and algorithm features are not included.  
Alghamdi \textit{et al.} overviewed optimization and performance analysis techniques of RISs, but it is limited in analyzing problem formulations \cite{rawa}. 
In \cite{kfai}, Faisal and Choi specialized in ML approaches for RIS-aided wireless networks, but model-based and heuristic approaches are not included. Besides, some state-of-the-art ML techniques, including graph learning and hierarchical learning, are not included in \cite{kfai}.  
By contrast, multiple model-based approaches are introduced in \cite{cunh} for signal processing of RISs, but many heuristic and ML techniques are not covered.    
Liu \textit{et al.} presented RIS beamforming, resource management and ML for RIS-aided wireless networks, but only RL is presented in detail \cite{yliu}. Supervised learning, unsupervised learning, and FL are briefly discussed in \cite{yliu}, while newer techniques, such as graph learning, transfer learning, and hierarchical learning, are not covered. In \cite{zheng2022survey}, Zheng \textit{et al.} surveyed the channel estimation and practical RIS control under imperfect/statistical/hybrid CSI, but some optimization techniques are not included.

This work is different from existing studies in the following aspects: 
\begin{itemize}
    \item  Control and optimization have been included in many surveys, but this work is the first to systematically investigate optimization techniques of RIS-aided wireless networks, ranging from problem formulations to steps, features, advantages, and difficulties of nearly 20 techniques.    
    \item We present in-depth analyses to apply these optimization techniques to RISs. For example, deep neural network (DNN) and deep reinforcement learning (DRL) are included in many existing surveys, but some important questions are not discussed, i.e., dataset acquisition for neural network training in RIS-aided environments, and customizing the state, action, and reward function definitions for RL-enabled RIS control. The answers to these questions are critical to taking full advantage of RISs.
    \item Finally, we present the most state-of-the-art ML techniques for optimizing RIS-aided wireless networks, e.g., graph learning, transfer learning, and hierarchical learning, which are not included in existing surveys, to the best of our knowledge. These novel techniques may bring new research directions.  
\end{itemize}
To summarize, this survey answers the following: what are the state-of-the-art techniques for optimizing RIS-aided wireless networks, and how do they cover different aspects with respect to each other?

%% file: 3_Background.tex
\section{Background and Problem Formulations for Optimizing RIS-aided Wireless Networks}
\label{sec-bac}

This section first introduces the fundamentals of RIS technology, and then it overviews the problem formulations of RIS-aided wireless networking solutions, including maximization of sum-rate/capacity, energy efficiency, user fairness, secrecy rate, and minimization of power consumption. For each objective, we summarize related works in terms of scenarios, phase-shift resolutions, channel settings, CSI, control variables, constraints, and algorithms. 
Additionally, we investigate problem formulations with integer control variables, such as discrete RIS phase shifts and resource allocation problems. Finally, imperfect CSI scenarios are discussed within deterministic and stochastic models.

\begin{figure}[!t]
\centering
\includegraphics[width=0.95\linewidth]{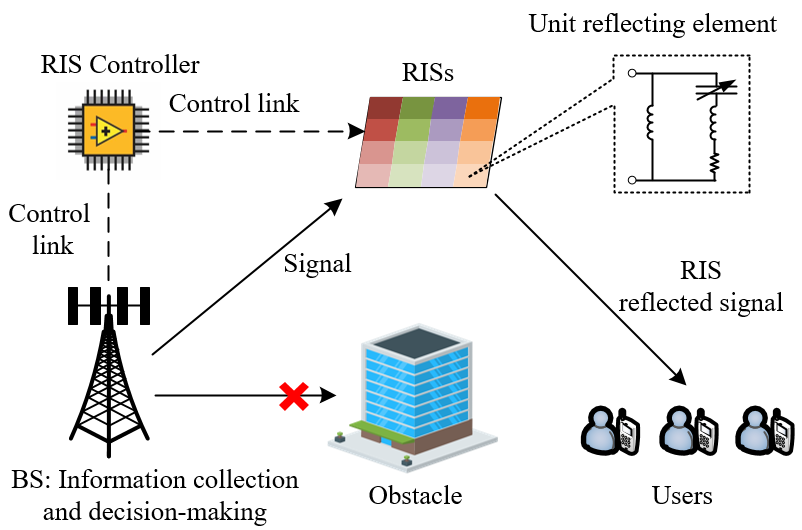}
\caption{Illustration of RIS operation and control principles.}
\label{fig-risdeploy}
\setlength{\abovecaptionskip}{-2pt} 
\vspace{0pt}
\end{figure}

\subsection{Fundamental Theories of RIS Technology}

This subsection introduces the fundamentals of RIS technologies, including RIS operation principles, RIS control, and RIS deployment. Note that there are a few studies that have explicitly introduced the fundamental principles of RISs\cite{yliu,ertu,marco}, and so this subsection serves as a brief background in our work. The reader is referred to those studies for further details.  

\subsubsection{RIS Operation Principles}
An RIS is a man-made two-dimensional reflecting surface, and the core feature of RISs is that the electromagnetic response can be intelligently reconfigured. Each RIS element can tune the phase of the incident signal, enabling a customized signal propagation environment. There have been various designs of RISs, which can be can be categorized in terms of power sources, energy consumption, tuning features, etc\cite{yliu}. For example, Fig. \ref{fig-risdeploy} shows an example of a varactor-based RIS, which applies varactor diodes with tunable biasing voltages to achieve desired phase shifts. By reflecting the incident signal from the BS, the RIS provides an alternative transmission link. 

\subsubsection{RIS Control}
A smart RIS controller is usually deployed using a device such as field-programmable gate array (FPGA). As shown in Fig. \ref{fig-risdeploy}, the RIS phase shift pattern is designed at the BS, and it will send control signals to the RIS controller for phase-shift configuration. In this case, the BS will collect the required information for decision-making. Specifically, CSI is one of the most important pieces of information for RIS phase-shift design, since RIS elements have to respond rapidly to channel dynamics. 
Meanwhile, other information may also be required, which varies between different RIS phase-shift design algorithms.
For example, Huang \textit{et al.} consider user positions as input for phase-shift design, which means the user location should be collected\cite{huang2019indoor}. 
Additionally, the communication frequency between the BS and RIS controller depends on the specific RIS design. For instance, the unit elements of passive metasurfaces will remain static during normal operation, and the controller has fewer communication demands with the BS\cite{marco}. By contrast, for RISs with active tunability, the controller may require frequent information sharing to make real-time responses to dynamic channel conditions.

\subsubsection{RIS Deployment}
Due to the low hardware cost and low energy consumption, RISs can be easily deployed on building walls and ceilings. These terrestrial RISs may be deployed in a centralized manner as a single large surface, or as decentralized surfaces that are closer to users. The centralized deployment means that more users can be covered, while the decentralized deployment has lower control complexity. In addition, RISs can also be placed on aerial platforms. For instance, UAVs with RISs can provide full space reflections with mobility and higher flexibility\cite{lu2021aerial,tekbiyik2022reconfigurable}.  

To realize the benefits of deploying RISs in wireless networks, RISs should be carefully configured, including locations, on/off status, amplitude, phase shifts, etc. In particular, phase-shift design is the key to RIS operation, which will directly affect network performance. Meanwhile, other network decisions, such as transmit beamforming, user association and resource allocation, should be jointly optimized. These joint optimization problems are usually non-convex and highly non-linear, requiring various optimization techniques for different scenarios, e.g., joint active and passive beamforming, RIS-related resource allocation, physical layer security, etc. In the following, we will introduce the problem formulations for optimizing RIS-aided wireless networks.

\begin{table*}[!t]
\caption{Summary of RIS-aided sum-rate/capacity maximization works under various constraints}
\centering
\small
\setstretch{1.15}
 \begin{threeparttable}  
\resizebox{1\textwidth}{!}{%
\begin{tabular}{|m{0.5cm}<{\centering}|m{2.1cm}<{\centering}|m{1.6cm}<{\centering}|m{1.8cm}<{\centering}|m{1.8cm}<{\centering}|m{4.8cm}<{\centering}|m{4.5cm}<{\centering}|m{3.8cm}<{\centering}|}
\hline 
Ref. & Scenario\tnote{1}  & Phase-shift resolution & Channel settings  & CSI & Control variables & Constraints & Algorithms  \\
\hline
\cite{ruoc} & MIMO-DL-MU irregular RISs &  Continuous & Rayleigh fading  & Perfect & BS beamforming matrix, RISs selection and phase shifts  &  Total transmit power, RISs selection and phase constraints  & AO, Tabu search, extraction based cross-entropy methods  \\
\hline
\cite{cunhua} & MIMO-DL-MU Multicell  &  Continuous & Rayleigh and Rician fading  & Perfect & BS beamforming vectors  and RIS phase shifts & Total transmit power and RISs phase constraints  & BCD, MM, Complex circle manifold  \\
\hline
\cite{shuowen}  &  MIMO-DL OFDM  & Continuous & Rayleigh and Rician fading &  Perfect   & Transmit covariance matrix and RIS phase shifts  & Total transmit power and RISs phase constraints  &   AO   \\
\hline
\cite{yu}  &  MIMO-UP/DL & Continuous/ Discrete &  Rayleigh fading  & Perfect &  Source precoders and RIS phase shifts  &  Total transmit power and RISs phase constraints  &  AO    \\
 \hline
\cite{ming} &  MISO-DL-MU  &  Discrete  & Correlated Rician fading  & Statistical/ Instantaneous  &  BS beamforming matrix and RIS phase shifts  &  Total transmit power and RISs phase constraints  & Penalty dual decomposition, Stochastic SCA   \\
 \hline
\cite{jiey} & MISO-DL-SU Cognitive Radio  & Continuous & Rician fading  & Perfect/ Imperfect  &  BS beamforming vectors  and RIS phase shifts   & Interference level, total transmit power, and RISs phase constraints  & BCD, SDP, SOCP \\
 \hline 
\cite{boya}  & MISO-DL-MU  & Discrete & Reflection-dominated  &  Imperfect &   BS beamforming vectors  and RIS phase shifts  & Total transmit power and RISs phase constraints  &  AO, BnB, SDR\\
 \hline
\cite{huayan}  & MISO-DL-MU & Continuous & Rayleigh and Rician fading  & Perfect/ Imperfect  & BS beamforming vectors  and RIS phase shifts  &  Total transmit power and RISs phase constraints & FP, BCD, complex circle manifold  \\
 \hline
 \cite{chang}  &  SISO-UL-SU   & Discrete  & Rayleigh fading  & Estimated &  RIS phase shifts   &  RISs phase constraint  & Successive refinement algorithm    \\
\hline
\cite{xidong} &  MISO-DL-MU NOMA   & Continuous/ Discrete & Rayleigh and Rician fading  &  Perfect  & BS beamforming vector, RIS phase shifts, and user decoding order  & Successive interference cancellation, total transmit power, and RISs phase constraints  & AO, SCA, sequential rank-one constraint relaxation    \\
 \hline
\cite{yuanbin}  &  MISO-UL-MU mmWave V2X  & Continuous &  Saleh-Valenzuela   & Imperfect &  Transmit power, multi-user detection matrix, and RIS phase shifts  &  Target SINR, total transmit power, and RISs phase constraints  &  AO, SCA, penalty CCP \\
\hline
\cite{dai2021reconfigurable}  &  MIMO-DL-MU  &  Continuous/ Discrete &  Rician fading  & Perfect &  RIS phase shifts  &  RISs phase constraint  &  Particle swarm optimization (PSO) \\
\hline
\cite{zhi2021statistical}  &  MIMO-DL-MU  & Discrete & Rayleigh and Rician fading  & Statistical &  RIS phase shifts  &  RISs phase constraint  & Genetic algorithm   \\
\hline
\cite{rivera}  & Point to point communication &  Discrete & Rician fading  & Imperfect & RIS phase shifts  & RISs phase constraint  & Greedy algorithm  \\
\hline
\cite{ni2021resource}  &  Multi-cell NOMA   & Continuous & Rayleigh and Rician fading  & Perfect  & User association, subchannel assignment, reflection matrix, power allocation, and decoding order.  & Target data rate, total transit power, association, decoding, and RISs phase constraints. &  Matching method, convex upper bound substitution, SCA, and SDR. \\
\hline
\cite{chen2021qos}  &  V2X communications    &  Continuous & Rayleigh and Rician fading    &  Large-scale 
 and slowly varying   &  Transmit power, multi-user detection matrix, spectrum sharing, and RIS phase shifts    &  Maximum transmit power, RISs phase, QoS requirement, and spectrum sharing protocol    &  BCD, SDR, CCP \\
\hline
\end{tabular}}

 \begin{tablenotes}    
        \footnotesize       
        \item[1]MISO: multiple input single output; DL/UL: downlink/uplink; SU/MU: single user/ multiple users.  
\end{tablenotes} 
      
\end{threeparttable}  
\label{tab-rate}
\vspace{-15pt}
\end{table*}

\begin{figure}[!t]
\centering
\includegraphics[width=0.9\linewidth]{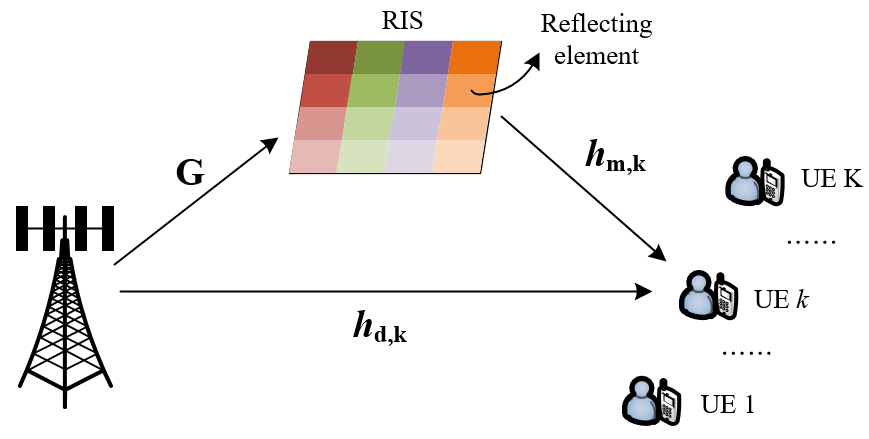}
\caption{Downlink channel of RIS-aided multi-user systems.}
\label{fig-ris}
\setlength{\abovecaptionskip}{-2pt} 
\vspace{0pt}
\end{figure}

\subsection{Sum-rate/Capacity Maximization} 
\label{sec-sub-rate}
Fig. \ref{fig-ris} shows a RIS-aided downlink transmission system, in which one base station (BS) with $M$ antennas serves $K$ single-antenna users, and the RISs have $N$ reflecting elements. The users can receive signals by direct transmission link BS-user and indirect transmission link BS-RIS-user. The signal-to-interference-plus-noise ratio (SINR) of user $k$ is:
\begin{equation}\label{eq7}
\gamma_{k}=\frac{|(\bm{h}^{R}_{k}\bm{\Theta}\bm{G}+\bm{h}^{D}_{k})p_{k}|^2}{\sum\limits_{j=1,j\neq k}^{K}|(\bm{h}^{R}_{k}\bm{\Theta}\bm{G}+\bm{h}^{D}_{k})p_{j}|^2+N_{0}^2},
\end{equation}
where $p_{k}$ is the transmit power at the BS for active beamforming, $\bm{G}\in \mathbb{C}^{N \times M}$ indicates the channel gain from BS antennas to RIS elements, $h^{R}_{k} \in \mathbb{C}^{1 \times N}$ indicates the channel gain from RIS elements to user $k$, $h^{D}_{k} \in \mathbb{C}^{1 \times M}$ indicates the channel gain from BS antennas to user $k$, and $N_{0}^2$ is the noise power. The RISs reflect the signal to users via a phase-shift vector $ \theta_{n}=e^{j\phi_{n}}$ for passive beamforming, and we define a diagonal matrix $
\bm{\Theta}=\text{diag}(\theta_{1},\theta_{2},...,\theta_{n},...,\theta_{N})\in \mathbb{C}^{N\times N}$. \blue{Here RISs may have different operation modes to change the phase shifts, which depend on their specific designs. For instance, RIS elements can be reconfigured electrically, mechanically, or thermally based on their tuning design, and a thorough survey can be found in \cite{marco}. However, note that RISs cannot be completely passive due to their inherent property of being configurable\cite{yliu}. In addition, note that the BS in Fig. \ref{fig-ris} can serve multiple users simultaneously due to multi-antenna BS beamforming, and RIS elements are used to reflect the incident signal and change the phases\cite{yu2007transmitter}.}

Sum-rate/channel capacity improvement is one of the most widely considered advantages of RIS. Compared with the direct transmission BS-user, RISs provide an indirect link that can be line-of-sight, leading to less path loss and higher SINR. To maximize the weighted sum-rate
\begin{subequations}\label{e-sum}
\begin{align}
\max\limits_{p,\bm{\Theta}}  \qquad & \sum_{k=1}^{K}w_{k} \log(1+\gamma_{k})  & \tag{\ref{e-sum}} \\
 \text{s.t.}  \qquad & \sum_{k=1}^{K}||p_{k}||^2\leq P_{max}, & \label{e-sum1}\\
  \qquad &|\theta_{n}|=1, n=1,2,...,N,  & \label{e-sum2}
\end{align}
\end{subequations}
where $P_{max}$ is the maximum transmission power of BS, $w_{k}$ is the weight of user $k$. Equation (\ref{e-sum}) aims to maximize the sum-rate of all $K$ users, and equation (\ref{e-sum1}) is the transmission power constraint. Equation (\ref{e-sum2}) is the RISs phase constraint which can be continuous or discrete.

Table \ref{tab-rate} summarizes sum-rate/channel capacity maximization works in RIS-aided wireless communications, including scenarios, phase-shift resolutions, channel settings, CSI availability, control variables, constraints and algorithms. This problem has been investigated in various scenarios, i.e., MIMO\cite{ruoc,cunhua,shuowen,yu}, MISO\cite{ming,jiey,boya,huayan }, SISO\cite{chang}, NOMA\cite{xidong}, mmWave\cite{yuanbin} and vehicle communications\cite{yuanbin}.
These works mainly consider BS beamforming vectors and RIS phase shifts as control variables, which are known as joint active and passive beamforming, and the total transmit power and RIS phases are included as constraints. However, the joint optimization problem is very challenging due to network dynamics and the large number of RIS elements. The fractional terms of SINR, logarithm introduced by Shannon theory, and non-convex constraint of RIS phase shifts lead to significant complexity for the joint optimization. Consequently, most existing studies decouple the control variables using AO\cite{ruoc,cunhua, shuowen,yu,boya}. For example, a widely applied scheme is to first optimize the BS beamforming vectors, then solve the RIS phase shifts sub-problem iteratively. Finally, as one of the core control variables, the RIS phase-shift control is supposed to be continuous in many studies. The first reason is that discrete control variables lead to integer constraints that are NP-hard; another reason is that the achieved results can be converted into the nearest discrete values using the rounding method\cite{gang}. 

\begin{table*}[!t]
\caption{Summary of RIS-aided power minimization works under various constraints}
\centering
\small
\setstretch{1.05}
\resizebox{1\textwidth}{!}{%
\begin{tabular}{|m{0.5cm}<{\centering}|m{2.1cm}<{\centering}|m{1.5cm}<{\centering}|m{1.8cm}<{\centering}|m{1.2cm}<{\centering}|m{3.9cm}<{\centering}|m{5cm}<{\centering}|m{3.5cm}<{\centering}|}
\hline 
Ref. & Scenario  & Phase-shift resolution & Channel settings & CSI & Control variables & Constraints & Algorithms  \\
\hline
\cite{qingqing} &   MISO-DL-MU & Continuous & Rician fading  & Perfect &  AP beamforming vector and RIS phase shifts  &  SINR and RISs phase constraints  & AO, SDR, two-stage algorithm  \\
\hline
\cite{qingqing2} & MISO-DL-MU &  Continuous/ Discrete  & Rician fading   &  Perfect  &   AP beamforming vector and RIS phase shifts  & SINR and RISs phase constraints  &  Successive refinement algorithm, BnB   \\
\hline
\cite{guizhou2} & MISO-DL-MU broadcast  &  Continuous  & Rayleigh and Rician fading  &  Imperfect   & BS precoding vector and RIS phase shifts  & SINR and RISs phase constraints  & AO, SDR  \\
\hline
\cite{huimei} & MISO-DL-MU broadcast  &  Continuous & Rayleigh fading   & Perfect  &  BS beamforming vectors  and RIS phase shifts  &  SNR and RISs phase constraints  &  SDR, SCA  \\
\hline
\cite{guiz3} &  MISO-DL-MU broadband  &  Continuous & Rayleigh fading  & Imperfect  & BS precoder matrix and RIS phase shifts  & Target transmission rate, outage probability, and RISs phase constraints   &  AO, SDR, penalty CCP   \\
\hline
\cite{beixiong} & SISO-DL-MU NOMA/OMA  & Discrete & Rayleigh fading  & Perfect   &  BS transmit power allocation and RIS phase shifts   &  Target transmission rate and RISs phase constraints   & AO, linear approximation,    \\
\hline
\cite{yiqing} & MISO-DL-MU NOMA & Continuous/ Discrete & Rician fading &  Perfect  & BS beamforming vectors and RIS phase shifts  & Target transmission rate and RISs phase constraints  &  SOCP, ADMM   \\
\hline
\cite{minf} &  MISO-DL-MU NOMA  &   Continuous & Rayleigh and Rician fading  & Perfect   & BS beamforming vectors  and RIS phase shifts  &  Target transmission rate and RISs phase constraints  &    AO, successive convex relaxations   \\
\hline
\cite{jianyue} & MISO-DL-MU NOMA  & Continuous & Rayleigh distribution &   Perfect  &  BS beamforming vectors  and RIS phase shifts  & Target transmission rate and RISs phase constraints   & SDR, FP, SCA \\
\hline
\cite{yang2021optimal} & Full-duplex transmission  & Continuous & Rayleigh fading  &  Perfect  & Power allocation of signal sources and RIS phase shifts  & Target transmission rate and RISs phase constraints   & Lagrangian dual method, SDP \\
\hline
\cite{zhiyang} &   MISO-DL-MU edge computing  &  Continuous & Rayleigh fading  &  Perfect  &  Offloading bits, transmission time, power allocation, and RIS phase shifts & Computation ability, NOMA rate region, time constraint, offloading bits, and total transmit power & AO, BCD \\
\hline
\cite{cao2022joint} &  MISO-DL-SU maritime communication   & Continuous & Rician fading  & Estimated  &  Mode selection, transmission power, BS beamforming vector, and RIS phase shifts   &  Rate threshold, binary selection, maximum transmit power, and RISs phase constraints   & AO, exhaustive search, matching method, SDP     \\
\hline
\end{tabular}}
\label{tab-power}
\vspace{-10pt}
\end{table*}

\subsection{Power Minimization} 
Power minimization is another widely investigated topic for RIS-aided wireless communications. Transmission power reduction not only saves the power consumption of wireless networks but also reduces the interference on adjacent cells. Power minimization problems with QoS constraints can be described by 

\begin{subequations}\label{e-min}
\begin{align}
\min\limits_{p,\bm{\Theta}}  \qquad &  \sum_{k=1}^{K}||p_{k}||^2 & \tag{\ref{e-min}} \\
 \text{s.t.}  \qquad & \gamma_{k} \geq \gamma_{min}, k=1,2,...,K,  & \label{e-pmin1}\\
  \qquad &|\theta_{n}|=1, n=1,2,...,N,  & \label{e-min2}
\end{align}
\end{subequations}
where $\gamma_{min}$ is the minimum SINR requirement of users.  The objective function is greatly simplified by minimizing power consumption $\sum_{k=1}^{K}||p_{k}||^2$, and the quality of service (QoS) requirements are balanced by SINR constraints as $\gamma_{k} \geq \gamma_{min}$. Here, the SINR or date-rate requirements introduce fractional or logarithmic terms in the constraints, which makes equation (\ref{e-pmin1}) a challenging non-convex problem. 

We summarize power minimization-related works in Table \ref{tab-power}. These studies aim to minimize the transmit power of BSs with SINR or data rate constraints. Similar to the sum-rate maximization problem, power minimization is also investigated in diverse scenarios, including MISO\cite{qingqing,qingqing2,guizhou2,huimei}, SISO\cite{beixiong}, NOMA\cite{yiqing,beixiong,minf,jianyue} and full-duplex antennas \cite{yang2021optimal}. 
Meanwhile, these studies still assume continuous RIS phase shifts and perfect CSI to reduce the optimization complexity\cite{qingqing,huimei,minf,jianyue}, and the main control variables are BS beamforming vectors and RIS phase shifts. BnB and successive refinement algorithm are applied in \cite{qingqing2} for discrete optimization, and Zheng \textit{et al.} obtain the optimal RIS phase shifts first and then finds the nearest discrete value \cite{beixiong}. In addition, imperfect CSI is investigated in \cite{guizhou2,guiz3}, in which \cite{guizhou2} applies penalty CCP to handle the CSI uncertainty, and S-procedure and Bernstein-Type inequality are used in \cite{guiz3} to transform the QoS constraints under CSI error. Perfect CSI is a common setting in many existing studies, but such strong assumptions may be impractical in real-world applications. 
On the one hand, more advanced channel estimation methods should be developed to reduce the CSI estimation error\cite{zheng2022survey}; on the other hand, robust optimization algorithms are expected to handle the channel uncertainty.

\begin{table*}[!t]
\caption{Summary of RIS-aided energy efficiency maximization works under various constraints}
\centering
\small
\setstretch{1.05}
\resizebox{1\textwidth}{!}{%
\begin{tabular}{|m{0.5cm}<{\centering}|m{2.1cm}<{\centering}|m{1.5cm}<{\centering}|m{1.4cm}<{\centering}|m{1cm}<{\centering}|m{3.8cm}<{\centering}|m{5cm}<{\centering}|m{3.5cm}<{\centering}|}
\hline 
Ref. & Scenario  & Phase-shift resolution & Channel settings & CSI & Control variables & Constraints & Algorithms  \\
\hline
\cite{chongwen} &  MISO-DL-MU  &  Continuous & 3GPP  &  Perfect  &  BS beamforming vectors  and RIS phase shifts  & Target data rate, total transmit power, and RISs phase constraints  &  AO, gradient descent, MM \\
\hline
\cite{linsong} &   MISO-DL-MU multicast  & Continuous & Rician fading & Perfect/ Estimated  &   BS covariance matrix and RIS phase shifts & Total transmit power and RISs phase constraints  & AO, FP  \\
\hline
\cite{shuaiqi} & SISO-MU D2D  &  Discrete  & Rician fading  &  Perfect   & D2D transmitter power and RIS phase shifts &  Target data rate, total transmit power, and RISs phase constraints   & AO, FP, Dinkelbach method  \\
\hline
\cite{zhaohui2} & MISO-DL-MU rate splitting   & Continuous & Rayleigh fading  & Perfect  &  BS beamforming vector, RIS phase shifts, and user message rate   &  Common message decoding, target data rate, total transmit power, and RISs phase constraints  &  AO, SCA  \\
\hline
\cite{zhaohui}  &   MISO-DL-MU  & Continuous  & Rayleigh fading  &  Perfect  & BS beamforming vector, RIS phase shifts and on/off  & Target data rate, total transmit power, RIS on/off status, and RISs phase constraints    & AO, SCA, greedy searching    \\
\hline
\cite{yutong} & Cell-free MIMO-DL-MU  & Discrete & Rician fading  & Perfect  & BS beamforming vectors  and RIS phase shifts  & Total transmit power and RISs phase constraints  &  AO \\
\hline
\cite{zhang2020joint} &  SISO-DL-MU NOMA   &  Continuous & Rician fading  &  Perfect   & Subcarrier allocation, BS beamforming vector, and RIS phase shifts  &  Subcarrier assignment, total transmit power, and RISs phase constraints   &  AO, matching method, DC programming, and univariate search. \\
\hline
\end{tabular}}
\label{tab-energy}
\vspace{-10pt}
\end{table*}

\subsection{Energy Efficiency Maximization} 
Energy efficiency is a critical metric for green 5G and 6G networks. Different from power minimization problems, the objective of energy efficiency maximization includes both transmission rate and energy consumption metrics, which can better evaluate the power utilization efficiency. The main benefit of RISs lies in the capability of reshaping the signal propagation path with extremely low power consumption, making RISs a promising technique to improve energy efficiency. To maximize energy efficiency, one can formulate
\begin{subequations}\label{e-ee}
\begin{align}
\max\limits_{p,\bm{\Theta}}  \qquad & \frac{\sum_{k=1}^{K}w_{k} \log(1+\gamma_{k})}{\sigma^{-1} \sum_{k=1}^{K}||p_{k}||^2+KP_{UE}+P_{BS}+NP_{R}(\varrho)}  & \tag{\ref{e-ee}} \\
 \text{s.t.}  \qquad & \sum_{k=1}^{K}||p_{k}||^2\leq P_{max}, & \label{e-ee1}\\
  \qquad &|\theta_{n}|=1, n=1,2,...,N,  & \label{e-ee2}\\
  \qquad & \gamma_{k} \geq \gamma_{min}, k=1,2,...,K,  & \label{e-ee3}
\end{align}
\end{subequations}
where $\sigma$ is the efficiency of the transmit power
amplifier, $P_{UE}$ is the hardware static power consumed by one user, $P_{BS}$ is the total hardware static power consumption in BS, and $P_{R}(\kappa)$ is the power consumption of one RIS reflecting element with resolution $\varrho$. To maximize energy efficiency in equation (\ref{e-ee}), the numerator is to maximize the sum-rate, while the denominator is to reduce power consumption. Constraint (\ref{e-ee1}) is the total transmit power limit, (\ref{e-ee2}) is the RIS phase constraint, and (\ref{e-ee3}) is the QoS requirement indicated by target SINR or data rate. 

Problem (\ref{e-ee}) is more complicated than sum-rate maximization or power minimization problems, since it includes both logarithm and fractional terms in the objection function and constraints. 
Energy efficiency maximization-related works are summarized in Table \ref{tab-energy}. Similar to former problem formulations, BS beamforming vectors and RIS phase shifts are main control variables, and constraints include target data rate, total transmit power, and RIS phases\cite{chongwen,linsong,yutong}. 
To solve the energy efficiency maximization problem, the key is to decouple the transmission rate and energy consumption items in equation (\ref{e-ee}). Dinkelbach method is applied in \cite{shuaiqi}, but these conventional methods can not be directly applied to sum-ratio problems.
Note that the power consumption definition in equation (\ref{e-ee}) may change case by case, which depends on the scenario of deploying RISs. For example, the total power consumption in most studies include BS transmit power, RIS energy consumption and user device power. By contrast, Zhou \textit{et al.} investigate the energy efficiency with BS sleep control, and the total BS power consumption becomes the denominator in the objective of equation (\ref{e-ee})\cite{zhou2023hierarchical}.

\begin{table*}[!t]
\caption{Summary of RIS-aided fairness maximization works under various constraints }
\centering
\small
\setstretch{1.05}
\resizebox{1\textwidth}{!}{%
\begin{tabular}{|m{0.5cm}<{\centering}|m{2.2cm}<{\centering}|m{1.5cm}<{\centering}|m{1.5cm}<{\centering}|m{1.3cm}<{\centering}|m{3.5cm}<{\centering}|m{5cm}<{\centering}|m{3.5cm}<{\centering}|}
\hline 
Ref. & Scenario  & Phase-shift resolution & Channel setting & CSI & Control variables & Constraints & Algorithms  \\
\hline
\cite{guiz} & MISO-DL-MU multicast  & Continuous /Discrete & Rayleigh and Rician fading   & Perfect &  BS precoding matrix and RIS phase shifts  & Total transmit power and RISs phase constraints  & SOCP, MM     \\
\hline 
\cite{gang} &  SISO/MISO-DL-MU NOMA  & Continuous & Rayleigh and Rician fading  &  Perfect  & BS power allocation and RIS phase shifts  &  Target SINR, channel strength, total transmit power, and RISs phase constraints   & BCD, SDR   \\
\hline
\cite{qurrat} & MISO-DL-MU   &  Continuous & Rayleigh fading  & Perfect  & RIS phase shifts  & RIS phase constraint  & Projected gradient ascent \\
\hline
\cite{hailiang} & MISO-DL-MU  & Continuous  & Rayleigh and Rician fading &  Perfect  & BS beamforming vectors  and RIS phase shifts  & Total transmit power and RISs phase constraints  & SOCP, SDR, SCA  \\
\hline
\cite{manij} & Cell-free MIMO-UL-MU  & Continuous & Rician fading  &  Estimated  & User transmit power and RIS phase shifts  & User transmit power and RISs phase constraints  & AO   \\
\hline
\cite{shiqi} & MIMO-DL-MU SWIPT   & Discrete  & Rician fading  &  Perfect  & BS beamforming vectors  and RIS phase shifts   & Total transmit power, target harvested energy, and RISs phase constraints  & AO, optimal BnB, reformulation-linearization  \\
\hline
\cite{zaid} & Full-Duplex relay networks & Continuous  & Rayleigh fading  & Perfect  &  Transmit power and RIS phase shifts  & Total transmit power and RISs phase constraints & AO, SDR \\
\hline
\cite{menghua} & MIMO-DL-MU coordinated multi-point  & Continuous & Rayleigh and Rician fading  & Perfect   &   BS beamforming vectors  and RIS phase shifts  & Total transmit power and RISs phase constraints  & SOCP, SDR, MM   \\
\hline
\cite{zhi2022power} & MIMO-UL  & Continuous /Discrete &  Rician fading  &   Statistical  &  RIS phase shifts  &  RIS phase constraint  &   Genetic algorithm.   \\
\hline
\cite{Atapattu} & Point to point communications &  Continuous & Rayleigh fading  &  Estimated  & RIS phase shifts  & Target SINR and RISs phase constraints    &  SDR, greedy-iterative method   \\
\hline
\end{tabular}}
\label{tab-fair}
\vspace{-10pt}
\end{table*}

\subsection{ User Fairness Maximization} 
\label{sec-sub-fair}
Former problem formulations usually consider the total or average network performance as objectives, but the fairness among multiple users is equally important. Such user fairness metrics can describe the experience of cell-edge users, guaranteeing the worst-case network performance.
User fairness maximization aims to maximize the minimum SINR or data rate, indicating that users can achieve target performance even in the worst case.  
For instance, the max-min SINR problem can be defined by
\begin{subequations}\label{e-fa}
\begin{align}
\max\limits_{p,\bm{\Theta}}  \qquad & \min\limits_{k \in K}\{\gamma_{k}\}  & \tag{\ref{e-fa}} \\
 \text{s.t.}  \qquad & \sum_{k=1}^{K}||p_{k}||^2\leq P_{max}, & \label{e-fa1}\\
  \qquad &|\theta_{n}|=1, n=1,2,...,N,  & \label{e-fa2}\\
  \qquad & \gamma_{k} \geq \gamma_{min}, k=1,2,...,K,  & \label{e-fa3}
\end{align}
\end{subequations}
where $\min\limits_{k \in K}\{\gamma_{k}\}$ is the minimum SINR among $K$ users. Maximizing the worst user experience will improve the fairness of the whole network.
  
Table \ref{tab-fair} summarizes existing works for fairness maximization problems in RIS-aided wireless networks. Although various scenarios have been investigated, the primary control variables are still BS beamforming vectors and RIS phase shifts\cite{guiz,gang,hailiang,menghua}, and the constraints focus on total transmit power and RIS phases\cite{guiz,hailiang,zaid,menghua}. The formulated problem (\ref{e-fa}) is more challenging than conventional max-min fairness beamforming problems. RISs not only introduce additional non-convex constraints but also makes the reflective beamforming vector coupled with the transmit beamforming vectors in the SINR term, thus making problem (\ref{e-fa}) highly nonlinear and non-convex\cite{hailiang}. 
Therefore, instead of optimizing the complicated objective functions directly, approximation-based algorithms are frequently applied. For example, MM and SCA algorithms construct surrogate functions as the upper bound of original objective functions, which are easier to be optimized than problem (\ref{e-fa}). Additionally, the genetic algorithm (GA) and greedy-iterative method are used in \cite{zhi2022power} and \cite{Atapattu} for RIS phase-shift control, respectively. Compared with model-based algorithms, these heuristic algorithms can obtain fast solutions efficiently, but the optimality cannot be guaranteed.

\subsection{Secrecy Rate Maximization} 
\label{sec-sub-sec}

Physical layer security is increasingly of interest for wireless communications, and various
techniques have been proposed to enhance physical layer security, e.g., artificial noise-aided beamforming and cooperative jamming. However, these approaches may lead to high hardware costs and power consumption, and RISs provide a novel low-cost solution by manipulating the signal propagation path. 

\begin{figure}[!t]
\centering
\includegraphics[width=0.8\linewidth]{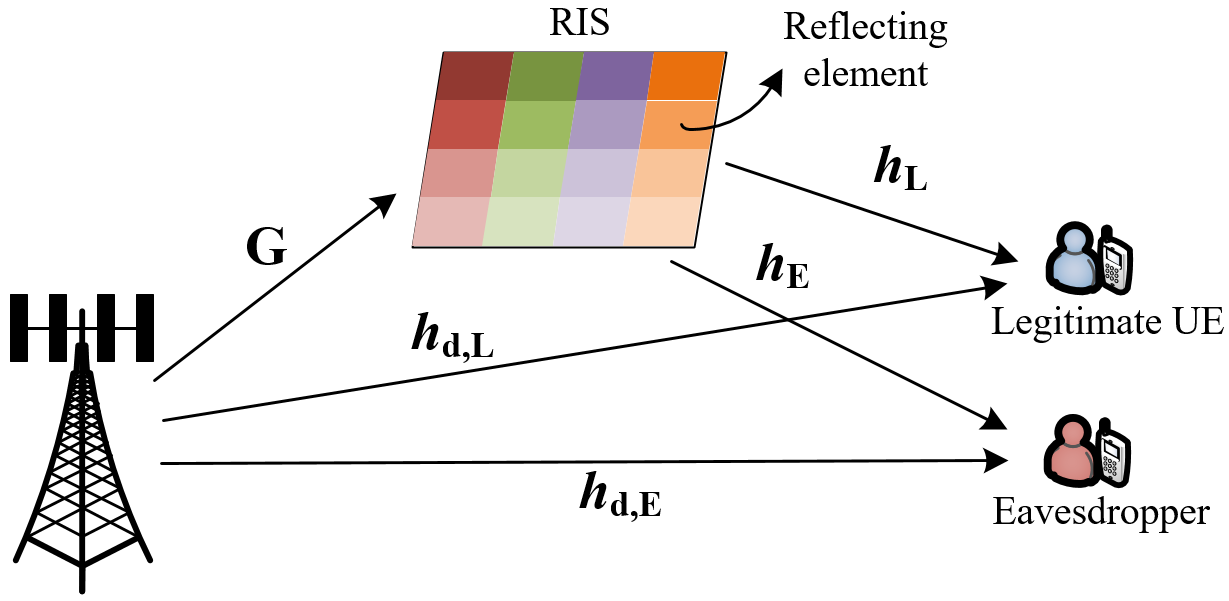}
\caption{The downlink channel model of RIS-aided secure transmission.}
\label{fig-secrecy}
\setlength{\abovecaptionskip}{-2pt} 
\vspace{-10pt}
\end{figure}

Fig. \ref{fig-secrecy} shows a RIS-aided downlink transmission system with one legitimate user and one eavesdropper.  
The data rate of the legitimate user is: 
\begin{equation}\label{eq-legi}
R_{L}=\log(1+\frac{|(\bm{h}_{L}\bm{\Theta}\bm{G}+\bm{h}_{d,L})p_{L}}|^2{N_{L,0}^2}),
\end{equation}
where $\bm{h}_{L}$ is the channel from RISs to the legitimate user, $\bm{h}_{d,L}$ is the direct transmission channel from RISs to the legitimate user, and $N_{L,0}$ is the Gaussian noise at the legitimate user.    

Similarly, the data rate of the eavesdropper is: 
\begin{equation}\label{eq-legi2}
R_{E}=\log(1+\frac{|(\bm{h}_{E}\bm{\Theta}\bm{G}+\bm{h}_{d,E})p_{L}}|^2{N_{E,0}^2}),
\end{equation}
where $\bm{h}_{E}$ is the channel from RISs to the eavesdropper, $\bm{h}_{d,E}$ is the direct transmission channel from RISs to the eavesdropper, and $N_{E,0}$ is the Gaussian noise at the eavesdropper. Finally, to maximize the secrecy rate, we have
\begin{subequations}\label{e-sec}
\begin{align}
\max\limits_{p,\bm{\Theta}}  \qquad & R_{L}-R_{E}  & \tag{\ref{e-sec}} \\
 \text{s.t.}  \qquad & ||p||^2\leq P_{max}, & \label{e-sec1}\\
  \qquad &|\theta_{n}|=1, n=1,2,...,N.  & \label{e-sec2}
\end{align}
\end{subequations}

Table \ref{tab-secure} summarizes the RIS-aided secure transmissions-related studies. It shows that most existing works apply continuous phase shifts with perfect CSI sharing, and BS beamforming and RIS phase shifts are still the major control variables. Most existing studies consider single-user cases\cite{zheng,hong,xianghao,biqian,miao,zijie,huiming2}, which significantly lower the optimization complexity without considering interference between multiple legitimate users. Moreover, the perfect CSI acquisition of legitimate users may be unrealistic in practice, indicating a gap between theoretical studies and real-world applications. 

On the other hand, RIS technique is also combined with conventional secure transmission strategies, i.e., artificial noise, to achieve secure transmissions\cite{wei,xianghao}, and the simulations demonstrate that integrating RIS with artificial noise can achieve a higher secure transmission rate. 
However, these coupled control variables also increase the optimization complexity.
The main difficulties of solving problem (\ref{e-sec}) are the non-convex objective function and RIS phase-shift constraints (\ref{e-sec2}). Subsequently, AO-based estimation methods are widely applied as low-complexity solutions. For instance, the original problem is decoupled into multiple sub-problems by optimizing one control variable and holding other variables fixed, and then applying approximation-based algorithms, e.g., SCA\cite{wei,xianghao,zijie} and MM\cite{limeng,huiming2}, to solve each sub-problem.

\begin{table*}[!t]
\caption{Summary of RIS-aided secure transmission-related works under various constraints}
\centering
\small
\setstretch{1.05}
\resizebox{1\textwidth}{!}{%
\begin{tabular}{|m{0.5cm}<{\centering}|m{2.2cm}<{\centering}|m{1.5cm}<{\centering}|m{1.5cm}<{\centering}|m{1.3cm}<{\centering}|m{3.9cm}<{\centering}|m{5cm}<{\centering}|m{3.5cm}<{\centering}|}
\hline 
Ref. & Scenario  & Phase-shift resolution & Channel setting  &  CSI & Control variables & Constraints & Algorithms  \\
\hline
\cite{limeng} \cite{limeng2} & MIMO wiretap channel  &  Continuous & Rayleigh fading   &  Perfect/ Imperfect  &  Transmit covariance matrix and RIS phase shifts  & Total transmit power and RISs phase constraints   & AO, one-by-one optimization, bisection search, MM.   \\
\hline
\cite{zheng} &  MISO-DL-SU &  Continuous & Rayleigh fading  & Estimated  &  Transmit
beamformer and RIS phase shifts & Secure transmission rate and RISs phase constraints   & AO, SDR  \\
\hline
\cite{wei} & MISO-DL-MU SWIPT & Continuous & Rayleigh fading  &  Perfect  & BS transmit beamforming and artificial noise covariance matrix, and RIS phase shifts.  & Harvested power, total transmit power, and RISs phase constraints  & AO, SDR, SCA  \\
\hline
\cite{hong} & MISO-DL-SU  & Continuous & Rayleigh fading  & Perfect  & Transmit beamformer and RIS phase shifts   & Total transmit power and RISs phase constraints   & AO  \\
\hline
\cite{xianghao} & MISO-DL-SU  &  Continuous & Rician fading  &  Perfect &  AP transmit beamforming and artificial noise covariance matrix, and RIS phase shifts.  &  Total transmit power, RISs phase, and secrecy rate constraints & AO, penalty-based approach, SCA, SDR   \\
\hline
\cite{biqian} & MISO-DL-SU  & Continuous  & Rician fading  & Perfect/ Imperfect  & BS transmit beamforming vector and RIS phase shifts & Secrecy capacity requirement and RISs phase constraints   &  SDR, projected gradient algorithm   \\
\hline
\cite{miao} & MISO-DL-SU  & Continuous  & Correlated Rician fading  & Perfect   & AP transmit beamforming
vector and RIS phase shifts  & Total transmit power and RISs phase constraints   &  AO \\
\hline
\cite{zijie} & SISO-DL-SU & Continuous  & Rician fading  &  Perfect  & RIS phase shifts  &  RISs phase constraint   &  AO, SDR, SCA  \\
\hline
\cite{huiming2} & MISO-DL-SU & Continuous  & Rayleigh fading  & Partial & Transmit beamforming vector and RIS phase shift  &  SNR and RISs phase constraints   &  Oblique manifold algorithm, MM  \\
\hline
\cite{jie} & MISO-DL-MU &  Continuous/ Discrete & Rician fading  & Perfect   & BS transmit
beamformer, and RIS phase shifts &  Total transmit power and RISs phase constraints  &  AO based path-following algorithm, heuristic algorithm \\
\hline
\cite{niu2021simultaneous} & MISO-DL-MU    &  Continuous & Rician fading  &  Perfect   & BS transmit beamforming and RIS phase shifts   &  Total transmit power, RIS phase, and energy conservation constraints  &  AO, CCP, one dimension search     \\
\hline
\end{tabular}}
\label{tab-secure}
\vspace{-10pt}
\end{table*}

\subsection{ Optimization with Integer Constraints: Discrete RIS Phase Shifts and Resource Allocation Problems } 
\label{sec-sub-dis}

Previous sections show that many existing studies assume continuous RIS phase shifts for simplicity, but practical RISs usually have limited phase-shift resolutions, indicating discrete phase shifts for the incident signal. However, such realistic settings will lead to discrete control variables along with mixed integer nonlinear programming (MINLP) problems, considerably increasing the difficulty of optimization. 
In addition, RIS element on/off control, resource allocation, and association problems will also involve integer control variables. 

Compared with former problem formulations, the main difference is that discrete control variables lead to integer constraints. For instance, discrete RIS phase shifts include a constraint $\theta_n \in \{0, \frac{2\pi}{2^\varrho},...,(2^\varrho-1)\frac{2\pi}{2^\varrho},2\pi\}$, where $\varrho$ is the RIS phase-shift resolution. RIS elements on/off, subchannel allocation, and user association problems will involve binary constraints as $\chi \in \{0,1\}$, where $\chi$ is the binary decision variable.  
We summarize three approaches to formulate optimization problems with integer control variables: 
\begin{itemize} 
    \item \textbf{Relaxation method}: This method is to relax the discrete RIS phase shifts $\theta_n \in \{0, \frac{2\pi}{2^\varrho},...,(2^\varrho-1)\frac{2\pi}{2^\varrho},2\pi\}$ into continuous phase shifts with $0\leq \theta_n < 2\pi$. Similarly, the binary control variables in resource allocation and association problems with $ \chi \in \{0,1\}$ are converted into $0\leq \chi \leq 1$. Such linear programming relaxation can transform NP-hard problems into related problems that may be solvable in polynomial time. In addition, the relaxation method may introduce penalties in the objective function by allowing violating constraints such as Lagrangian relaxation. Then, the reformulated problem formulations are solved by using algorithms such as SCA or MM. These methods will be included in Section \ref{sec-model}.
    \item \textbf{Quantization approach}: The quantization method is mainly used to simplify the RIS phase-shift control. It considers continuous RIS phase shifts when solving the problem, and then the achieved optimal RIS phase shifts are quantized into the closest discrete values as $\theta_n \in \{0, \frac{2\pi}{2^\varrho},...,(2^\varrho-1)\frac{2\pi}{2^\varrho},2\pi\}$. For example, the authors in \cite{gang} maximize the received signal strength by using BCD and SDR, and the achieved continuous RIS phase shifts are easily converted into the nearest discrete values. Compared with the relaxation method, the quantization approach has much lower complexity with few reformulations. However, the solution quality may be considerably degraded when quantizing the continuous values into discrete solutions. For example, the quantization approach may have difficulty handling binary decision problems since all continuous solutions between 0 and 1 can only be quantized into values 0 or 1. In this case, the relaxation method is more appropriate for binary decision-making.   
    \item \textbf{Heuristic and ML techniques}: Heuristic and ML techniques also provide attractive solutions for MINLP problems. Discrete control variables are directly optimized without relaxation and transformation, which will be introduced in Section \ref{sec-heu} and \ref{sec-ml}. 
\end{itemize}

\begin{table*}[!t]
\caption{Summary of discrete control variables for optimizing RIS-aided wireless networks}
\centering
\setstretch{1.05}
\small
\resizebox{1\textwidth}{!}{
\begin{tabular}{|m{2.5cm}<{\centering}|m{11.5cm}<{\centering}|m{5cm}<{\centering}|}
\hline 
Discrete control variables &  Main features   &     Solution algorithms     \\
\hline
 Discrete RIS phase shifts &  Discrete phase shift indicates that the phase of each RIS element can only be selected from multiple fixed values. It is more practical than continuous phase changing, but increases the optimization complexity as MINLP problems.   &  Relaxation method, quantization method\cite{gang}, BnB\cite{boya}, heuristic and ML algorithms\cite{zhi2022power,liu2020machine}.   \\
\hline
 RIS on/off decision & RIS on/off decision is to decide the on/off status of each RIS element, which can better save energy consumption and improve energy efficiency. &     Greedy algorithms \cite{zhaohui}, dual method.\\
\hline
 Association decisions &  User-RIS-BS associations are key control variables for optimizing multi-cell and multi-RIS network performance. In addition, these association decisions can apply to RIS-UAV systems for the user-UAV association.   &   \multirowcell{2}[10pt][c]{Matching theory is the most widely \\ used method for solving association \\ and resource allocation problems\\ \cite{wu2022resource, zuo2020resource}. Meta-heuristic and \\ ML algorithms also present  \\ promising solutions. }       \\
\cline{1-2}
 Resource allocation \qquad decision &  Resource allocation is a key decision for optimizing RIS-aided wireless network performance, which will directly decide the user experience.    &         \\
\hline
 D2D-user pairing in RIS-aided D2D networks &  RIS provides a promising opportunity for the interference control of D2D communications, and the pairing of celluar users and D2D links is important to reuse the subchannels allocated to users.  &  Hungarian algorithm\cite{cao2021sum}, heuristic pairing\cite{yang2021reconfigurable2}.    \\
\hline
 Task offloading in RIS-aided MEC &  RIS could change the channel condition for MEC services to increase channel efficiency, and task offloading decision is one of the core decisions for MEC services.       &  BCD and SCA \cite{peng2022active}, DRL\cite{li2021joint}.      \\
\hline
\end{tabular}}
\label{tab-discrete}
\vspace{-5pt}
\end{table*}

\begin{table*}[!t]
\caption{Summary of different CSI error models for optimizing RIS-aided wireless networks}
\centering
\setstretch{1.05}
\small
\resizebox{1\textwidth}{!}{
\begin{tabular}{|m{2.5cm}<{\centering}|m{11.5cm}<{\centering}|m{5cm}<{\centering}|}
\hline 
CSI error \qquad model &  Main features   &     Solution algorithms     \\
\hline
 Deterministic model      &  Deterministic model defines an upper bound of the CSI error. Most existing studies convert the optimization problem into max-min formulations to guarantee the worst-case performance, i.e., secrecy rate maximization \cite{lu2020robust}, weighted sum-rate maximization \cite{hao2021robust}.    &  AO, SDR, SCA\cite{hao2021robust}, penalty CCP, MM\cite{limeng} \cite{limeng2}.     \\
\hline
Stochastic model   &  Stochastic model indicates that CSI error follows certain distributions without a predefined upper bound. Therefore, an outage probability constraint is applied to guarantee network performance, i.e., power minimization and secrecy rate maximization with SINR and secrecy rate outage probability constraints\cite{zhao2021outage,guo2021learning}.       &   SCA\cite{zhao2021outage}, DDPG\cite{guo2021learning}     \\
\hline
\end{tabular}}
\label{tab-imper}
\vspace{0pt}
\end{table*}

In summary, the quantization approach has the lowest complexity by transforming continuous RIS phase shifts into the nearest discrete values, but such brute-force transformation may degrade the network performance. The relaxation method converts discrete control variables into continuous optimization problems. It may guarantee optimality but require case-by-case analyses and complicated design. By contrast, heuristic and ML techniques can better handle discrete optimization problems by using heuristic rules and ML algorithms. For instance, discrete RIS phase shifts are defined as actions in \cite{liu2020machine}, and then the DRL agent interacts with the wireless environment directly to maximize the long-term benefit.

Table \ref{tab-discrete} overviews integer control variables that are involved in the optimization of RIS-aided wireless networks, including discrete RIS phase shifts, RIS on/off control, resource allocation and association, D2D-user pairing for RIS-D2D communications, and task offloading decisions for RIS-aided MEC. Table \ref{tab-discrete} shows that handling these integer control variables is critical to optimize network performance, and heuristic and ML algorithms are regarded as appealing approaches to solve these NP-hard problems.

\subsection{Optimization Constraints for Imperfect CSI}
CSI availability is critical for properly optimizing RIS-aided wireless networks. Note that there are many advanced channel estimation methods to provide accurate CSI\cite{zheng2022survey}, and then most existing studies assume perfect CSI as shown from Table \ref{tab-rate} to \ref{tab-secure}. 
However, obtaining perfect and instant CSI is impractical due to limited feedback overhead, noise, and interference. As shown in Table \ref{tab-imper}, the CSI estimation error can be described by deterministic or stochastic models.

\begin{itemize} 
\item \textbf{Deterministic model}: The deterministic model indicates an upper bound of the CSI error as $||e_{Ch}|| \leq e_{Ch,max}$, where $e_{Ch}$ is the estimation error and $\leq e_{Ch,max}$ is the upper bound. Then $||e_{Ch}|| \leq e_{Ch,max}$ will be included in the problem formulations shown from Section \ref{sec-sub-rate} to \ref{sec-sub-sec}. In addition, problem formulations with the deterministic error model will become a max-min problem to guarantee the worst-case performance, which is similar to Section \ref{sec-sub-fair}.
\item \textbf{Statistical model}: On the other hand, the statistical model considers the CSI error as a random variable with specific distributions such as complex Gaussian distribution. Without CSI error bound, an extra constraint is required to guarantee the network performance $Pr(\gamma \geq \gamma_{min} ) \geq Pr_{min}$, where $\gamma_{min}$ is the minimum SINR requirement, and $Pr_{min}$ is the minimum probability requirement that the SINR is higher than the target value. 
$Pr(\gamma \geq \gamma_{min} ) \geq Pr_{min}$ is a probabilistic constraint because there is no upper limit on the CSI error, and a large error will unavoidably lead to system outages. \end{itemize}

Finally, note that there are many algorithms that can be used to optimize problems with imperfect CSI constraints, including AO, SCA, SDR, and DRL, which will be introduced in the following sections.

\subsection{Discussions and Analyses}
Sections III-B to III-H have investigated various problem formulations for RIS-aided networks, and then this subsection aims to analyze the common features of these formulations. Especially, identifying the main challenges of solving these problems can motivate us to find more efficient solutions.

1) \textbf{Non-convex Objectives and Constraints}: One common feature of RIS-related optimization problems is that the objectives and constraints are usually non-convex and highly non-linear. For instance, fractional terms are frequently involved with SINR, and the logarithm is usually included due to Shannon's formula. \blue{These fractional and logarithmic terms lead to non-convex terms in objectives and constraints. Thus dedicated transformation and relaxation are needed to reformulate the problem for convexity, which requires case-by-case analyses for each problem formulated. }

2) \textbf{Highly Coupled Control Variables}: RIS passive beamforming is often combined with other techniques, e.g., BS active beamforming, NOMA, and UAVs. This results in highly coupled control variables, e.g., RIS phase-shift design, user decoding order in NOMA, BS transmit power control, UAV trajectory design, and so on. \blue{For instance, in RIS-UAV systems, when changing the UAV altitude, the RIS phase shifts must be simultaneously optimized to maintain the network performance. Compared with optimizing RIS phase shifts solely, such correlation between network functions and control variables is much more complicated.}
The ideal solution is to jointly optimize all variables simultaneously, but this can be extremely difficult due to the interactions between these techniques.

3) \textbf{Large Solution Spaces}: RIS-related optimization problems usually involve a large solution space due to the considerable number of RIS elements. Meanwhile, the integration with other techniques also contributes to the size of the solution space. \blue{For instance, compared with RIS passive beamforming, joint active and passive beamforming problems are more complicated by including BS beamforming vectors, leading to a much larger solution space.}
Such large solution spaces indicate extra difficulty when exploring the optimal solution, e.g., large datasets and action spaces for deploying ML algorithms.

4) \textbf{Integer Control Variables}: As introduced in Section III-G, integer control variables are frequently involved in optimizing RIS-aided networks, e.g., resource allocation and user-RIS-BS association. \blue{These integer control variables will result in NP-hard problems, which cannot be efficiently solved in polynomial time. In addition,} optimization problems become more complicated when both integer and continuous control variables are included, such as joint RIS phase-shift design and elements on/off control.

Finally, given the above features, it is critical to investigate efficient optimization algorithms to handle these challenges and realize the full potential of RIS-aided networks. In the following sections, we will introduce model-based, heuristic, and ML optimization approaches.

%% file: 5_Model_based.tex
\section{Model-based Optimization Algorithms for RIS-aided Wireless Networks}
\label{sec-model}

This section introduces model-based algorithms and applications for optimizing RIS-aided wireless networks, including AO, MM, SCA, BCD, SDR, SOCP, FP, and BnB. In addition, we summarize the features, advantages, drawbacks, difficulties, and applications of these techniques.   

\iffalse
\begin{figure}[!t]
\centering
\includegraphics[width=0.9\linewidth]{Image/fig-AO.jpg}
\caption{The alternating optimization procedure for joint active and passive beamforming.}
\label{fig-AO}
\setlength{\abovecaptionskip}{-2pt} 
\vspace{-10pt}
\end{figure}
\fi

\subsection{Alternating Optimization} 

AO has been widely applied for RIS-related control and optimizations. The main reason is the high complexity of joint optimization problems that include multiple control variables such as the BS beamforming matrix and RIS phase shifts. \blue{For instance, the joint active and passive beamforming is usually decoupled into an active BS beamforming sub-problem and passive RIS beamforming sub-problem, and then each sub-problem is alternatively optimized.}
\blue{In particular,} for an optimization problem with $I$ control variables $\Vec{x} = (x_1,x_2,...,x_i,...,x_I) $ and  $x_i \in X_{i}$, to minimize objective function $f(\Vec{x})$, AO method is summarized by
\blue{
\begin{itemize}
    \item Step 1: Initializing the control variables by setting \qquad $\Vec{x^{0}} = (x^0_1,x^0_2,...,x^0_i,...,x^0_I)$. Defining control variable number $i=1$, iteration number $l=1$, maximum iteration number $L$, and termination threshold $\delta$. 
    \item Step 2: In the $l^{th}$ iteration, for control variable $x^{l}_{i}$, 
    optimizing $f(\Vec{x})$ by finding $x_{i}$ that satisfies 
    \begin{equation} \label{eq-ao}
     x^{l}_i \leftarrow \argmin\limits_{x_{i} \in X_i} f(\underbrace{x^{l}_1,x^{l}_2,...,x^{l}_{i-1}}_{\textbf{done}}\underbrace{,x_{i},}_{\textbf{current}}\underbrace{x^{l-1}_{i+1},...,x^{l-1}_{I}}_{\textbf{todo}}),    
    \end{equation}
    while holding all the other control variables $(x^{l}_1,x^{l}_2,...,x^{l}_{i-1},x^{l-1}_{i+1},...,x^{l-1}_{I})$ constant. 
    \item Step 3: $i=i+1$ and repeating from Step 2 until $i\leq I$. 
    \item Step 4: If $f(\Vec{x_{l}})-f(\Vec{x_{l-1}}) \leq \delta $, stopping all iterations and $\Vec{x_{l}}$ is the optimal solution; if not, moving to step 5. 
    \item Step 5: If $l\leq L$, then $l=l+1$, and setting control variable number $i=1$ and repeat from Step 2; if not, outputting $\Vec{x_{L}}$ as the optimal solution.
\end{itemize}} 
In equation (\ref{eq-ao}), AO simplifies joint optimization problems by optimizing single control variables alternatively while holding other variables unchanged\cite{jcbe}. Each iteration is time-efficient by optimizing one individual variable, which is easily implemented. In addition, it does not require step size parameter tuning and extra storage vectors. AO provides an iterative optimization scheme, but it still relies on other techniques to solve each sub-problem. Also, having each variable monotonically decrease at each iteration does not guarantee the algorithm will converge to a global minimum, and moreover, the convergence may slow down near an optimum point\cite{jcbe}.

The RIS is often combined with other techniques for joint optimization, such as joint active and passive beamforming, RIS-related resource allocation, RIS-NOMA, and RIS-MEC, leading to coupled control variables and large solution spaces.    
AO is particularly useful in solving such joint optimization problems. For example, 
the RIS-MEC system can be decoupled into RIS phase-shift control sub-problem and task offloading sub-problem, and these two sub-problems will be iteratively optimized to reduce the overall complexity. Joint active and passive beamforming is another example that has been widely investigated, which applies AO to generate BS active beamforming and RIS passive beamforming sub-problems\cite{qingqing,qingqing2}.

\subsection{Block Coordinate Descent}  
Coordinate descent is a very useful method to solve large-scale optimization problems, and BCD is considered a generalized version to improve computation efficiency.
Compared with AO, each block in the BCD algorithm may include several control variables, enabling dynamic block generation, selection and updating. Therefore, BCD method is more suitable than AO for optimizing a large number of control variables simultaneously, which has been widely applied to RIS-related optimization problems. 

BCD method sequentially minimizes the objective function $F(\Vec{x})$ in each block $x_{i}$ while the other blocks are held fixed. Specifically, it minimizes $ x^{l}_{i} \leftarrow \argmin\limits_{x_i \in \mathscr{X_{i}}} (f(x_{i})+f_{i}(x_i))$ while holding other blocks $x_{1},x_{2},...,x_{i-1},x_{i+1},...x_{I}$ fixed. However, it is worth noting that each block consists of multiple control variables, and the block selection and updating method will affect the BCD performance. An ideal block selection method is expected to maximize the improvement by choosing the blocks that decrease $F(\Vec{x})$ by the largest amount\cite{Julie}. On the other hand, there are many alternatives for the block updating method such as block proximal updating
     \begin{equation} \label{eq-bcd2}
        x^{l}_{i} \leftarrow \argmin\limits_{x_i \in \mathscr{X_{i}}} (f(x_{i})+f_{i}(x_i)+\frac{L_{i}^{l-1}}{2}\|x_{i}-x^{l-1}_{i}\|^2),
    \end{equation}
where $L_{i}^{l-1}>0$. Equation (\ref{eq-bcd2}) is more stable than conventional BCD by including $\frac{L_{i}^{l-1}}{2}\|x_{i}-x^{l-1}_{i}\|^2$. 
The BCD algorithm is easily deployed with low memory requirements and iteration costs, allowing parallel or distributed implementations. But the block selection may affect the algorithm performance, and block updating is difficult in some cases. 

Similar to AO, BCD is considered as an iteration-based scheme to reduce problem-solving complexity. BCD has been applied to sum-rate maximization \cite{cunhua,jiey,huayan}, user fairness maximization \cite{gang}, and power minimization \cite{zhiyang}. 
As an example, a two-block BCD is used to maximize the sum-rate in \cite{huayan}, in which the first block is for BS active beamforming and the second is for RIS passive beamforming, then these blocks are iteratively optimized.

\begin{figure}[!t]
\centering
\setlength{\abovecaptionskip}{-2pt} 
\includegraphics[width=0.95\linewidth]{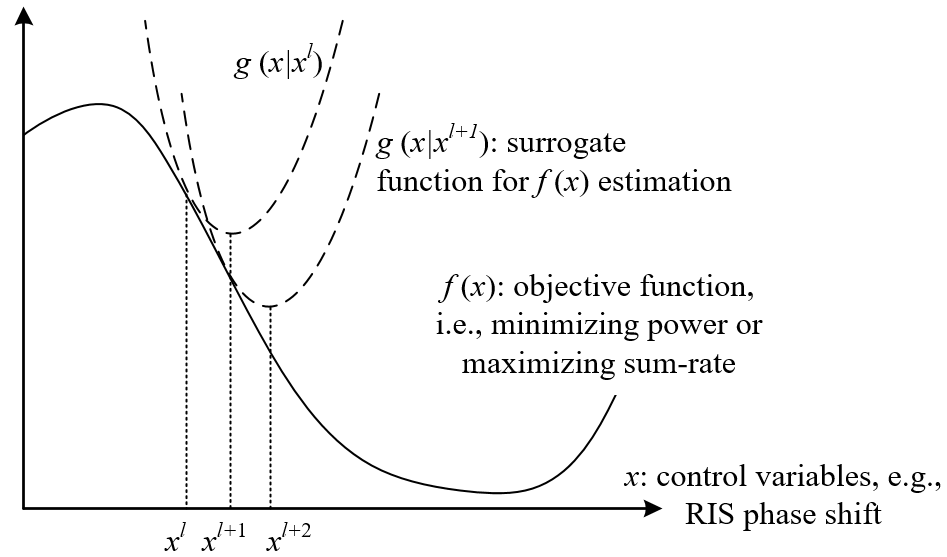}
\caption{MM method for RIS-related optimizations.}
\label{fig-mm}
\vspace{-10pt}
\end{figure}

\subsection{Majorization-Minimization Method} 

MM is an iterative optimization method that has been applied for RIS control and optimizations. Consider an optimization problem $\min\limits_{x}  f(x) $ and $x \in \mathscr{X}$, where $f(x)$ is a continuous objective function and $\mathscr{X}$ is a convex closed set. In RIS-related control problems, the $f(x)$ is usually complicated to solve directly due to fractional and logarithmic terms.
As shown in Fig. \ref{fig-mm}, the main idea of the MM algorithm is to construct a surrogate function $g(x)$ that can locally approximate the objective function $f(x)$, e.g., power minimization or sum-rate maximization. $g(x)$ is considered an upper bound of $f(x)$, which is easier to be optimized. Therefore, optimizing $g(x)$ can either improve the objective function value or leave it unchanged with $g(x) \geq f(x)$\cite{ying}.   

Constructing a surrogate function $g(x)$ is the first step of applying the MM algorithm, since $g(x)$ will be optimized directly instead of the original objective $f(x)$. The $g(x)$ construction rules include:\\
    A1): $g(x^{l-1}|x^{l-1})=f(x)$; \quad A2): $g(x|x^{l-1}) \geq f(x)$;\\
    A3): $ g'(x|x^{l-1};d)|_{x=x^{l-1}}=f'(x^{l-1};d)|_{x=x^{l-1}}$;\\
    A4): $g(x|x^{l-1})$ is continuous in $x$ and $x^{l-1}$, \\
where $x^{l-1}$ is the produced point at iteration $l-1$. 
$g(x|x^{l-1})$ is an approximation function of $f(x)$ at the iteration $l$, and "$|$" in $g(x|x^{l-1})$ means that the point $x^{l-1}$ is already on this function. 
$d$ indicates the distance from a point $x$ to a set $\mathscr{X}$ and $d=\inf\limits_{x'\in \mathscr{X}}||(x-x')||$. $f'(x;d)$ is the directional derivative of $f(x)$ in direction $d$.
Assumptions (A1) and (A2) indicate that $g(x|x^{l-1})$ is a tight upper bound of the original objective $f(x)$. It guarantees that optimizing $g(x|x^{l-1})$ can meanwhile find an improved objective value for $f(x)$. 
Note that surrogate function may be defined in various ways, e.g. Jensen's inequality, Convexity inequality, Cauchy–Schwarz inequality.     
Then, the surrogate function $g(x|x^{l-1})$ is iteratively minimized and updated by
$ x^{l} \leftarrow \argmin\limits_{x \in \mathscr{X}} g(x|x^{l-1})$ until convergence. 
 
As an estimation-based method, MM is considered a low-complexity solution for many RIS-related optimizations, including sum-rate maximization\cite{cunhua}, fairness maximization\cite{menghua,guiz}, secure transmission\cite{huiming2} and so on.
For example, the joint active and passive beamforming problem is decoupled into BS transmit power control and RIS phase-shift optimization in both \cite{cunhua} and \cite{menghua}. Then, the RIS phase-shift optimization problem is first converted into a non-convex quadratically constrained quadratic program (QCQP) problem\footnote{A QCQP problem example is given by equation (\ref{eq-qcqp}) in Section \ref{section_sdr}, which is frequently formulated in wireless networks.}, and a MM algorithm is applied to obtain locally optimal solutions by $\min_{\theta} g(\theta|\theta^l)$ with constraint $|\theta_n|=1$. After that, the optimal phase shift $\theta$ in current iteration $l$ is obtained as $\hat{\theta}^l$, and then $l=l+1$ and $\hat{\theta}^l$ becomes a new $\theta^l$ in $\min\limits_{\theta} g(\theta|\theta^l)$.  

The MM applies surrogate functions to avoid the complexity of optimizing the non-convex objective function directly, transforming non-differentiable problems into smooth optimizations. The MM method requires that the surrogate function $g(x)$ must be a global upper bound for $f(x)$, which is a fundamental assumption for using MM. However, defining such a tight upper bound can be impractical in some cases, which may prevent the application of the MM method.

\subsection{Successive Convex Approximation} 

\begin{figure}[!t]
\centering
\setlength{\abovecaptionskip}{-2pt} 
\includegraphics[width=1.02\linewidth]{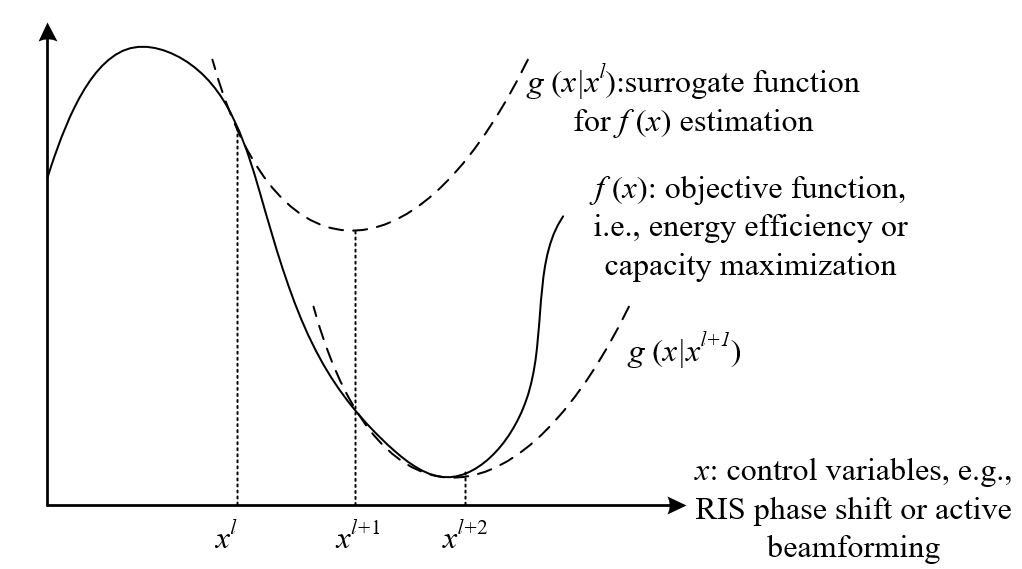}
\caption{ Using SCA algorithm for RIS-related optimization.}
\label{fig-SCA}
\vspace{-10pt}
\end{figure}

Similar to the MM algorithm, SCA applies a surrogate function $g(x)$ to approximate the original objective function $f(x)$, which is shown in Fig. \ref{fig-SCA}. However, the $g(x)$ in the SCA algorithm does not have to be a tight upper bound for $f(x)$, reducing the complexity of the surrogate function design \cite{palomar}. Therefore, SCA is more flexible and easier to be implemented for RIS-related optimization problems.

The SCA method first constructs a surrogate function $g(x)$, and the assumptions are similar to the MM algorithm:  \\ 
A1): $g(x|x^{l-1})$ is continuous in $\mathscr{X}$;\\
A2): $g(x^{l-1}|x^{l-1})=f(x)$;\\
A3): $g(x)$ is differentiable with $\nabla_{x} g(x|x^{l-1})|_{x=x^{l-1}}=\nabla_{x} f(x)|_{x=x^{l-1}}$. \\
SCA relaxes the upper bound condition for the surrogate function, but $g(x|x^{l-1})$ must be strongly convex in $\mathscr{X}$.
Then, solving the constructed surrogate problem 
$\mathbf{\hat{x}}(x^{l}) \leftarrow \argmin\limits_{x \in \mathscr{X}} g(x|x^{l-1})$,
and smoothing the next point by 
\begin{equation}
    x^{l}=x^{l-1}+\beta^{l-1}(\mathbf{\hat{x}}(x^{l})-x^{l-1}),
\end{equation} 
where $\beta^{l-1}$ is the step size for value updating.
Finally,  $g(x)$ construction and solving are repeated until meeting the convergence criteria. 
In SCA, the surrogate function $g(x)$ does not have to be a tight upper bound for $f(x)$. Therefore, the step size in each iteration requires dedicated designs to guarantee an accurate approximation. The factor $\beta^{l-1}$ is used to control the $x^{l}$ updating step size. Meanwhile, the MM algorithm updates the whole control variable $x$ at each iteration, but SCA can be naturally implemented in a distributed manner when the constraints are separable. 

Compared with MM, SCA is more frequently applied in RIS-related optimizations due to the relaxed upper bound, e.g., sum-rate maximization in \cite{ming,yuanbin,xidong} and power minimization in \cite{huimei,jianyue}.
Defining a surrogate function is the key to using the SCA method, which depends on specific objective functions and constraints in RIS-related applications.
For instance, the non-convex BS transmit power constraint in \cite{huimei} is replaced by a first-order Taylor approximation to apply the SCA algorithm. By contrast, Pan \textit{et al.} in \cite{cunh} claim that the unit modulus constraint of the RIS phase shift $|\theta_n|=1$ can be relaxed as a series of convex constraints, e.g., $1 \leq 2 {\rm Re}\{\theta^*_n\theta_n^l\}-|\theta_n^l|^2$, where ${\rm Re}\{\cdot\}$ denotes the real part of a complex argument and $\theta^*$ is the conjugate of $\theta$.

\subsection{Semidefinite Relaxation}  
\label{section_sdr}

Many RIS-related signal processing problems can be described by QCQP formulations, and SDR is an efficient solution to solve QCQP problems\cite{shuzhong}. The QCQP problem is defined by
\begin{equation}\label{eq-qcqp}
\begin{aligned}
\min\limits_{x\in \mathscr{X}}  \qquad & x^{T}C x  \\
 \text{s.t.}  \qquad & x^{T}D_{i} x \geq b_{i}, i=1,2,3,,,n,  
\end{aligned}
\end{equation} 
where the "$\geq$" in the constraint can also be replaced by "$\leq$". Note that $x^{T}C x$ produces an $1\times 1$ matrix, and therefore $x^{T}Cx=Cx^{T}x=Tr(Cx^{T}x)$. Similarly, $x^{T}D_{i}x=D_{i}x^{T}x=Tr(D_{i}x^{T}x)$ is achieved. By introducing $X=xx^T$, then  
\begin{equation}\label{eq-qcqp2}
\begin{aligned}
\min\limits_{x\in \mathscr{X}}  \qquad & Tr(CX)  \\
 \text{s.t.}  \qquad & Tr(DX_{i}) \geq b_{i}, i=1,2,3,,,I,  \\
 \qquad & X\succeq 0,     \\
  \qquad & rank(X)=1,   
\end{aligned}
\end{equation}
where $Tr$ indicates the trace operation, and $X\succeq 0$ indicates that $X$ is positive semidefinite with $X=xx^T$. Then, the non-convex constraint $rank(X)=1$ is relaxed and achieve 
\begin{equation}\label{eq-qcqp3}
\begin{aligned}
\min\limits_{x\in \mathscr{X}}  \qquad & Tr(CX)    \\
 \text{s.t.}  \qquad & Tr(DX_{i}) \geq b_{i}, i=1,2,3,,,I,   \\
 \qquad & X\succeq 0.  
\end{aligned}
\end{equation}

Equation (\ref{eq-qcqp3}) is an SDR of (\ref{eq-qcqp2}), which can be efficiently solved by semidefinite programming (SDP)\cite{zhiquan}. 
SDR has been very generally applied to RIS-related optimization problems, since the $rank(x)=1$ is frequently formulated for phase control. Specifically, the RIS phase shift constraint $|\theta_n|=1$ is non-convex with $\theta\theta^T=1$. 
Then we can define $\mathcal{V}=\theta\theta^T$ with $\mathcal{V}\succeq1$ and $rank(\mathcal{V})=1$, which can be then transformed and relaxed as shown by equations (\ref{eq-qcqp2}) and (\ref{eq-qcqp3}). 

However, the main obstacle to applying SDR is to transform a globally optimal solution $\hat{\mathcal{V}}$ into a feasible solution $\hat{\theta}$. An ideal solution is that $\hat{\mathcal{V}}$ is rank-one, and then $\hat{\theta}$ is easily obtained by solving $\hat{\mathcal{V}}=\hat{\theta}\hat{\theta}^T$. Otherwise, if $rank(\hat{\theta})>1$, a rank-one approximation may be used to obtain a sub-optimal solution $\widetilde{\theta}$. There are multiple methods to find a feasible $\widetilde{\theta}$ from $\hat{\mathcal{V}}$, leading to various solution qualities. 
For instance, Mu \textit{et al.} propose a penalty-based method to relax the rank-one constraint, finding a sub-optimal solution by introducing penalties if $rank(\hat{x})>1$ \cite{mu2021simultaneously}. 
SDR has been used for sum-rate maximization \cite{boya, peilan, ni2021resource}, power minimization \cite{guizhou2,huimei,guiz3,jianyue}, fairness maximization \cite{gang,hailiang,zaid,menghua}, and secure transmission \cite{zheng,wei,xianghao,biqian,zijie}.

\begin{figure}[!t]
\centering
\setlength{\abovecaptionskip}{-3pt} 
\includegraphics[width=0.7\linewidth]{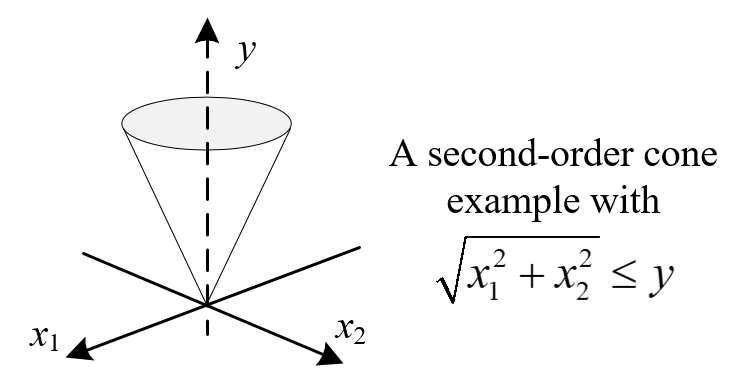}
\caption{An example of a second-order cone in 3D space.}
\label{fig-socp}
\vspace{-15pt}
\end{figure}

\subsection{Second-order Cone Programming}  
SOCP is another method that is used to efficiently solve optimization problems in wireless networks, especially for QCQP and fractional problems. 
Fig. \ref{fig-socp} presents a second-order cone example in 3D space. SOCP is a generalization of linear and quadratic programming that allows for affine combinations of variables to be constrained inside a second-order cone
\begin{equation}\label{eq-scop}
\begin{aligned}
\min\limits_{x\in \mathscr{X}}  \qquad & C^{T} x   \\
 \text{s.t.}  \qquad & ||A_{i}x+b_{i}||\leq c_{i}^{T}x+d_{i}, i=1,2,3,,,I, 
\end{aligned}
\end{equation} 
where $A \in \mathbb{R}^{n_i\times n}$, $b_i \in \mathbb{R}^{n_i}$, $c_i \in \mathbb{R}^n$, and $d_i \in \mathbb{R}$. The $x$ in equation (\ref{eq-scop}) may be RIS phase shifts, BS beamforming vectors, and so on, which depends on specific application scenarios. Consider the inverse image of the unit second-order cone with an affine mapping
\begin{equation} \label{eq-scop3}
 ||A_{i}x+b_{i}||\leq c_{i}^{T}x+d_{i} \leftrightarrow 
 \left[ \begin{array}{c}  
    A_{i} \\  
    c_{i}^{T} \\ 
  \end{array} \right] x +  
\left[ \begin{array}{c}  
    b_{i} \\  
    d_{i} \\ 
  \end{array}
\right] \in \mathscr{C}_{n_i+1}.       
\end{equation}
Therefore, SOCP is a convex optimization problem with a convex objective function and convex constraints. Equation (\ref{eq-scop3}) indicates the core properties of SOCP problems, and hence many problems are converted into SOCPs and solved efficiently\cite{miguso}. 

For instance, sum and fractional problems are frequently defined in RIS-related problems to maximize the sum-rate or total throughput regarding the SINR
\begin{equation}\label{eq-scop7}
\begin{aligned}
\min\limits_{x\in \mathscr{X}}  \qquad & \sum_{i=1}^{I}\frac{||C_{i}^T+D_{i}||^2}{A_{i}^Tx+B_{i}}  \\
 \text{s.t.}  \qquad & A_{i}^Tx+B_{i} \geq 0, i=1,2,3,,,I, 
\end{aligned}
\end{equation} 
which is converted into a SOCP by
\begin{equation}\label{eq-scop8}
\begin{aligned}
\min\limits_{x\in \mathscr{X}}  \qquad & \sum_{i=1}^{I}t_{i} \\
 \text{s.t.}  \qquad & (C_{i}^T+D_{i})^T(C_{i}^T+D_{i}) \leq t_{i}(A_{i}^Tx+B_{i}),    \\
  \qquad & A_{i}^Tx+B_{i} > 0, i=1,2,3,,,I.  
\end{aligned}
\end{equation} 

SOCP can be efficiently solved by the interior point method. Meanwhile, SOCP is less general than SDP since equation (\ref{eq-scop}) may be transformed into an SDP problem. However, the complexity of solving SOCP is $O(n^2\sum_{i}{n_i})$, while the complexity for SDP is $O(n^2\sum_{i}{n_i}^2)$\cite{nest}. Such complexity difference is crucial for large-dimension problems. 

Finally, to apply SOCP for RIS-aided optimizations, the first step is to utilize AO or BCD scheme to decouple the control variables into multiple sub-problems, e.g., BS precoding matrix and RIS passive beamforming\cite{guiz,yiqing,menghua}, coordinated transmit beamforming and RIS passive beamforming\cite{hailiang}. For example, the max-min data rate problem in \cite{menghua} is decoupled into SOCP-based BS beamforming and SDR-based RIS phase-shift control, and the data rate maximization problem in \cite{jiey} is converted into a SOCP-based BS active beamforming and SDR-based RIS passive beamforming.

\subsection{Fractional Programming} 
FP refers to optimization problems involving ratios or fractional terms.
FP is particularly useful for wireless network optimizations due to the fractional terms in communication systems, especially for SINR and energy efficiency\cite{zappone}. 

Consider a single-ratio FP problem to maximize the SINR of single UE by $\max\limits_{x\in \mathscr{X}} {f(x)}/{g(x)}$, where $f(x)$ is the signal strength and $g(x)$ is the interference and noise. There are many classic methods to solve FP problems, such as Charnes-Cooper transform and Dinkelbach’s transform\cite{Dinke}. Dinkelbach’s method reformulates the problem into $\max\limits_{x\in \mathscr{X},y\in \mathbb{R}}  f(x)-yg(x)$, where $y$ is the auxiliary variable that is updated iteratively $y^{(l+1)}={f(x)^l}/{g(x)^l}$, 
and $l$ is the iteration number. Then, alternatively updating $y$ and $x$ will lead to a converged solution with non-decreasing $y^l$. However, instead of the single-ratio problem, sum-ratio FP problems are more frequently involved in wireless networks, i.e., maximizing sum-rate or total channel capacity as 
$\max\limits_{x\in \mathscr{X}}  \ \sum_{i=1}^{I} f_{i}(x)/g_{i}(x) $.

However, classic methods can not be directly generalized to sum-ratio cases, since maximizing single ratios cannot guarantee the convergence and maximization for sum-ratio cases. An equivalent transform proposed by \cite{shenk} is 
\begin{equation}\label{eq-fp5}
\max\limits_{x\in \mathscr{X},y\in \mathbb{R}}  \quad 2yf(x)^{0.5}-y^{2}g(x),  
\end{equation} 
which can be readily converted into sum-ratio problems.
In addition, equation (\ref{eq-fp5}) is further generalized to sum-ratio problems as
\begin{equation}\label{eq-fp7}
\max\limits_{x\in \mathscr{X},y\in \mathbb{R}}  \quad \sum_{i=1}^{I}F_{i}(2y_{i}C_{i}(x)^{0.5}-y_{i}^{2}D_{i}(x)),  
\end{equation}
where $F_{i}$ is a non-decreasing function. Equation (\ref{eq-fp7}) is particularly useful given the frequently used term $\sum log(1+SINR)$ in wireless communications. 

The FP method can significantly lower the problem-solving complexity by eliminating fractional items. This transformation is very useful for RIS-related optimization problems, especially considering that RIS phase shifts will affect the received signal strength and interference simultaneously.
In addition, the FP method can be particularly useful for RIS-related max-min fairness problems, which are usually formulated as
$\max\limits_{x\in \mathscr{X}}  \  \min\limits_{1\leq i \leq I} \ {f_{i}(x)}/{g_{i}(x)}$, where $x$ indicates the control variables, e.g., RIS phase shifts and BS transmit power. $f_{i}(x)$ can be the signal strength of user $i$, and $g_{i}(x)$ indicates the interference and noise. Then the max-min fairness problems can be reformulated as 
\begin{equation}\label{eq-fp9}
\begin{aligned}
\max\limits_{x\in \mathscr{X}, y,z\in \mathbb{R}}  &  \quad z \\
 \text{s.t.}  \quad & 2y_{i}f_{i}(x)^{0.5}-y_{i}^{2}g_{i}(x) \geq z; i=1,2,3,,,I.
\end{aligned}
\end{equation}
where $z$ is an intermediate objective function that is included in the constraint. A detailed proof of obtaining equation (\ref{eq-fp9}) can be found in \cite{shenk}.

The FP method significantly reduces the optimization complexity by decoupling the fractional terms. Therefore, it has been widely used in wireless network optimizations, including power control, beamforming, energy efficiency, and so on\cite{huayan,shuaiqi}. However, note that FP is usually used for transformation, and then the reformulated problems still need to be solved by other techniques. A widely considered method is first to apply FP to eliminate the fractional terms in objective functions, e.g., throughput and power consumption for energy efficiency maximization, received signal strength and interference for SINR maximization. And then, AO is used to separate the coupled control variables, e.g., RIS phase-shift design and BS transmit power control, and optimize each sub-problem iteratively.

\begin{figure}[!t]
\centering
\setlength{\abovecaptionskip}{-3pt} 
\includegraphics[width=0.8\linewidth]{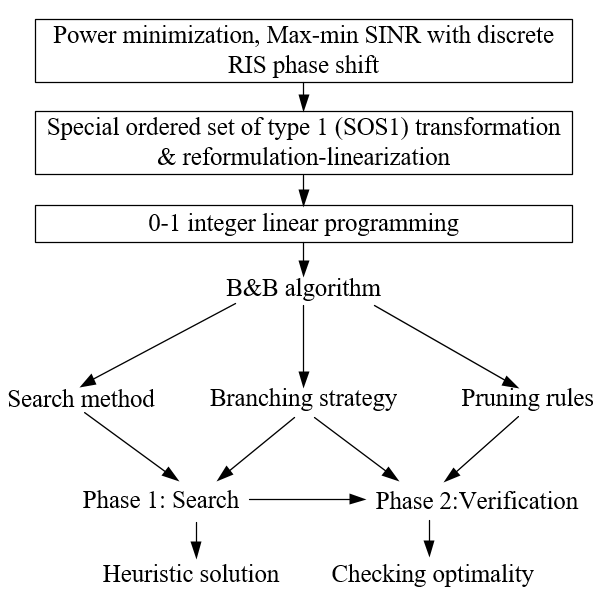}
\caption{Using BnB for RIS control with discrete phase shifts.}
\label{fig-bb}
\vspace{-15pt}
\end{figure}

\begin{table*}[!t]
\caption{Summary of model-based optimization algorithms for RIS-aided wireless networks }
\centering
\setstretch{1.15}
\small
\resizebox{1\textwidth}{!}{%
\begin{tabular}{|m{1cm}<{\centering}|m{3.7cm}<{\centering}|m{3.3cm}<{\centering}|m{3cm}<{\centering}|m{3cm}<{\centering}|m{4.9cm}<{\centering}|}
\hline 
Methods &  Main features   &     Advantage    &  Drawbacks    &  Difficulties   &  Application scenarios  \\
\hline
AO & Decoupling the joint optimization into multiple sub-problems, and alternatively optimizing each sub-problem.   &  The problem-solving complexity is greatly reduced. Each sub-problem may be easier to solve.  & Iterative optimization may lead to sub-optimal results; the convergence must be proved. &  The complexity is high when each sub-problem is still complicated.  & AO is the most widely applied optimization scheme for RIS-aided networks. It decouples the joint optimization into multiple sub-problems and then optimizes them iteratively \cite{chongwen}.  \\
\hline
BCD &  The control variables are divided into multiple blocks. Minimizing one block in each iteration and keeping other blocks fixed.  & Cheap iteration costs; low memory requirements; potential for parallel implementation   &  Block selection may affect the BCD performance, and block updating is difficult in some cases.  &  Block selection and updating methods are complicated.  & BCD employs alternating schemes to reduce joint optimization complexity, e.g., sum-rate maximization \cite{cunhua} and power minimization\cite{zhiyang}, and one block in BCD may include multiple control variables.     \\
\hline
MM & Iteratively constructing and optimizing an upper bound surrogate function that can locally approximate objective functions. &  Avoiding the complexity of optimizing non-convex objective functions directly.  &  The surrogate function must be a strict tight upper bound for objective functions, which is hard to achieve in practice.   &  The surrogate function must follow the shape of objective functions and meanwhile be easy to optimize.  &  The MM applies surrogate functions as low-complexity solutions for many RIS-aided optimizations, i.e., sum-rate maximization\cite{cunhua}, fairness maximization\cite{menghua,guiz}, secure transmission\cite{huiming2}.    \\
\hline
SCA & Constructing and optimizing surrogate functions iteratively to estimate the objective function.  & Low computational complexity; the tight upper bound is not required for the surrogate function; naturally implemented in a distributed manner.  &  The step size selection is critical for an accurate approximation.  &  Surrogate function and step size selection.  &   SCA relaxes the tight upper bound constraint on surrogate function design. Such an estimation-based approach is easier to be implemented in RIS-related optimization, e.g., sum-rate maximization \cite{ming,yuanbin} and power minimization\cite{huimei,jianyue}.     \\
\hline
SDR & SDR is used to solve QCQP problems by relaxing the rank constraint. Then the reformulated problem is efficiently solved by SDP. & Given the objective problem, SDR is easily implemented without extra parameters or settings. & Approximation is required if the relaxed solution is not rank one.  &  The reformulated problem is complicated if the achieved solution is not rank one. & SDR is particularly useful to solve $rank(x)=1$ constraints, such as RIS phase-shift constraint $|\theta_{n}|=1$ \cite{boya} \cite{peilan,qingqing}.\\
\hline
SOCP & SOCP utilizes the property of the second-order cone, and many problems are reformulated into SOCP, which is much easier to be solved.  & SOCP can be efficiently solved by many existing algorithms. It has a lower complexity $O(n^2\sum_{i}{n_i})$ than SDR.  &  Problem reformulation into SOCP is complicated.  &  The main difficulty lies in how to reformulate the original problem into SOCP.  & SOCP can be very useful if the RIS-related problems can be easily formulated as a second-order cone, which has been used for power minimization \cite{yiqing} and user fairness \cite{guiz,hailiang,menghua}.\\
\hline
FP & FP refers to optimization problems that involve fractional terms, which is very useful for wireless communications considering the form of SINR and energy efficiency.  & FP is easily implemented without extra parameters or problem formulation requirements.  &  The reformulated problems generally require iterative optimization to approximate the solution of original FP problems.   & Compared with single-ratio problems, wireless networks are more related to sum-ratio problems, which are more complicated to solve.   & FP is particularly useful when decoupling the fractional terms in RIS-related problems, e.g., SINR and energy efficiency. It is widely applied for optimizing RIS-aided networks \cite{huayan,jianyue,chongwen}.    \\
\hline
BnB & BnB is mainly designed for combinational optimization problems. It applies a tree to enumerate all possible subsets and sub-problems.  &  Lower complexity than direct optimizations. The solution quality is controlled by customized search, branching, and pruning rules.  &  The algorithm is slow when constantly searching or branching in the worst case.   &  The algorithm performance relies on the searching and pruning method, which is hard to select in some cases.  &  Different from aforementioned technique, BnB is mainly applied for discrete and combinational optimization problems, i.e., RIS on/off and discrete phase shift\cite{shiqi,qingqing2,boya}.     \\
\hline
\end{tabular}}
\label{tab-comparison}
\vspace{-13pt}
\end{table*}

\subsection{Branch-and-Bound } 

BnB is a classic scheme for combinatorial and discrete optimization problems\cite{clau}. To minimize $f(x)$ with $x\in \mathscr{X}$, BnB applies a tree scheme to enumerate all possible subsets $X_{i} \subseteq \mathscr{X}$, and each subset $X_{i}$ indicates a sub-problem $f_{i}(x)$. Solving sub-problems $f_{i}(x)$ will generate and prune branches based on the estimated lower and upper bounds.  

A BnB algorithm consists of three basic operations: branching, bounding, and pruning.
Considering a non-linear integer programming problem, and the BnB scheme is summarized as Fig. \ref{fig-bb}, including the search method, branching strategies, and pruning rules. In particular, the search method indicates the order of sub-problem exploration in the tree, e.g., which RIS phase-shift combination is first explored. The branching strategy specifies how to generate new sub-problems from the solution space, e.g., how to generate a new set of phase-shift designs. Finally, pruning rules can prevent exploring specific regions of the tree, which will eliminate sub-optimal RIS phase-shift solutions.
BnB produces a series of sub-problems $f_{i}(x)$ that are equivalent to the original $f(x)$, which is much more efficient than brute-force enumeration.  It provides an alternative solution for challenging problems that cannot be solved directly. An important advantage is that the quality of the solution is controlled by customized searching, branching, and pruning rules. 

BnB is mainly applied for RIS control with discrete phase shift, including sum-rate maximization \cite{boya}, power minimization \cite{qingqing2}, and max-min SINR \cite{shiqi}. The main reason is that the problem formulations are usually MINLP problems, which are NP-hard and intractable. As shown in Fig. \ref{fig-bb}, the MINLP is converted into an 0-1 integer linear programming using the special ordered set of type 1 (SOS1) transformation \cite{qingqing2} and reformulation-linearization \cite{shiqi}. BnB performance is very dependent on search and pruning rules, and defining these rules can be difficult in some cases. In addition, the algorithm may converge slowly when constantly searching and branching for new solutions, which may be caused by the considerable number of RIS elements.

\subsection{Discussions and Numerical Results}

Table \ref{tab-comparison} summarizes model-based algorithms for RIS-aided wireless networks, including main features, advantages, disadvantages, difficulties, and application scenarios. 

Firstly, considering the high complexity of RIS-related optimization, AO is regarded as the primary scheme to decouple the joint optimization problem into several sub-problems. Then, each sub-problem is alternatively solved by using different algorithms, e.g., SCA, SDR, and BnB. Compared with AO, the BCD algorithm applies a similar iterative optimization scheme, but one block may include multiple control variables. When there are a large number of control variables, the BCD algorithm can be more efficient, e.g., coupled optimization problem with a considerable number of control variables.

MM and SCA are two estimation-based algorithms that avoid the complexity of direct optimizations. However, the MM algorithm requires a tight upper bound when designing the surrogate function. Such requirements can be impractical in some cases, especially considering non-convex and highly non-linear RIS phase-shift design problems.
By contrast, the SCA method relaxes the upper bound requirement for surrogate functions, which is more flexible and easier for design and implementation.
However, without the upper bound constraint, the updating step size in SCA may affect the solution quality, which should be carefully selected. MM and SCA are usually considered low-complexity solutions for RIS-aided wireless network optimizations. 

SDR and FP are usually combined with other techniques for optimizations. In particular, SDR is mainly used to relax the RIS phase constraints, while FP can decouple the numerator and denominator for SINR and energy efficiency terms. These two techniques reformulate the original problems into low-complexity or even convex forms, then other techniques can be applied. Meanwhile, SOCP takes advantage of the property of the second-order cone, which is efficiently solved by many existing methods. But the main difficulty is how to transform the problem with logarithm and fractional terms into a second-order cone. BnB is mainly designed for combinational and discrete optimization problems, e.g., RIS control with discrete phase shifts and elements on/off.

\begin{figure}[!t]
\centering
\subfigure[Average throughput comparison under various peak traffic loads. AOFP: combining AO and the FP algorithm; surrogate method: using a surrogate function to approximate the objective function in a black-box manner. ]{ \label{fig_result1}
\includegraphics[width=7.2cm,height=5.2cm]{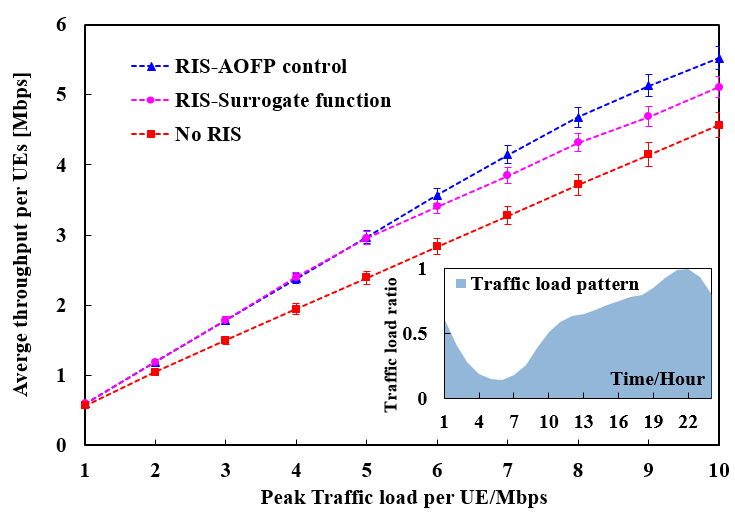}
}
\,
\subfigure[Convergence performance of AOFP under various numbers of RIS elements.]{ \label{fig_result2}
\includegraphics[width=7.2cm,height=5.2cm]{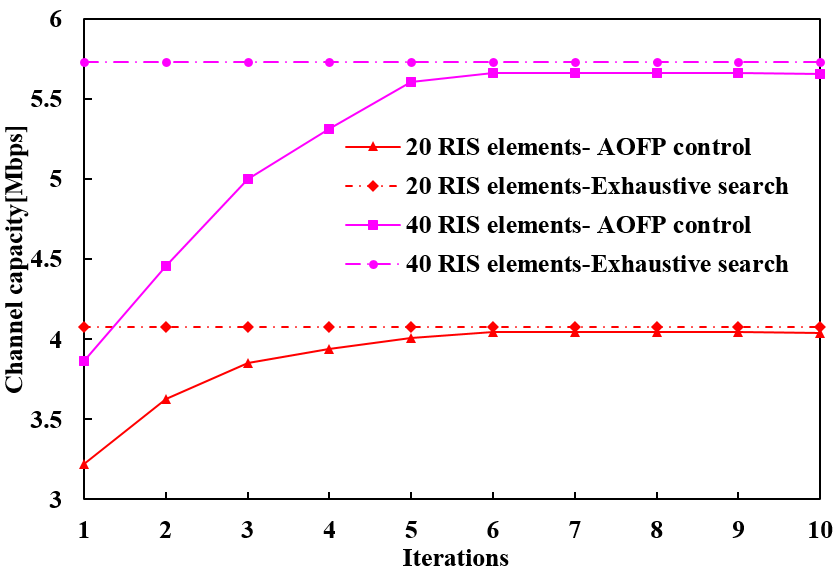}
}
\setlength{\abovecaptionskip}{0pt} 
\caption{Simulation results by combining AO and FP to maximize the channel throughput. We consider a MISO system with one BS and multiple UEs, and the daily traffic load pattern is shown in \ref{fig_result1}. Detailed simulation parameters and algorithms can be found in \cite{zhou2023cooperative}.}
\vspace{-15pt}
\label{fig-re-model}
\end{figure}

Finally, it is worth noting that these algorithms are not independent, and multiple algorithms are usually combined for transformation and optimizations. The main objective of Table \ref{tab-comparison} is to analyze the feasibility of these problems for various RIS-related optimizations, and the most efficient solution for specific scenarios requires case-by-case analyses. For instance, Fig. \ref{fig-re-model} shows an example of combining AO and FP for RIS phase-shift control in an MISO system with multiple UEs. Specifically, it applies FP to decouple the received signal strength with interference and noise, and then uses AO to optimize multiple control variables alternatively\cite{zhou2023cooperative}. Fig. \ref{fig_result1} presents the average throughput under various peak traffic loads, which involves a daily traffic load pattern as shown by the blue shade in Fig. \ref{fig_result1}. Meanwhile, we consider surrogate optimization as a baseline, which applies surrogate functions to approximate the objective function in a black-box manner. When the peak traffic load is light, one can observe that AOFP and surrogate function have comparable performance, which means that the channel capacity can already satisfy the traffic demand. However, when peak traffic load increases, AOFP attains higher throughput than baselines, which demonstrates that RIS control and deployment strategy should consider dynamic UE traffic demand. Additionally, Fig. \ref{fig_result2} presents the convergence performance of AOFP, in which the objective function is improved with increasing iterations and finally converges. This reveals the basic features of AO, which is to guarantee the objective function will be improved iteration-by-iteration, and such a scheme has been widely used in RIS-related optimization studies.

%% file: 5_Heuristic_based.tex
\section{Heuristic Algorithms for RIS-aided Wireless Networks }
\label{sec-heu}

As presented in Section \ref{sec-model}, model-based algorithms have specific requirements for problem formulations, especially for convexity and continuity. 
Meanwhile, the large number of RIS elements, dynamic channel conditions, and various CSI levels further contribute to the overall complexity.  
Therefore, transformations and relaxations are required to convert the original problem into specific forms. 
Moreover, these transformations are usually problem-specific, requiring case-by-case analyses and dedicated design. 

By contrast, heuristic algorithms have fewer requirements for objective functions and constraints, which will significantly reduce the complexity. 
Compared with model-based methods, heuristic algorithms are usually considered low-complexity solutions. In the following, we will introduce four heuristic algorithms, including CCP, meta-heuristic algorithms, greedy algorithms, and matching-based algorithms.

\subsection{Convex-concave Procedure}

The CCP algorithm uses the local heuristic to solve difference of convex (DC) problems, which is considered a low-complexity solution for complicated wireless network optimization\footnote{The main reason that CCP is considered a heuristic algorithm is that it applies a simple heuristic rule for optimization, which is iteratively finding two points with the same tangent vectors\cite{lipp}. Hence, CCP fits well with our defined classifications of heuristic algorithms.}. 

DC problems are frequently formulated in many fields, representing many scenarios that cannot be solved in polynomial time. The DC problem is defined as
\begin{equation}\label{eq-ccp}
\begin{aligned}
\max\limits_{x\in \mathscr{X}}  &  \quad f_{0}(x)-g_{0}(x) \\
 \text{s.t.}  \quad & f_{i}(x)-g_{i}(x) \leq 0; \ i=1,2,3,,,I,
\end{aligned}
\end{equation}
where $f(x)$ and $g(x)$ are both convex. DC problems are usually non-convex unless $g_{i}(x)$ are affine, which is generally hard to solve.

\begin{figure}[!t]
\centering
\includegraphics[width=1\linewidth]{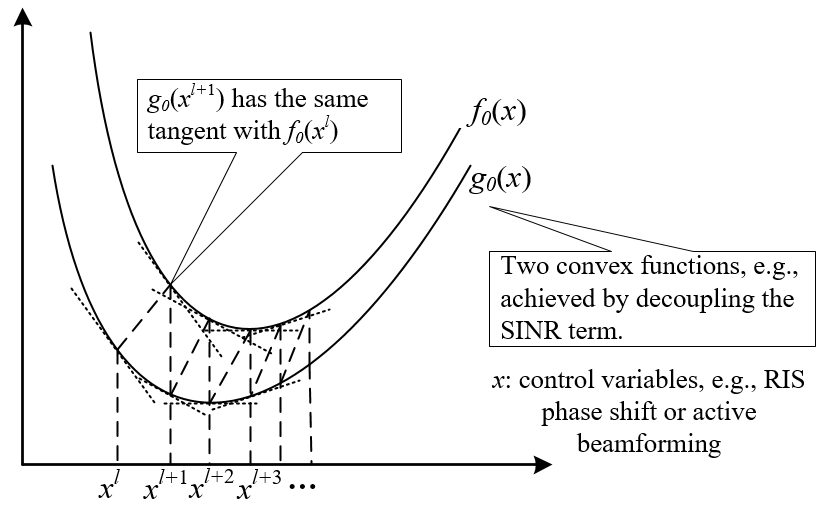}
\caption{Convex concave procedure for RIS-related optimization.}
\label{fig-ccp}
\setlength{\abovecaptionskip}{-2pt} 
\vspace{-10pt}
\end{figure}

As shown in Fig. \ref{fig-ccp}, the core idea of CCP is to find $x^{l+1}$ in the $l+1$ iteration that satisfies $\nabla_x f(x^{l+1})=\nabla_x g(x^{l})$, indicating a point on $f(x)$ that has the same tangent with $g(x^{l})$\cite{yulie}. 
The CCP algorithm will first form 
    \begin{equation} \label{eq-ccp112}
        \overline{g}_{i}(x|x^l)=g_{i}(x^{l})+ \nabla_x g(x)(x-x^{l}),\ i=0,1,2,3,,,I. 
    \end{equation}
Then it solves the following problem to get $x^{l+1}$
    \begin{equation}\label{eq-ccp2}
    \begin{aligned}
    \max\limits_{x\in \mathscr{X}}  &  \quad f_{0}(x)-\overline{g}_{0}(x|x^l)\\
    \text{s.t.}  \quad & f_{i}(x)-\overline{g}_{i}(x|x^l) \leq 0; \ i=1,2,3,,,I.
    \end{aligned}
    \end{equation}
Equation (\ref{eq-ccp2}) is equivalent to  $\nabla_x f(x^{l+1})=\nabla_x g(x^{l})$ by deriving the objective function, and equations (\ref{eq-ccp112}) and (\ref{eq-ccp2}) are iteratively repeated until reaching the stop criteria. 
CCP algorithm does not require a dedicated step size design, and the main reason is that the estimator $f_{0}(x)-\overline{g}_{0}(x|x^l)$ is global. It retains all the information from the convex component $f(x)$ and only linearizes the concave portion $g(x)$.

Moreover, there are multiple extensions of the CCP algorithm.  For instance, the penalty CCP includes a penalty term for violations, which removes the requirements for feasible initial points. The RIS phase shift optimization problem is reformulated as \cite{cunh}
\begin{equation}\label{eq-ccp11}
\begin{aligned}
\max\limits_{\theta, \tau >0}  &  \quad f(\theta)-\tau^{l} \sum_{i=1}^{2I} v_{i} \\
\text{s.t.}  \quad & {g}(\theta) \geq  D,\\
\quad & |\theta_i^l|^2 - 2 Re\{\theta^*_i\theta_i^l\} \leq v_i-1,\\
\quad & |\theta_i|^2 \leq 1+v_{i+I}, \ i=1,2,3,,,I,
\end{aligned}
\end{equation}  
where $\tau^{l} \sum_{i=1}^{2I} v_{i}$ is the penalty term, $v_i$ are slack variables, and $\tau^{l}$ is a coefficient that will decline in each iteration for convergence. After some transformations, problem (\ref{eq-ccp11}) can be solved by using the CVX toolbox, and the detailed procedure is included in \cite{cunh}. 

CCP has been applied in \cite{guiz3,yuanbin,chen2021qos,niu2021simultaneous,xu2022reconfigurable} for controlling RIS phase shifts. In these works, the joint optimization problem is first decoupled into multiple sub-problems using BCD or AO, then penalty CCP is used to solve the RIS phase shifts sub-problem. The main motivation is the high complexity of solving non-convex RIS control problems. For example, 
the sum-rate maximization problem in \cite{chen2021qos} is converted into three sub-problems: joint optimization of the transmit power and spectrum sharing, SDR-based multi-user detection, and CCP-based RIS phase shifts. However, note that CCP is a heuristic algorithm that will find a locally optimum solution, and the initial point $x^0$ may affect the final output. In particular, there may exist multiple locally optimal solutions, and CCP can easily get stuck in a sub-optimal one.

\subsection{Meta-heuristic Algorithms}

One of the main difficulties of controlling RISs is the large number of RIS elements, leading to huge solution spaces. Therefore, it is hard to achieve exact solutions by finding a closed-form expression, and hence model-based approximation algorithms such as SCA and MM are applied. 
However, these methods have stringent requirements for objectives and constraints, especially for convexity, continuity, and differentiability. By contrast, meta-heuristic algorithms can search significantly large solution spaces with few or no additional requirements on problem forms \cite{bozorg2017meta}. 
It usually contains intelligent policies to guide the heuristic exploration, producing high-quality solutions efficiently. 
Meta-heuristic algorithms have been extensively developed, e.g., genetic algorithm (GA), particle swarm algorithm (PSO), ant colony optimization, simulated annealing, and tabu search \cite{beheshti2013review}. 

\begin{figure}[!t]
\centering
\includegraphics[width=0.7\linewidth]{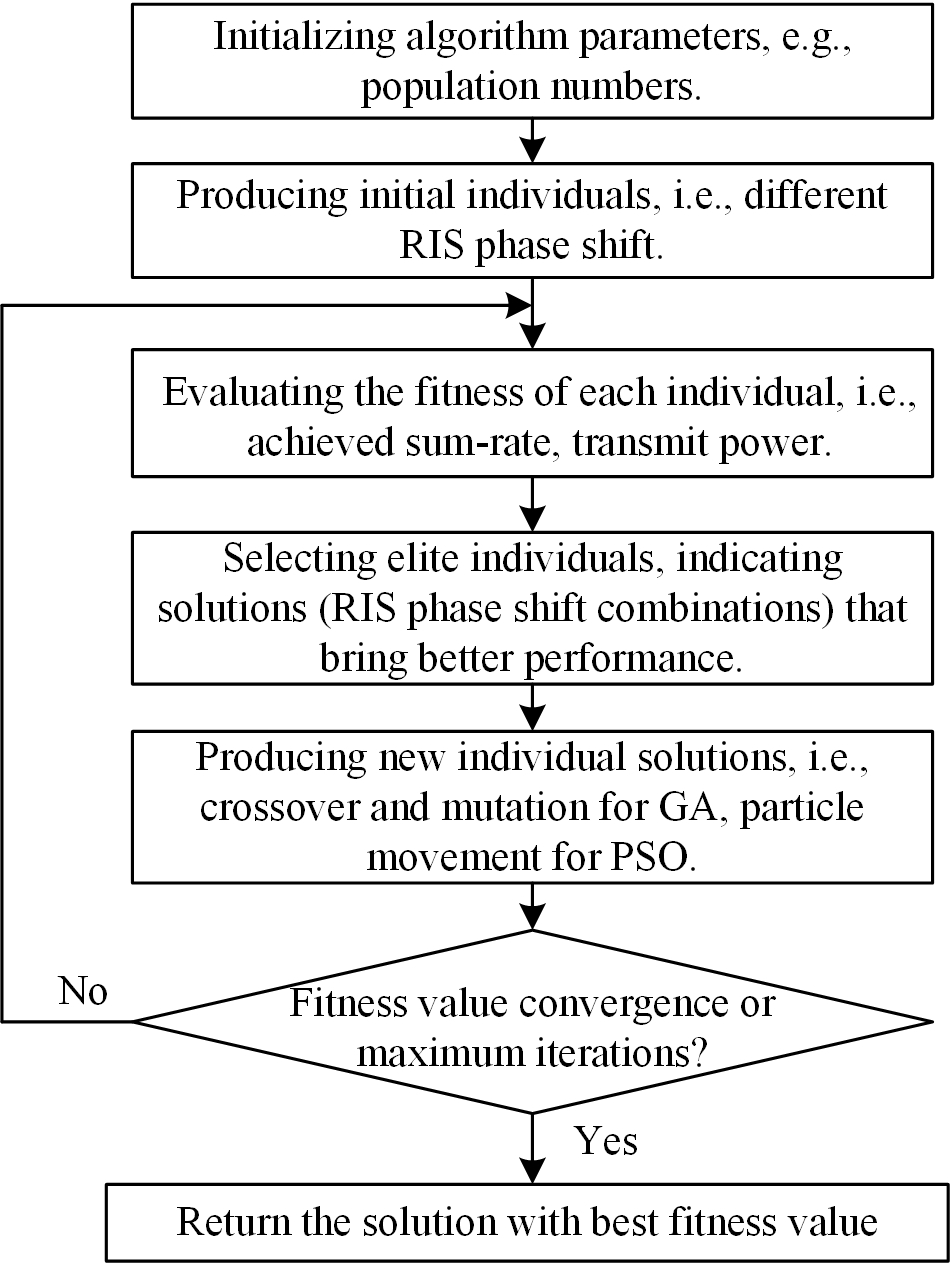}
\caption{Population-based meta-heuristic algorithms for RIS phase-shift control.}
\label{fig-pso}
\setlength{\abovecaptionskip}{-2pt} 
\vspace{-10pt}
\end{figure}

Fig. \ref{fig-pso} shows the steps of using population-based meta-heuristic algorithms for RIS phase-shift design. The first step is to initialize the algorithm parameters such as population numbers and crossover rate in a genetic algorithm.
Then, the algorithm will produce initial individuals, which indicates various RIS phase-shift designs. The objective function is converted into a fitness function, e.g., the sum-rate or energy efficiency. After that, the algorithm will constantly search for better solutions using heuristic rules iteratively, such as evolution strategy in a genetic algorithm, and particle movement for PSO. Finally, the heuristic exploration will stop if the fitness function values converge or reach maximum iteration numbers.

Compared with model-based methods, the main advantage is that meta-heuristic algorithms can easily adapt to both continuous and discrete RIS phase shifts without relaxation and transformation. PSO and GA are used for RIS phase shifts in \cite{dai2021reconfigurable} and \cite{zhi2022power} to maximize the data rate. Statistical CSI is investigated in \cite{zhi2021statistical} to obtain a closed-form expression of the uplink ergodic data rate, then GA is deployed for phase control to maximize the data rate. In addition, Tabu search is applied to irregular RIS to decide the element design in \cite{ruoc}. 

The simulations in \cite{dai2021reconfigurable,zhi2022power,zhi2021statistical,ruoc} show that meta-heuristic algorithms can significantly reduce the optimization complexity, especially for MINLP problems. 
However, it may be trapped in local optima, and the algorithm performance relies on the parameter settings. For example, the phase shifts of hundreds of RIS elements require a large number of populations in GA, leading to high exploration costs. By contrast, reducing the population numbers may lower the probability of finding optimal solutions.         

\subsection{Greedy Algorithms} 
Most former algorithms are designed to find global optima of the objective function. However, many problems are NP-hard and non-convex, and the solutions are usually problem-specific with a series of transformations. To this end, greedy algorithms are proposed as low-complexity alternatives. In particular, greedy algorithms refer to the problem-solving heuristic that makes locally optimal decisions at each stage regardless of global optima\cite{martins2006metaheuristics}.
Consider an minimization problem $\min f(\Vec{x})$, and the control variables $\Vec{x}$ include $\Vec{x}=\{x_1,x_2,...,x_i,...,X_I\}$. 
At each stage, the greedy algorithm will optimize only one control variable $x_i$ by $\hat{x_i}=\argmin_{x_i\in X_i} f(\Vec{x})$, while holding the rest of variables unchanged. Then it moves to the next stage until $i=I$.

\begin{figure}[!t]
\centering
\includegraphics[width=0.85\linewidth]{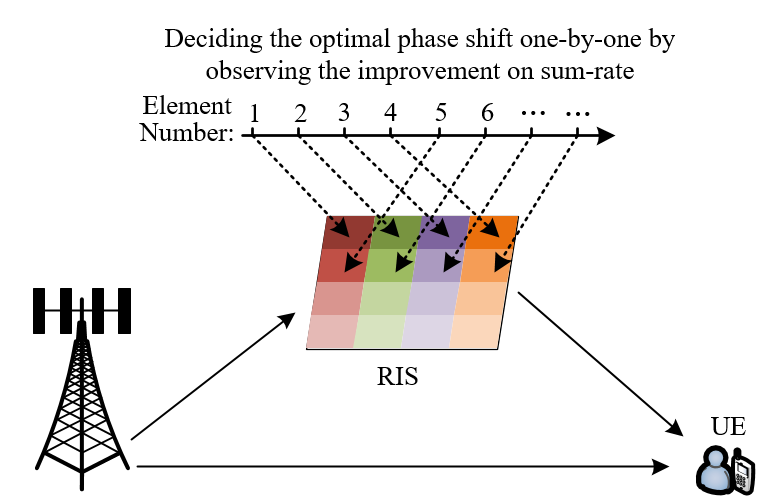}
\caption{Greedy method for RIS phase shift control.}
\label{fig-greedy}
\setlength{\abovecaptionskip}{-2pt} 
\vspace{-10pt}
\end{figure}

RIS elements' on/off control is a non-convex problem with discrete constraints. It may be solved by relaxing the integer constraint $ \chi \in \{0,1\}$ into $0\leq \chi \leq 1$, but the reformulated problem can still be complicated. 
A low-complexity solution is a greedy element-by-element control. Specifically, it evaluates the on/off decision of one RIS element at each stage by observing the changes in objective functions. If the performance is improved by achieving a higher sum-rate and lower power consumption, then the on/off status will be updated\cite{zhaohui}. Similarly, as shown in Fig. \ref{fig-greedy}, this greedy scheme can also be applied to control RIS phase shifts, indicating that one element is optimized at a time by observing the improvement of objective functions, e.g., achieving higher data rate or energy efficiency. Then, the next RIS element is optimized sequentially\cite{rivera,Atapattu,tewes}. 

In addition, greedy algorithms are used to relax the constraints. For example, the RIS phase shift is allowed to violate the stringent constraints in \cite{angli}, then the achieved objective values are compared with the theoretical optimal results to find a feasible solution. Greedy schemes may be combined with AO to handle problems with multiple sub-objectives. For instance, a greedy scheme is applied in \cite{muham} to maximize the served users by controlling RIS phase shifts, then it schedules the users to minimize the age of information.  
The main advantage of the greedy algorithm is the low complexity by decoupling the joint optimization into multiple stages. However, instead of global optima, it can only achieve local optima. The simple greedy policy means that there is no guarantee for the algorithm's performance, which may lead to poor output in some cases.

\subsection{Matching Theory-based Methods} 
The matching theory is useful for optimizing resource allocation problems, i.e., subcarrier assignment, user-BS and user-RIS association, and mode selection. These problems are formulated as MINLPs, and a possible solution is to relax the zero-one constraints and reformulate the problem. A linear conic relaxation method is proposed in \cite{zhang2021joint} for the user-RIS association, but it further requires SDP to solve the relaxed formulation, leading to high computational complexity. In addition, the problem becomes even more complicated when RIS on/off control and phase shifts are involved. Consequently, the primary motivation for applying matching theory is to achieve low-complexity solutions efficiently. 

Consider the most widely applied many-to-one matching problem with two finite and disjoint sets of players $\mathcal{U}$ and $\mathcal{B}$. $\mathcal{U}$ represents users, and $\mathcal{B}$ may be BS, RIS or subchannels.
\begin{definition}
The considered many-to-one matching problem is defined by\\
(a) Matching relationship function $f:=\mathcal{U}\times\mathcal{B}$ with $u \in \mathcal{U}$ and $b \in \mathcal{B}$, e.g., the many-to-one association relationship between multiple users and one BS;\\
(b) $|f(u)|=1$ with $\forall u \in \mathcal{U}$, indicating that one $u$ can only be matched with at most one $b$ in many-to-one matching problem, e.g., one user can be associated with at most one BS. \\
(c) $|f^{-1}(b)|\geq K$ with $\forall b \in \mathcal{B}$, which means that $b$ has a capacity limit for the connection with $u$. For example, one BS has a maximum service capability for users.\\ 
(d) $b=f(u)$ $\leftrightarrow$ $u=f^{-1}(b)$. This means the matching is bidirectional and mutual. 
\end{definition}
To describe the exchange operation between different matching, the swap matching is considered 
\begin{definition}
Given $u\in f^{-1}(b)$ and $u' \in f^{-1}(b')$ with $u,u' \in \mathcal{U} $ and  $b,b' \in \mathcal{B} $, the swap matching is defined by $f_{u,b,u',b'}=\{f \backslash \{ (u, b)(u',b') \} \} \cup  \{ (u, b')(u',b) \}$. 
\end{definition} 
Swap matching allows $u$ and $u'$ to exchange their matched $b$ and $b'$, while other players remain unchanged. For instance, two users can exchange their associated BSs without changing other association pairs.
\begin{definition} \label{defi-block}
$u$ and $u'$ become a swap blocking pair if and only if\\
(a) For all players in $\{ u,b,u',b' \}$, $F(f_{u,b,u',b'}) \geq F(f)$, where $F$ is the utility function of players. This means that the utility functions of all involved players will not decrease. \\
(b) At least one player in $\{ u,b,u',b' \}$ has $F(f_{u,b,u',b'}) > $ $ F(f)$, indicating that at least one player's utility is improved, e.g., at least one user achieves higher channel capacity or data rate by switching pairs.
\end{definition}
Definition \ref{defi-block} shows that the overall utility can be improved by finding swap matching pairs. Then the stable matching is defined by
\begin{definition} \label{defi-stable}
The matching relationship between two sets $\mathcal{U}$ and $\mathcal{B}$ is two-sided exchange-stable if there is no swap blocking pairs. This means that the overall utility such as channel capacity or sum-rate cannot be improved by switching user-BS associations.      
\end{definition}

\begin{table*}[!t]
\caption{Summary of heuristic algorithms for RIS-aided wireless networks}
\centering
\small
\setstretch{1.1}
\resizebox{1\textwidth}{!}{%
\begin{tabular}{|m{1.1cm}<{\centering}|m{3cm}<{\centering}|m{4.1cm}<{\centering}|m{3cm}<{\centering}|m{3.8cm}<{\centering}|m{4.4cm}<{\centering}|}
\hline 
Methods &  Main features   &     Advantage    &  Drawbacks    &  Difficulties   &  Application \qquad \qquad \qquad scenarios  \\
\hline
CCP & CCP aims to obtain local optima of DC problems by iteratively finding points with the same tangent values.  &  CCP algorithm does not require a dedicated step size design, and it retains all of the information from the convex
component and only linearizes the concave portion.  & The selection of initial points may affect the final results, and hence it requires an initialization method. &  The problem has to be reformulated into the DC form; initial points may need to be selected several times due to local optima.  & Penalty CPP is used for imperfect CSI in \cite{guiz3}, and statistical CSI in \cite{yuanbin}. Other applications include non-convex RIS control problems in \cite{chen2021qos,niu2021simultaneous,xu2022reconfigurable}. \\
\hline
Meta-heuristic algorithms & Meta-heuristic algorithms apply intelligent policies to guide the heuristic exploration iteratively, producing high-quality solutions efficiently. &  Meta-heuristic algorithms can search huge solution spaces with few or no additional assumptions.  & It can be trapped in local optima, and these algorithms require many iterations. The algorithm performance is sensitive to parameter selections.  & Selecting the best parameters is complicated in meta-heuristic algorithms. For instance, crossover probability in GA and inertia weight in PSO may decide the algorithm performance.   & Meta-heuristic algorithms are mainly deployed for RIS phase-shift control, including GA for rate maximization in \cite{zhi2022power}, PSO for phase-shift optimization in \cite{dai2021reconfigurable, zhi2021statistical}, tabu search for irregular RIS in \cite{ruoc}.     \\
\hline
Greedy algorithms &  Instead of global optima, greedy algorithms make locally optimal decisions at each stage of solving the problem.    & Greedy algorithms can greatly reduce the complexity by decoupling the original 
 problem into multiple stages.   &  It may present poor global performance, since the local optima in one stage may lead to bad results for the next stage.  &  Finding the trade-off between low complexity and good algorithm performance is critical to using the greedy heuristic.   & Greedy algorithms are used for RIS phase control in \cite{rivera,Atapattu,tewes} and on/off control in \cite{zhaohui} as low-complexity solutions, providing low-complexity alternative solutions for NP-hard problems.  \\
 \hline
Matching theory-based method &  Matching-based method is designed to solve matching or association problems with two sides of players.  &  Compared with direct optimization methods, matching-based methods have lower complexity for large-scale problems.  &  Matching-based method may require exhaustive searches to find matching pairs.  &  Defining the utility function for two sides of players with peer effects is difficult.  & Matching theory is mainly applied to resource allocation and association problems in RIS-aided networks, e.g., channel assignment and user-BS-RIS association\cite{wu2022resource, zuo2020resource}.    \\
 \hline
\end{tabular}}
\label{tab-heuristic}
\vspace{0pt}
\end{table*}

\begin{figure}[!t]
\centering
\includegraphics[width=0.9\linewidth]{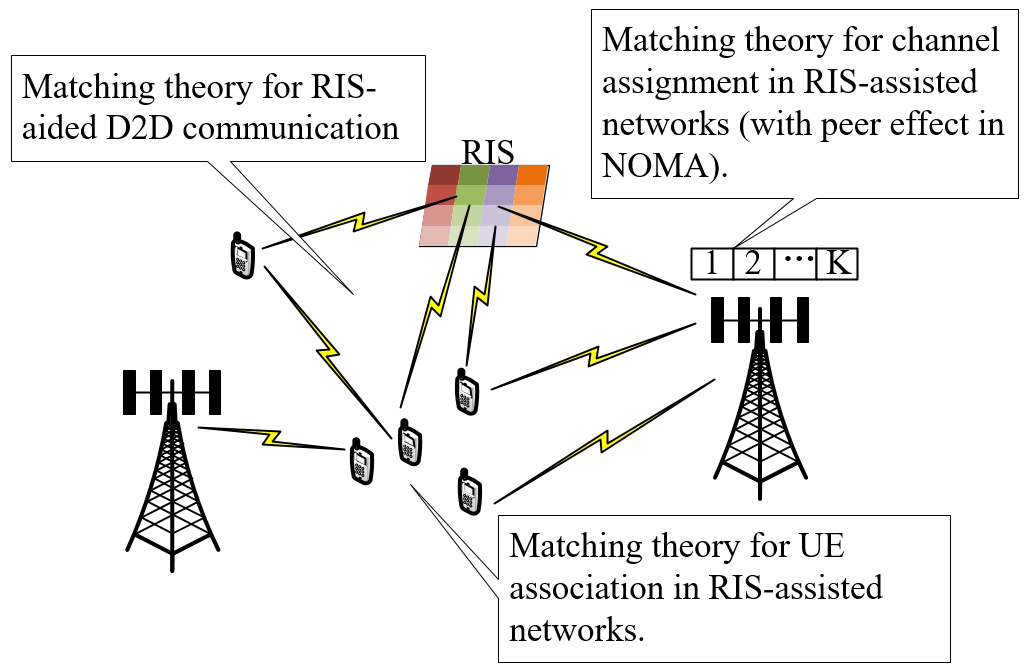}
\caption{Matching theory applications in RIS-aided wireless networks.}
\label{fig-match}
\setlength{\abovecaptionskip}{-2pt} 
\vspace{-10pt}
\end{figure}

%The existence of stable matchings has been proved in \cite{bodine2011peer}. 
Definition \ref{defi-stable} is very useful in matching theory, since it provides locally optimal criteria, and it is easily achieved by searching and eliminating all the swap blocking pairs. Fig. \ref{fig-match} presents the applications of matching theory in RIS-aided wireless networks, including D2D-user pairing, user-BS-RIS association, channel assignment, etc. It shows that matching theory provides an efficient solution for overcoming these NP-hard problems. For instance, a RIS-aided maritime communication system is investigated in \cite{cao2022joint}, in which many-to-one matching was applied for the joint mode selection and power control of BSs. In RIS-assisted NOMA system, many-to-one matching is used for channel assignment \cite{wu2022resource, zuo2020resource} and user clustering \cite{zhang2020joint}, while many-to-one and many-to-many matching are jointly considered in \cite{ni2021resource} for the UE association and channel assignment. Moreover, matching theory is applied in \cite{el2022latency} for edge computation offloading in RIS-aided networks, and a deferred acceptance matching game is formulated in \cite{taghavi2021user} for user association in mmWave networks with RISs.
The simulations in \cite{cao2022joint,wu2022resource, zuo2020resource,zhang2020joint, ni2021resource, el2022latency, taghavi2021user} demonstrate that matching theory is a low-complexity solution for resource allocation and association problems in RIS-aided communication systems. 
However, matching theory relies on iterative searching to eliminate swap blocking pairs, and the searching cost may increase exponentially with more players. In addition, the wireless network players will affect each other, changing the overall interference level. Such peer effects may increase the complexity of applying matching theory.

\begin{figure}[!t]
\centering
\subfigure[Sum-rate comparison under various numbers of RIS elements. The greedy algorithm applies element-by-element RIS phase-shift control. It decides the phase-shift of one element at each time by observing the improvement in sum-rate, and then moves to the next element. The genetic algorithm considers different phase shift designs as individuals, and uses evolutionary strategies to find near-optimal solutions.]{ \label{fig_result3}
\includegraphics[width=7.2cm,height=5.2cm]{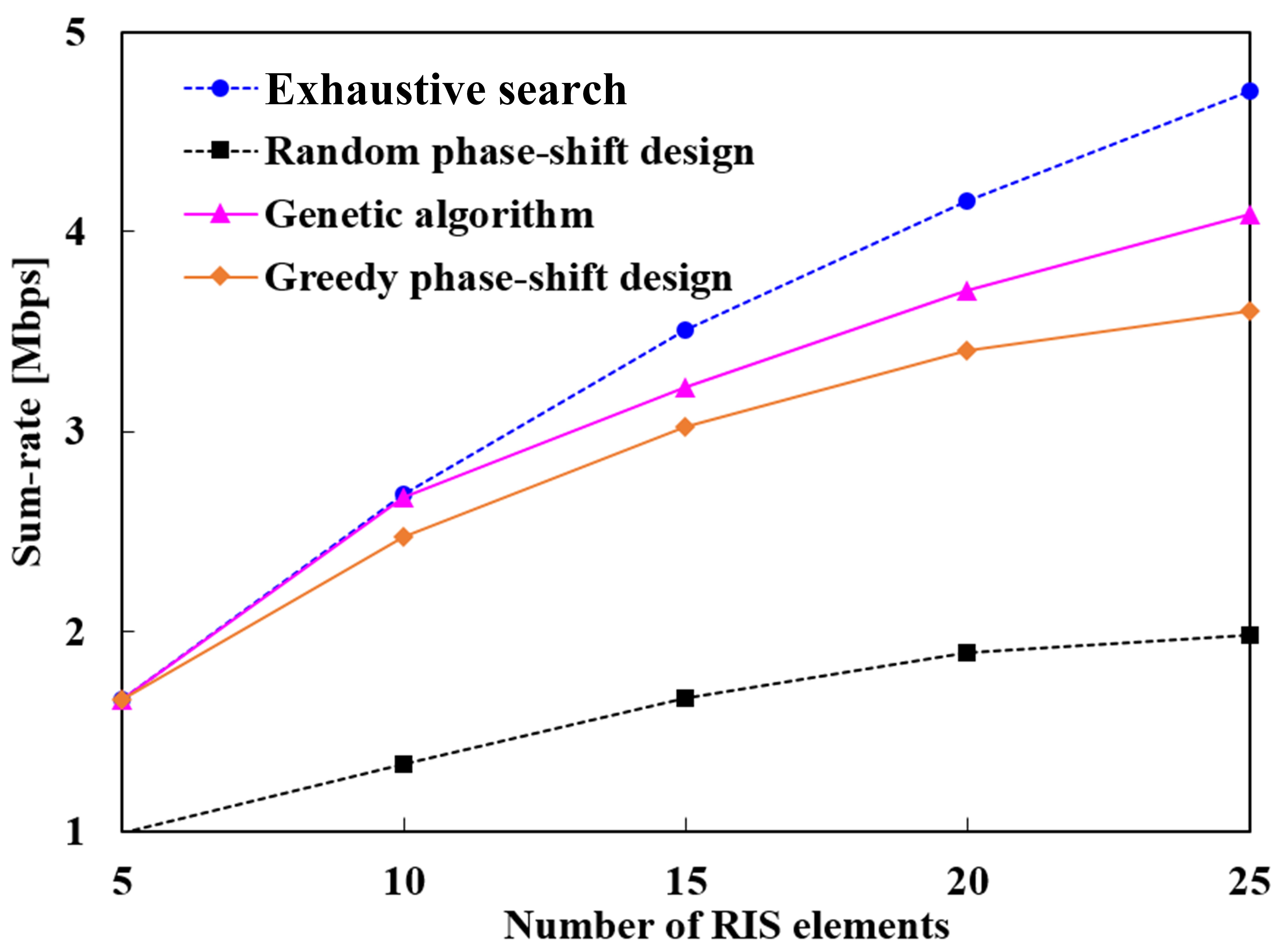}
}
\,
\subfigure[Convergence performance of the genetic algorithm ]{ \label{fig_result4}
\includegraphics[width=7.2cm,height=5.2cm]{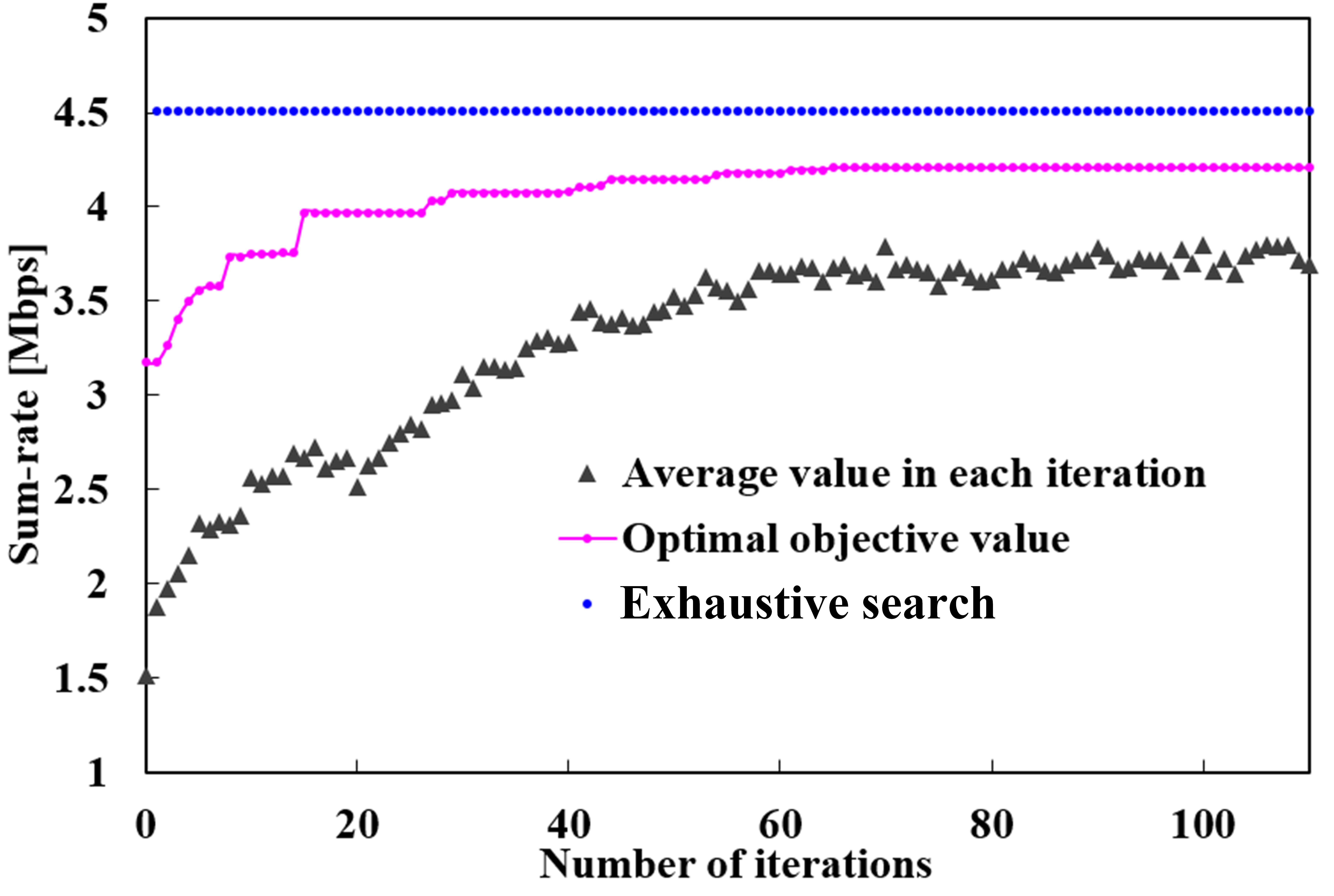}
}
\setlength{\abovecaptionskip}{0pt} 
\caption{\blue{Simulation results of greedy and genetic algorithms. We consider a MISO system with one BS and multiple UEs, and detailed simulation parameters and algorithms can be found in \cite{zhou2023heuristic}}}.
\vspace{-15pt}
\label{fig-re-heu}
\end{figure}

\subsection{Discussions and Numerical Results}

Table \ref{tab-heuristic} compares heuristic algorithms in terms of main features, advantages, drawbacks, difficulties, and applications.  
Compared with model-based algorithms, a common advantage of heuristic algorithms is their low complexity. 

RIS-related optimization problems may involve summation, logarithm, fractional terms, and discrete constraints in problem formulations, which are non-convex and highly non-linear. Applying model-based algorithms generally require a series of transformation and relaxation to achieve a convex or a concave reformulation, but this complexity is avoided in heuristic algorithms. For instance, many problems are easily converted into DC forms, and the CCP algorithm can be applied by iteratively finding two points with the same tangent vectors. Compared with other estimation-based methods such as MM or SCA, the CCP method has much lower complexity, since no extra surrogate function is required. However, the initial point selection may affect the solution quality of the CCP algorithm.  

Greedy algorithms employ a simple greedy policy for decision-making. They aim to maximize the current benefit, disregarding the effect on future stages. Greedy algorithms can efficiently solve problems in near-linear time complexity. However, greedy algorithms can only generate locally optimal results, and the increasing number of control variables may lead to poor performance.

By contrast, meta-heuristic algorithms apply more advanced heuristic rules for iterative exploration, e.g., genetic algorithm, tabu search, and PSO. 
Similar to greedy algorithms, meta-heuristic algorithms have no requirements for problem formulations and constraints, and objective functions can be easily converted into fitness functions. 
However, compared with greedy algorithms, meta-heuristic algorithms can better guarantee the solution quality by using heuristic rules for iterative optimization.

Different from previous approaches, matching theory specializes in solving resource allocation and association problems.
In matching-based methods, the control variables are considered as matching operations, and the objective function is improved by searching swap matching pairs. Therefore, when handling these allocation problems, matching theory is more efficient than other heuristic algorithms due to its dedicated design.

Note that heuristic algorithms may be combined with model-based algorithms. For instance, the energy-efficiency maximization problem in \cite{zhaohui} is decoupled into the beamforming optimization, phase control, and RIS on/off optimization, in which beamforming and phase control are solved by SCA, and RIS on/off is optimized by the greedy algorithm. Other combinations can be found in \cite{Atapattu} by combining SDR with greedy heuristic, and in \cite{ni2021resource} by combining SCA and SDR with matching methods. 

Finally, Fig. \ref{fig-re-heu} shows an example with greedy and genetic algorithms. In particular, the greedy algorithm applies element-by-element RIS phase-shift control. It decides the phase-shift of one element at each time by observing the improvement in sum-rate, and then moves to the next element. Meanwhile, genetic algorithm considers different phase shift combinations as individuals, and uses evolutionary strategies to find near-optimal solutions. Fig. \ref{fig_result3} provides the sum-rate under various numbers of RIS elements. It shows that heuristic algorithms can achieve satisfactory performance with a limited number of RIS elements. However, when the number of RIS elements increases, both greedy and genetic algorithms present sub-optimal results. In addition, Fig. \ref{fig_result4} illustrates the convergence performance of the genetic algorithm. It shows that the average values of individuals increase with iterations, and finally the optimal objective value converges. The main reason is that the genetic algorithm applies evolutionary policies, which will select elite individuals to produce new solutions, and therefore the solution quality is constantly improved.

%% file: 6_Machine_learning.tex
\section{ML-enabled Optimization for RIS-aided Wireless Networks }
\label{sec-ml}

ML has achieved great success in various fields, and this section investigates ML applications for the control and optimization of RIS-aided wireless networks, including supervised learning, unsupervised learning, RL, FL, graph learning, transfer learning, hierarchical learning, and meta-learning.

A variety of algorithms have been developed to optimize RIS-aided wireless networks. Early studies mainly considered model-based methods, and some heuristic algorithms are deployed as low-complexity solutions. However, there are several challenges for these conventional optimization techniques:

1) \textbf{Highly dynamic wireless environment}: Wireless networks are highly dynamic due to frequently changing channel conditions, traffic demands, and user conditions. These dynamics lead to great difficulty for conventional optimization schemes. As an example,  model-based methods need full knowledge of the formulated problem, but some sensitive information, e.g., real-time user locations, may be unknown in practice.

2) \textbf{Evolving network architecture}: The wireless network architecture is constantly evolving from RAN to cloud RAN, virtual RAN, and Open RAN. Consequently, these new architectures increase the complexity of network management, and conventional algorithms may have difficulty modelling and optimizing such complicated systems.

3) \textbf{Diverse user requirements}: Wireless network user types are not limited to enhanced Mobile Broad Band, Ultra Reliable Low Latency Communications, and massive Machine Type Communications. Some newly emerged applications, such as virtual and augmented reality, have more stringent requirements on network metrics, leading to a great burden for conventional optimization methods. 

Given these challenges, ML-enabled control and optimization techniques have become appealing approaches for wireless communications in general, as well as for  RIS-aided wireless networks. In the following, we will introduce the fundamentals and applications of various ML techniques. \blue{It is worth noting that ML algorithms can be applied to optimize RIS-aided networks in various ways, e.g., controlling RIS elements directly or jointly optimizing the whole RIS-aided network scenario. Here we focus on the application of using ML algorithms to optimize RIS elements directly, e.g., supervised learning-based sum-rate prediction, unsupervised learning-enabled RIS phase-shift optimization, RL-enabled RIS phase-shift control, and so on. }

\begin{table*}[!tbhp]
\caption{Summary of supervised learning for RIS-aided wireless networks}
\centering
\small
\setstretch{1.1}
\resizebox{1\textwidth}{!}{%
\begin{tabular}{|m{0.6cm}<{\centering}|m{2cm}<{\centering}|m{1.5cm}<{\centering}|m{2cm}<{\centering}|m{1.1cm}<{\centering}|m{1.7cm}<{\centering}|m{1cm}<{\centering}|m{1.5cm}<{\centering}|m{2.5cm}<{\centering}|m{2.2cm}<{\centering}|m{2.7cm}<{\centering}|}
\hline 
Ref. &  Scenario  & Phase-shift resolution & Channel settings & CSI & Objectives  &  Model &  Layer  &  Data acquisition & Input data  & Output data \\
\hline
\cite{zhang2021deep} & \multirow{2}*{\makecell{Point-to-point\\ SISO}}  &  \multirow{2}*{Discrete} & \multirow{2}*{\makecell{DeepMIMO \\ dataset}}  &  \multirow{2}*{Predicted} & Maximizing data rate  &  CNN  & 9 Conv2D layers  & Collected from fully active model & Estimated partial channels   &  Full channel information  \\
\cline{6-10}
 &   &  & &  &  &  FNN  & 5 Layers  &  Codebook &  CSI  & RIS phase shifts  \\
\hline
\cite{stylianopoulos2022deep} &  Rich-scattering Point-to-point &  Binary &  Generated by simulator    &  Predicted  & Maximizing data rate  &  DNN  &  4 layers  & Obtained from simulators & RIS phases   &  Second-order moments of CSI. \\
\hline
\cite{taha2021enabling} & Point-to-point SISO  & Continuous &  Wideband geometric & Partially  &  Maximizing data rate  &   DNN  &  6 layers   &  Exhaustive generation  &  CSI  & Estimated data rate  \\
\hline
\cite{aygul2021deep} & Point-to-point SISO   & Continuous  &   Wideband geometric   &  Perfect    & Maximizing data rate    & DNN   & 5 layers & Exhaustive search beamforming  &  Pilot signals   &  RIS phase shifts  \\
\hline
\cite{alexandropoulos2020phase} & Point-to-point SISO  &  Discrete &  Rician fading   &  Perfect & Maximizing data rate & DNN  &   5 layers  &  Simulation generated   & Transmit power, and positions    & RIS phase shifts   \\
\hline
\cite{ozdougan2020deep} & MISO-DL-SU  & Continuous &   Quasi-static flat-fading   &   Estimated  &  Maximize energy efficiency  &  DNN  &  5 layers; 6 layers  &  Separately generated  & Pilot signal  & RIS phase shifts and BS beamforming vector  \\
\hline
\cite{huang2019indoor} &  MISO-DL-SU  & Continuous &   Rayleigh fading   & Perfect     &  Maximizing data rate  & DNN    & 5 layers      &  Generated by estimation   &  User positions  & RIS phase shifts  \\
\hline
\cite{yang2021intelligent} & SISO-UL-MU   & Continuous &   Quasi-static  &  Perfect  & Maximizing SINR   &   CNN    &  5 layers   &  Collected using by USRP2 testbed  &  Incident RF signal  & Interfering user set  \\
\hline
\cite{song2021truly} & MISO-DL-MU  &  Continuous &  Rician fading  &  Perfect  & Maximizing secrecy rate  &  DNN    & 6 layers    &  Generated by AO algorithm     & Channel coefficients   & RIS phase shifts   \\
\hline
\cite{hu2021reconfigurable} & \multirow{2}*{\makecell{ MISO-DL-MU \\MEC }}  &  \multirow{2}*{Continuous} &   \multirow{2}*{\makecell{ Rician fading }}   &   \multirow{2}*{Predicted} &  \multirow{2}*{\makecell{Maximizing \\ data rate}}  & DNN   & 7 layers  & \multirow{2}*{\makecell{Generated by \\ BCD algorithm}} & CSI  &  \multirow{2}*{\makecell{RIS phase shifts; \\ offloading decision}}   \\
\cline{7-8} \cline{10-10}
 &   &  & &  &  &  DNN  & 7 layers  &   &  UE positions  &  \\
\hline
\end{tabular}}
\label{tab-supervised}
\vspace{0pt}
\end{table*}

\begin{figure*}[!t]
\centering
\setlength{\abovecaptionskip}{-5pt} 
\includegraphics[width=0.95\linewidth]{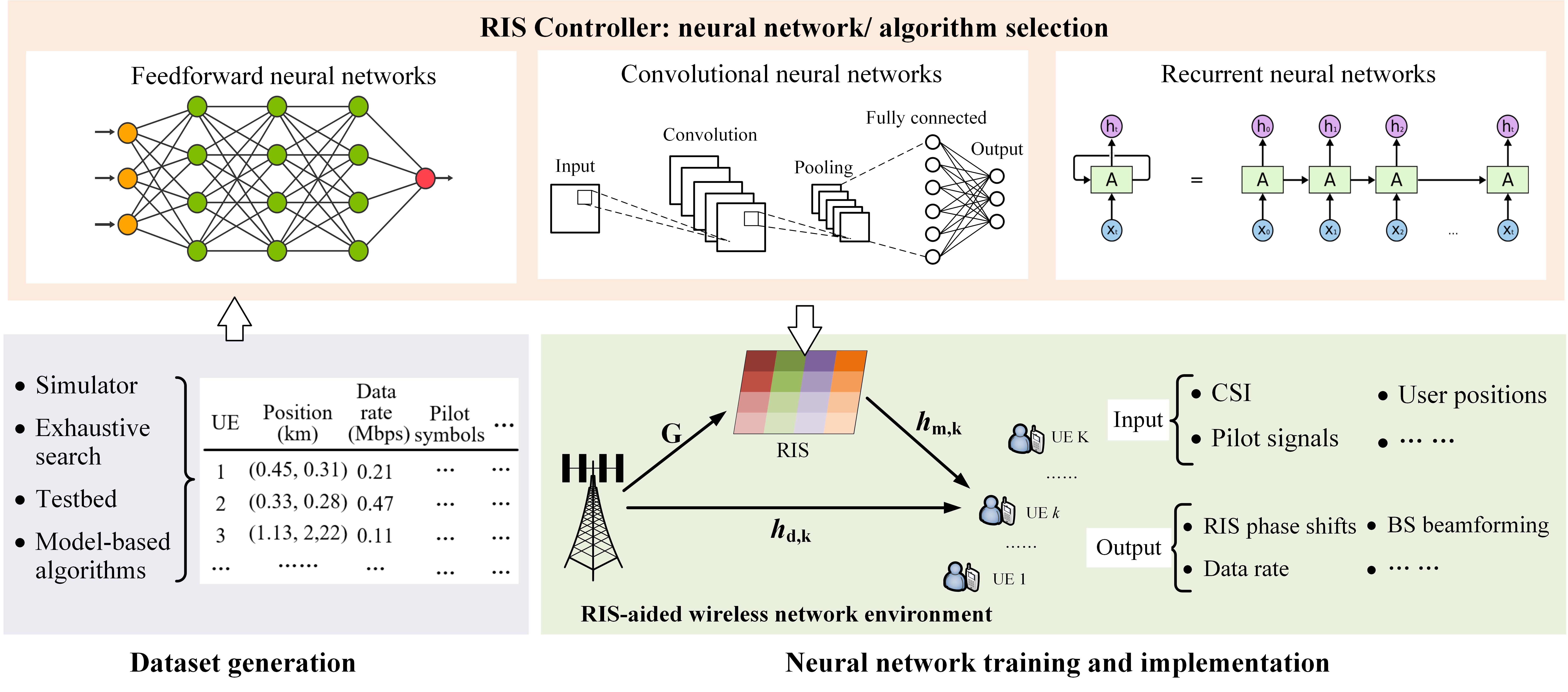}
\caption{Supervised learning for RIS-aided wireless networks.}
\label{fig-nn}
\vspace{0pt}
\end{figure*}

\subsection{Supervised Learning-enabled Optimization}

Supervised learning is designed to find the hidden relationships between inputs and labeled outputs. Supervised learning algorithms adjust their parameters to map the input to the expected output, and this relationship is used for the prediction and classification of unseen data. Table \ref{tab-supervised} summarizes supervised learning-based control and optimization studies for RIS-aided wireless networks. 
It shows that most studies consider partial CSI or pilot signals as input to predict full CSI or RIS phase shifts, and then utilize the prediction results to maximize the data rate. Meanwhile, there are various approaches for neural network model selection, dataset acquisition, input and output data definitions, etc.
This subsection will discuss how to apply supervised learning for optimizing RIS-aided wireless networks, including data acquisition, neural network architecture, loss functions and algorithm training. 

\subsubsection{Dataset Acquisition in RIS-aided Environments}
\label{sec-dataset}
A fine-grained dataset is the prerequisite for deploying supervised learning, since it relies on the labeled output for validation. Table \ref{tab-supervised} indicates that the dataset is generated in various ways: simulators, exhaustive searches, codebook, model-based optimization algorithms or live networks. For example, the exhaustive searches mean trying different solutions and then collecting the corresponding output to form labeled datasets\cite{taha2021enabling}\cite{aygul2021deep}. By contrast, a more efficient method is to reuse the data produced by AO\cite{song2021truly} and BCD\cite{hu2021reconfigurable} as model-based optimization algorithms. 

In addition, the algorithm performance also depends on the dataset size, ranging from 5000 \cite{alexandropoulos2020phase} and 30000\cite{aygul2021deep,taha2021enabling} to 200000 \cite{hu2021reconfigurable} in several studies. The simulation results in \cite{aygul2021deep,taha2021enabling} demonstrate that the achievable data rate is significantly improved when the number of training samples increases from 5000 to 30000. Note that the complex entry of the input data, especially the channel coefficient, is usually split into real and imaginary parts, increasing the dimension of the neural network input. 
Although there are several ways to generate the data for supervised learning, most existing datasets are simulation-based. Realistic datasets that are produced in real-world RIS-aided environments are still very rare. 

\subsubsection{Loss Functions and Algorithm Training}
Given the huge number of training samples, supervised learning models are trained to produce the expected output. Suppose that the prediction output is the RIS phase shifts \cite{zhang2021deep,aygul2021deep,alexandropoulos2020phase}, and   
the loss function is defined to minimize the mean square error (MSE) of algorithm training
\begin{equation}\label{ml-nnloss}
Loss (\omega)=\frac{1}{N}\sum^N_{i=1}(\theta_i-\hat{\theta}_{i}(\omega))^2,    
\end{equation}
where $N$ is the total number of outputs, i.e., the number of RIS elements, $\theta_i$ is the desired phase shift given by the dataset, $\omega$ is the neural network weight, and $\hat{\theta}_{i}(w)$ indicates the RIS phase shifts predicted by neural networks. The desired phase shift $\theta_i$ can be obtained in various ways, such as \blue{exhaustive} search or model-based approaches\cite{taha2021enabling,song2021truly}, which have been introduced in Section \ref{sec-dataset}. For example, Song \textit{et al.} apply AO to produce a dataset with desired targets for DNN training\cite{song2021truly}, and Hu \textit{et al.} apply BCD algorithm to generate target phase shift to train DNN models\cite{hu2021reconfigurable}. Meanwhile, note that the dataset must be divided into training and validation samples, since the objective of algorithm training is to predict unseen data. For example, the authors in \cite{song2021truly} include 10000 samples to predict the RIS phase shifts, of which 90\% is used for training and the remaining 10\% for testing purposes.

\subsubsection{Neural Network Architecture and Overfitting}
Table \ref{tab-supervised} shows that DNN is used in most studies to predict CSI or RIS phase shifts, and the network architecture ranges from 4 to 9 layers. It is known that more hidden layers may provide a better performance, but the computational complexity and training time will increase. Hence, the network architecture selection should consider the trade-off between performance and training costs.  

Overfitting is another important issue for neural network training. It means that the algorithm fits exactly to the current training data, but cannot achieve satisfactory prediction for unseen data, which should be carefully prevented. One solution is to add a random dropout layer with probabilities, ignoring the contribution of some neurons \cite{hu2021reconfigurable}. Multiple methods are provided by \cite{song2021truly} to suppress overfitting in predicting RIS phase shifts, including larger datasets (CSI and RIS phase shift pairs), decreasing hidden layers, and early stopping. 

Fig. \ref{fig-nn} summarizes how to apply supervised learning for RIS-aided wireless networks. Firstly, the datasets can be produced by various methods, including simulators, exhaustive searches, testbed, and model-based methods. The collected dataset may include UE positions, data rates, and pilot signals received at the transmitter and receiver, which mainly depends on the designed prediction algorithms. Then, one specific model will be selected, i.e., FNN, convolutional neural networks (CNNs), and recurrent neural networks (RNNs). Note that each neural network model has unique features and advantages, e.g., RNNs are suitable for handling sequential data, and CNNs can better handle spatial data. The selection of neural network models requires case-by-case analyses of the dataset size, quality, and data-processing demands.  The number of nodes and hidden layers of neural networks should be carefully designed, which will affect the network training time and accuracy. Finally, selected models are trained and implemented, and the algorithm output includes RIS phase shifts, achieved data rate, BS beamforming vectors and so on, which are further used to optimize network performance.
Supervised learning has been widely used for wireless networks. However, note that it relies on high-quality labeled datasets for model training, which may be inaccessible in practice. In addition, the algorithm performance is sensitive to hyperparameters, and the fine-tuning of parameters requires considerable experience.

\begin{table*}[!t]
\caption{Summary of unsupervised learning for RIS-aided wireless networks }
\centering
\small
\setstretch{1.05}
\resizebox{1\textwidth}{!}{%
\begin{tabular}{|m{0.7cm}<{\centering}|m{2cm}<{\centering}|m{1.5cm}<{\centering}|m{1.9cm}<{\centering}|m{1.3cm}<{\centering}|m{1.7cm}<{\centering}|m{1cm}<{\centering}|m{1.5cm}<{\centering}|m{2.7cm}<{\centering}|m{2.4cm}<{\centering}|m{2.2cm}<{\centering}|}
\hline 
Ref. &  Scenario  & Phase-shift resolution & Channel settings & CSI & Objectives  &  Model &  Layer  &  Data generation & Input data  & Output data \\
\hline
\multirow{3}*{\cite{song2020unsupervised}} & \multirow{3}*{\makecell{MISO-DL-MU}} &   \multirow{3}*{Continuous}  & \multirow{3}*{\makecell{Rician \\ fading}}  &\multirow{3}*{Perfect} &  \multirow{3}*{\makecell{Maximizing \\ sum-rate}}  &  CNN  &  6 layers &  \multirow{3}*{\makecell{Generated by \cite{huayan}}}  &  CSI  &  RIS phase shifts  \\
\cline{7-8} \cline{10-11}
 &   &  & & &  &  FNN & 5 layers  &   & Effective channel matrix   &  BS beamforming vector  \\
\hline
\cite{nguyen2021machine} & MIMO-DL-SU  &  Continuous &   Rician fading   &  Perfect   &  Maximizing spectral efficiency  & DNN  &  4 layers  & Generated by random exploration  & CSI   & RIS phase shifts   \\
\hline
\cite{dinh2022unsupervised} & Broadcasting
Communications for IoTs  &  Continuous &  Rician fading   & Statistical  &  Maximizing spectral efficiency & DNN  & 4 layers   &  Generated by random exploration  &  CSI  & RIS phase shifts  \\
\hline
\cite{gao2020unsupervised} & MISO-DL-SU  &  Continuous &  Rayleigh fading   &  Perfect & Maximizing data rate  & FNN   &   7 layers    & Generated as \cite{huang2019indoor}   &  CSI   & RIS phase shifts  \\
\hline
\multirow{4}*{\cite{lopez2022deep}} & \multirow{4}*{\makecell{MISO-DL-MU}}  &  \multirow{4}*{\makecell{Continuous/  \\ discrete}} &  \multirow{4}*{\makecell{ Geometry-based \\ clustered\\ delay line}} &  \multirow{4}*{Estimated} &  \multirow{4}*{\makecell{Maximizing \\ sum-rate}} &  FNN &  6 layers  &  Obtained from \cite{bjornson2021configuring}  &  CSI  &  RIS phase shifts \\  
\cline{7-11} 
&   &  &  &  & &  \multicolumn{3}{c|}{ \makecell{k-means is used to cluster RIS elements \\ based on their estimated cascaded channel \\ coefficient without dataset.}}  & Estimated cascaded channel of RIS elements.  &  RIS element clusters.  \\
\hline
\end{tabular}}
\label{tab-unsuper}
\vspace{-10pt}
\end{table*}

\subsection{Unsupervised Learning-based Optimization}

Supervised learning is data-demanding, but fine-grained labeled datasets may be inaccessible in practice, preventing the application of supervised learning algorithms. On the contrary, unsupervised learning can find hidden patterns of unlabeled data without predefined targets or human intervention. Table \ref{tab-unsuper} summarizes unsupervised learning algorithms for RIS-aided wireless networks. It shows that neural networks are used in unsupervised manners for RIS phase-shift configuration. Meanwhile, some classic unsupervised learning methods, such as k-means, can also be applied for clustering RIS elements. This subsection will introduce unsupervised neural networks and clustering algorithms.

\begin{figure}[!t]
\centering
\setlength{\abovecaptionskip}{-5pt} 
\includegraphics[width=0.8\linewidth]{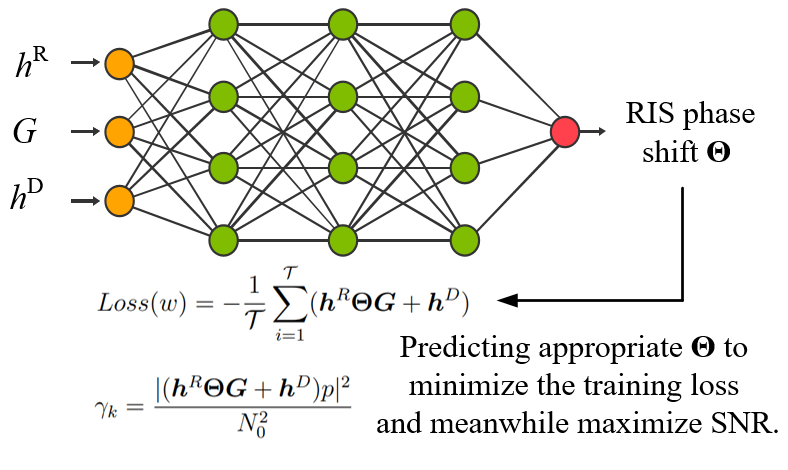}
\caption{Unsupervised neural networks for optimizing RIS phase shifts.}
\label{fig-unsupnn}
\vspace{0pt}
\end{figure}

\subsubsection{Algorithm Training and Network Architecture of Unsupervised Neural Networks} \label{s-ml-un} 
Supervised neural networks aim to minimize the loss between predicted results and desired target, i.e., predicted and target data rate in the dataset. However, in unsupervised neural networks, the loss function is directly related to optimization objectives. 
Specifically, we consider a single-user scenario as an example, and the user SNR is
\begin{equation}\label{eq-unsnr}
\eta_{k}=\frac{|(\bm{h}^{R}\bm{\Theta}\bm{G}+\bm{h}^{D})p|^2}{N_{0}^2},
\end{equation}
where $p$ is the transmit power at the BS, $\bm{G}$ indicates the channel gain from BS antennas to RIS elements, $h^{R}$ indicates the channel gain from RIS elements to the user, $h^{D}$ indicates the channel gain from BS antennas to the user, $N_{0}^2$ is the noise power, and $\bm{\Theta}$ is the matrix of RIS phase shifts. As shown in Fig. \ref{fig-unsupnn}, the neural network considers the channel state information as input, including $\bm{G}$, $\bm{h}^{R}$ and $\bm{h}^{D}$. Then, the output is the predicted RIS phase shifts $\bm{\Theta}$. The loss function is defined by
\begin{equation} \label{eq-unsup}
Loss (w)=-\frac{1}{\mathcal{T}}\sum^{\mathcal{T}}_{i=1}(\bm{h}^{R}\bm{\Theta}\bm{G}+\bm{h}^{D}),     
\end{equation}
where $\mathcal{T}$ is the minibatch size. To minimize the loss function equation (\ref{eq-unsup}), $\bm{h}^{R}\bm{\Theta}\bm{G}+\bm{h}^{D}$ must be maximized. This means that the neural network must predict appropriate RIS phase shifts $\bm{\Theta}$ to maximize $\bm{h}^{R}\bm{\Theta}\bm{G}+\bm{h}^{D}$, and the SNR will be maximized accordingly. 

Table \ref{tab-unsuper} shows that most existing works apply 2 to 5 hidden layers. In particular, the hidden layer numbers are related to the problem's complexity. The authors in \cite{nguyen2021machine} used 1 hidden layer with 40 nodes for $8 \times 2$ MIMO, and 2 hidden layers for $16 \times 2$ MIMO, achieving satisfactory simulation results without overfitting or underfitting. In addition, similar to supervised learning, early stop is applied in \cite{song2020unsupervised} to prevent overfitting.   

\subsubsection{Clustering Algorithms} 
Clustering algorithms are usually unsupervised ML algorithms, i.e., k-means and Density-based spatial clustering of applications with noise (DBSCAN). These algorithms are designed to partition objects into multiple sets to minimize the within-cluster sum of squares. Specifically, it aggregates objects with the same hidden patterns. For instance, k-means is used in \cite{lopez2022deep} to group RIS elements according to estimated channel coefficients, and then each group has the same RIS configurations to reduce the computational complexity. 

The main advantage of unsupervised learning is that it has no requirement on predefined targets, which is more practical in real-world applications. However, the absence of targets means that the model output is hard to validate or verify, and the solution quality cannot be guaranteed.

\begin{table*}[!t]
\caption{Summary of reinforcement learning for RIS-aided wireless networks }
\centering
\small
\setstretch{1.05}
\resizebox{1\textwidth}{!}{%
\begin{tabular}{|m{0.6cm}<{\centering}|m{2.1cm}<{\centering}|m{1.5cm}<{\centering}|m{1.7cm}<{\centering}|m{1.3cm}<{\centering}|m{1.7cm}<{\centering}|m{1.8cm}<{\centering}|m{2.9cm}<{\centering}|m{3.1cm}<{\centering}|m{3.3cm}<{\centering}|}
\hline 
Ref. &  Scenario  & Phase-shift resolution & Channel settings   & CSI & Objectives &  Algorithm  &   State definition  &  Action definition  & Reward function  \\
\hline
\cite{yang2020deep} & MISO-DL-MU NOMA  & Discrete &  Rayleigh fading  & Perfect  & Maximizing sum-rate  &  DDPG  & Current RIS phases &  RIS phase shifts    &  Sum-rate    \\
\hline
\cite{taha2020deep} & Point-to-point communications  &  Discrete   & Wideband geometric   & Estimated   &  Maximizing data-rate &  DRL & CSI  &  RIS phase shifts   & Data rate   \\
\hline
\cite{guo2021learning} &  MISO-DL-MU UAV  &  Continuous  & Saleh-Valenzuela  &  Imperfect  & Maximizing secrecy rate  &  DDPG  &  CSI  & RIS phase shifts and BS beamforming vector  &  Secrecy rate with penalty   \\
\hline
\cite{yang2020intelligent} & MISO-DL-MU with jammer & Continuous  &  Quasi-static flat-fading &  Perfect    & Maximizing sum-rate  &  Fast-policy hill-climbing learning  &  Previous jammer power and SINR, current CSI  & BS transmit power and RIS phase shifts   &   Maximizing data-rate, decreasing BS power and SINR penalty\\
\hline
\cite{yang2020deep2} & MISO-DL-MU  &  Continuous   & Rayleigh fading  &  Delayed &  Maximizing secrecy rate  &  DRL  &  CSI, previous secrecy rate and transmission rate, QoS level  & BS beamforming vector and RIS phase shifts   &  Maximize the system secrecy rate, guaranteeing QoS requirements.  \\
\hline
\cite{feng2020deep} & MISO-DL-SU  & Continuous  &  Rayleigh fading  &  Perfect  & Maximizing SNR  &  DDPG  &  SNR and current RIS phases   &  RIS phase shifts     &  Received SNR  \\
\hline
\cite{huang2020hybrid} & MISO-DL-MU THz &  Continuous  &  Rayleigh distribution  &  Perfect     &    Maximizing sum-rate  &  DDPG  &  Current BS and RIS beamforming vectors, CSI   &  BS beamforming vectors and RIS phase shifts   &  Throughput and the penalty of adjusting the beamforming direction.  \\
\hline
\cite{lee2020deep} & MISO-DL-MU  & Continuous  & Quasi-static flat-fading  &  Perfect  & Maximizing energy efficiency &  DRL  & CSI and energy level of RIS   & BS beamforming vector, RIS phase shifts and on/off   &  Energy efficiency  \\
\hline
\cite{lin2020deep} & MISO-DL-SU  &  Continuous  &  Quasi-static flat-fading   &  Perfect  & Power minimization  & DDPG   & CSI, previous outage events  & BS beamforming vector and RIS configurations  &  Energy efficiency  \\
\hline
\cite{huang2020reconfigurable} & MISO-DL-MU  & Continuous  &  Frequency flat fading   & Perfect & Maximizing channel capacity & DDPG & Transmit and received power, previous action, and CSI    & BS beamforming vector and RIS phase shifts  & Channel capacity    \\
\hline
\cite{zhang2021millimeter} & MISO-DL-MU mmWave &  Continuous  & 3GPP model  & Perfect/ Imperfect  &   Maximizing sum-rate   &   Distributed RL   &  CSI   &  RIS phase shifts  &  Data rate   \\
\hline
\cite{liu2020ris} &  MISO-DL-MU NOMA  & Continous & Rayleigh fading   &  Perfect  &  Maximizing energy efficiency  & Decaying DDQN   & RIS phases and positions, UE positions, and current BS power allocations & RIS phase and position and BS beamforming changes   &   Energy efficiency with penalty  \\
\hline
\cite{kim2021multi} &  Multi-cell communications  & Continuous  & Rayleigh fading   & Imperfect  &  Maximizing sum-rate &  Multi-agent DRL  &  Local and neighbor CSI, local sum-rate   & RIS phase shifts, BS beamforming vector, and UE power changes     & Sum-rate with interference penalties \\
\hline
\cite{samir2021optimizing} & MISO-UL-MU IoT UAV  &  Continuous  & Rician fading   & Perfect  & Minimizing sum AoI   &   DRL    & SNR and UAV height  & UAV  altitude changes  & Negative summation of age of information   \\
\hline
\multirow{2}*{\cite{yang2021machine}} & \multirow{2}*{\makecell{MISO-DL-MU \\ NOMA}}  &  \multirow{2}*{Continuous}  & \multirow{2}*{ \makecell{Rayleigh \\ fading}}  & \multirow{2}*{Perfect} &  \multirow{2}*{\makecell{Maximizing \\ sum-rate}}  & Object migration automation   &  Current RIS phase  &  Power allocation coefficient  &  Sum-rate  \\
\cline{7-10}
 &   &  &  & &   & DDPG  & Current RIS phase  &  RIS Phase changes  & Sum-rate difference \\
\hline
\end{tabular}}
\label{tab-rl}
\vspace{0pt}
\end{table*}

\subsection{Reinforcement Learning-based Optimization}

\begin{figure*}[!t]
\centering
\includegraphics[width=0.85\linewidth]{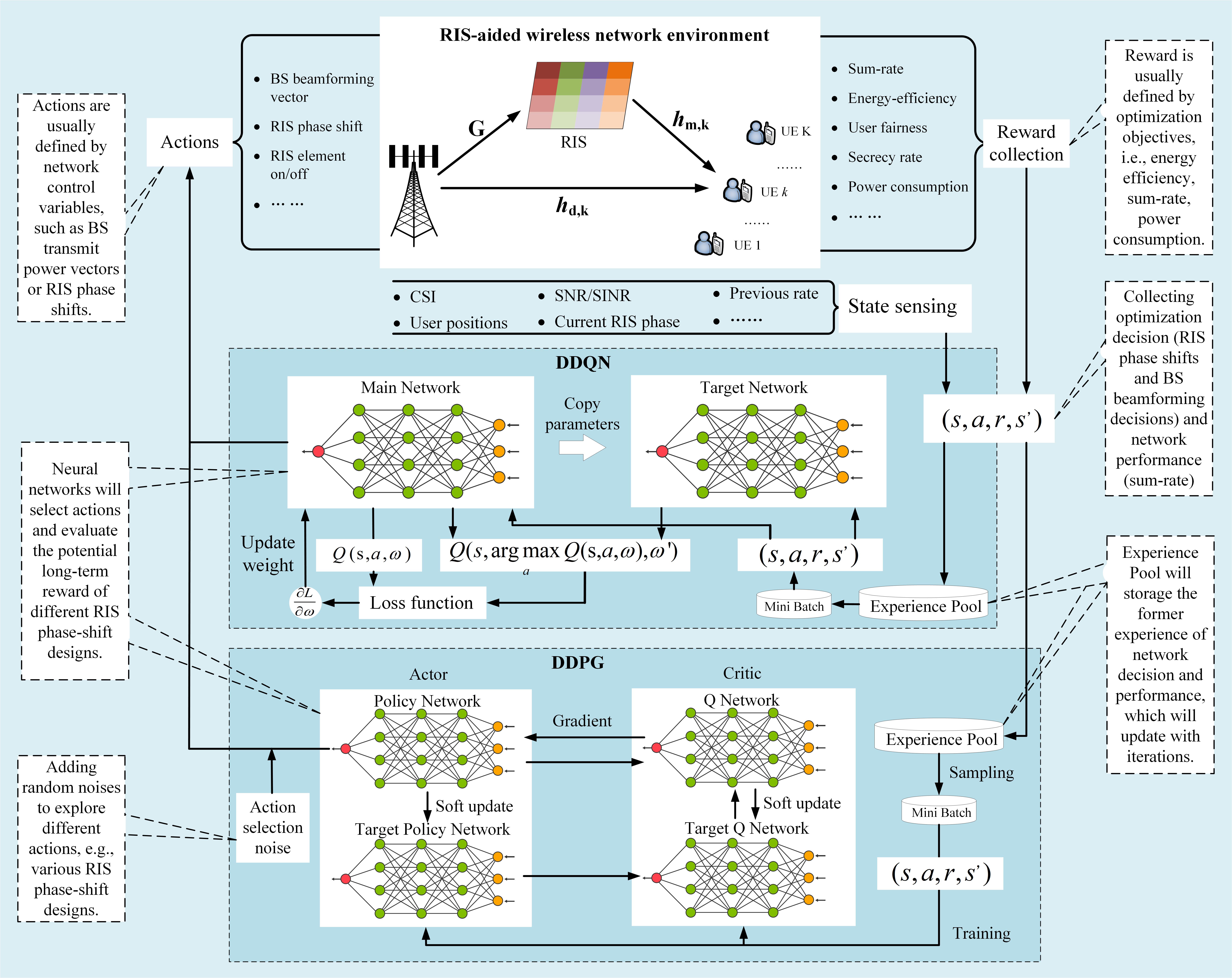}
\caption{\blue{\textbf{DRL-empowered RIS-aided wireless networks}}}
\label{fig-rl}
\setlength{\abovecaptionskip}{-2pt} 
\vspace{-10pt}
\end{figure*}

\blue{ RL is the most widely applied ML technique for optimization, including model-free (e.g., Q-learning, DQN, actor-critic learning) and model-based (i.e., dynamic programming) algorithms. However, defining the Markov decision process (MDP) is fundamental to applying model-based or model-free RL algorithms\cite{moerland2023model}\cite{sutton2018reinforcement}, and the RL agent interacts with the environment under an MDP scheme to learn the best long-term policy.} Given the current system state $s$, the agent selects an action $a$ for implementation and receives a reward $r$, and then the environment will move to the next state $s'$. An MDP model is critical to transforming the optimization problem into an RL context. Specifically, environmental status, control variables, and optimization objectives are defined as states, actions, and rewards, respectively. Then RL algorithms can be used subsequently to maximize the reward and improve the objective function.
Table \ref{tab-rl} summarizes existing studies that apply RL to RIS-aided wireless networks. This subsection will first analyze the state, action, and reward function definitions of these existing studies, and then present the algorithm architecture and training methods. 

\subsubsection{State Definition}
Table \ref{tab-rl} shows that the state may be defined in various ways, e.g., CSI\cite{taha2020deep,guo2021learning,yang2020intelligent}, current RIS phase\cite{yang2020deep,huang2020hybrid,liu2020ris}, position\cite{liu2020ris,samir2021optimizing}, energy level\cite{lee2020deep}, previous transmission rate\cite{yang2020deep2}.    
Specifically, the state refers to the environment status that should be considered for decision-making. For example, the CSI has a great effect on the RIS phase shifts, and therefore CSI is involved in the state definition of many studies \cite{taha2020deep,guo2021learning,yang2020intelligent}. Similarly, RIS-aided UAVs are investigated in \cite{samir2021optimizing}, and the UAV altitude is included in the state definitions because the height will directly affect the channel conditions.

\subsubsection{Action Definition}
In the context of MDP, the action indicates control variables that will change the state, such as RIS phase shifts\cite{yang2020deep}-\cite{yang2021machine}, 
BS beamforming \cite{guo2021learning,yang2020intelligent,yang2020deep2,huang2020hybrid,lee2020deep}, RIS positions\cite{liu2020ris} and elements on/off\cite{lee2020deep}. The control variables in problem formulations are easily converted into actions. However, note that many RL algorithms require discrete action spaces, but the control variables in problem formulations are usually continuous as shown in Section \ref{sec-bac}. The first solution is to quantize the control variables. For instance, the BS transmit power is quantized with an interval of 1 W\cite{lee2020deep}, and the RIS phase changes $\bigtriangleup \theta \in \{-\frac{\pi}{10},0,\frac{\pi}{10}\}$ in \cite{liu2020machine}. Another solution is to apply the deep deterministic policy gradient (DDPG) algorithm, which can handle continuous action-space problems\cite{yang2020deep,guo2021learning,huang2020hybrid,lin2020deep,yang2021machine}.

\subsubsection{Reward Functions}
The reward function is a crucial part of RL. As shown in Table \ref{tab-rl}, the reward function definition mainly depends on the optimization objectives, including data rate\cite{yang2020deep,taha2020deep,zhang2021millimeter,kim2021multi}, energy efficiency \cite{lee2020deep,lin2020deep,liu2020ris}, channel capacity\cite{huang2020reconfigurable}, and SNR\cite{feng2020deep}. Moreover, the reward function can include multiple objectives and constraints to balance the overall performance. As an example, the reward function in \cite{yang2020intelligent} has data rate as a positive term to maximize the data rate, while BS power consumption is a negative term to reduce power consumption. RL focuses on the long-term accumulated reward, which means it can better adapt to highly dynamic wireless environments without requiring full knowledge of the defined problem.

\subsubsection{Algorithm Architecture and Training}
\label{ss}

\blue{In Q-learning, the state-action values are updated by}
\begin{equation} \label{eq-qvalue}
Q^{new}(s,a)= Q^{old}(s,a)+\alpha(r+\eta \max\limits_{a} Q(s',a) -Q^{old}(s,a)),
\end{equation}
where $Q^{old}(s,a)$ and $Q^{new}(s,a)$ are old and new Q-values, respectively. $\alpha$ is the learning rate ($0< \alpha < 1 $), \blue{and $\eta$ is the discount factor $(0<\eta<1)$}. 

Equation (\ref{eq-qvalue}) indicates that a Q-table is used to record all the state-action values, leading to slow convergence for problems with large state-action space. To this end, DQN is proposed to use neural networks for Q-value estimation: 
\begin{equation} \label{eq-dqn}
Loss(\omega)=\mathscr{E}(r+\eta \max\limits_{a} Q(s',a,\omega')-Q(s,a,\omega)),
\end{equation}
where $\mathscr{E}$ represents the error between the predicted Q-value $Q(s_,a,\omega)$ and target Q-value $r+\eta \max\limits_{a} Q(s',a,\omega')$. $\omega$ and $\omega'$ are the weight of the main and target networks, respectively. The main network is used to predict current Q-values by $Q(s_,a,\omega)$, and the target network estimates target Q-values by $Q(s',a,\omega')$. 

In DQN, $\max\limits_{a} Q(s',a,\omega')$ indicates that the target network will select the action and meanwhile evaluate the action, and the maximizing operator will result in over-optimistic Q-value estimation. Then double deep Q-learning (DDQN) is proposed to mitigate Q-value over-estimation by
\begin{equation} \label{eq-ddqn}
\resizebox{0.88\hsize}{!}{$Loss(w)=\mathscr{E}(r+\eta Q(s',\arg \max\limits_{a}Q(s',a,\omega),\omega') - Q(s,a,\omega))$},
\end{equation}
\blue{where} $\arg \max\limits_{a}Q(s',a,\omega)$ means action selection of the main network, and $Q(s',\arg \max\limits_{a}Q(s',a,\omega),\omega')$ indicates the action evaluation of the target network. Decoupling the action selection and evaluation can provide more accurate Q-value prediction and prevent over-estimation.
 
DRL has been used for RIS phase-shift optimization in \cite{taha2020deep,feng2020deep, yang2020deep2}. In these studies, continuous phase shifts are quantized to form discrete action spaces for DQN or DDQN. On the contrary, DDPG can handle continuous action spaces directly without quantization, which has been used for continuous RIS phase-shift control in \cite{yang2020deep,guo2021learning,huang2020hybrid}.

DDPG is considered a combination of actor-critic learning and DQN, in which the actor network selects actions, and the critic network evaluates the state-action values. The loss function of the critic network is defined as
\begin{equation} \label{eq-ddpg1}
\resizebox{0.88\hsize}{!}{$Loss(w^C)=\mathscr{E}(r+\eta Q(s',a(s',\omega^{A'}),\omega^{C'}) - Q(s,a,\omega^{C}))$},
\end{equation}
where $a(s',\omega^{A'})$ indicates that action $a$ is selected by the target actor network with weight $\omega^{A'}$, and $Q(s',a(s',\omega^{A'}),\omega^{C'})$ means the state-action value is evaluated by the target critic network with weight $\omega^{C'}$. For the actor network, the policy gradient is 
\begin{equation} \label{eq-ddpg2}
\begin{aligned}
\nabla_{\omega^{A}} J \approx \frac{1}{\mathcal{T}}\sum^{\mathcal{T}}_{i=1} (\nabla_{a}Q(s,a,\omega^{C})&|_{s=s_{i},a=a(s_i,\omega^{A})} \\ 
& \cdot\nabla_{\omega^{A}}a(s_i,\omega^{A})|_{s=s_i}),
\end{aligned}
\end{equation}
In equation (\ref{eq-ddpg2}), the critic network provides the Q-value  $Q(s,a,\omega^{C})$, and it represents the expected accumulated reward for a given pair $(s,a)$. 
The actor network is trained to produce actions that can result in the maximum state-action value as predicted by the critic network. Therefore, a common approach to calculate the loss function of the actor network is
\begin{equation} \label{eq-ddpg3}
Loss(w^A)=-\frac{1}{\mathcal{T}}\sum^{\mathcal{T}}_{i=1}(Q(s_i,a_i,w^C)),
\end{equation}
which is computed by using the negative mean of the Q-values predicted by the critic network.

\begin{figure*}[!t]
\centering
\includegraphics[width=0.95\linewidth]{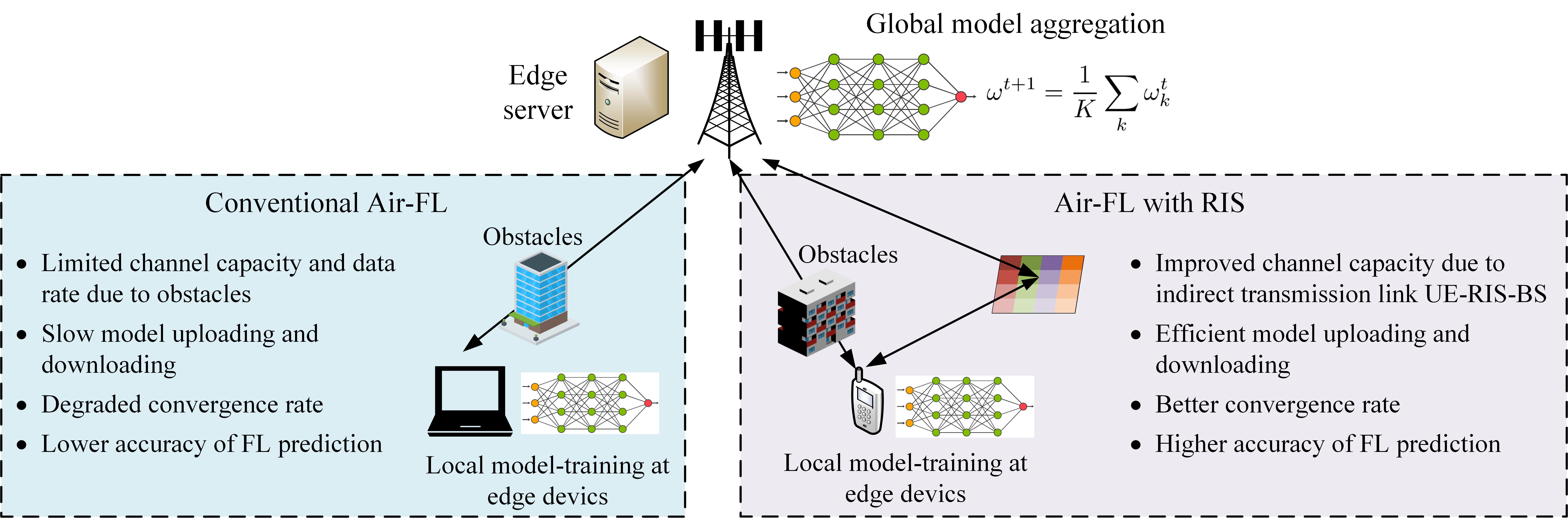}
\caption{Comparison between conventional Air-FL and Air-FL with RISs}
\label{fig-fl}
\setlength{\abovecaptionskip}{-2pt} 
\vspace{-10pt}
\end{figure*}

Fig. \ref{fig-rl} shows DRL-empowered RIS-aided wireless networks, which include DDQN and DDPG as two DRL examples. Based on the current state $s$, the agent selects BS beamforming vectors and RIS phase shifts as the action $a$. Then the action $a$ is implemented and rewards $r$ are collected, e.g., sum-rate, energy efficiency, or power consumption. The system will arrive at a new state $s'$ that is indicated by CSI, user positions, or SNR. The experience tuple $<s,a,r,s'>$ is saved in the experience pool, and a mini-batch is sampled for network training. For the DDQN algorithm, the main network is trained as equation (\ref{eq-ddqn}), and the target network will copy the weight of the main network, providing a stable reference. By contrast, the actor and critic networks are trained by equations (\ref{eq-ddpg1}) and (\ref{eq-ddpg2}) in DDPG, and it applies slow update strategies for target networks. 

\blue{Fig. \ref{fig-rl} presents the application of DDQN and DDPG to joint active and passive beamforming problems. Note that here the DDQN and DDPG algorithms can be easily generalized to many other RL algorithms without loss of generality.
This scheme can also be applied to other RIS-related scenarios. For instance, for the UAV-RIS joint optimization problem, one can include the UAV control variables in the action definition, and add UAV altitude in the state.} 
Finally, there have been various reinforcement learning algorithms, but one common deficiency is the low sampling efficiency. It requires substantial numbers of interactions for agent training, leading to large costs in real-world applications, e.g., hundreds of millions of samples.

\begin{table*}[!t]
\caption{Summary of federated learning and RIS-aided wireless networks }
\centering
\small
\setstretch{1}
\begin{threeparttable} 
\resizebox{1\textwidth}{!}{%
\begin{tabular}{|m{0.8cm}<{\centering}|m{1.7cm}<{\centering}|m{1.5cm}<{\centering}|m{1.8cm}<{\centering}|m{1cm}<{\centering}|m{3.3cm}<{\centering}|m{3.4cm}<{\centering}|m{3.8cm}<{\centering}|m{2.2cm}<{\centering}|}
\hline 
Ref. &  Scenario  & Phase-shift resolution  &  Channel settings & CSI &  FL-related objectives &   Control variables  &  Constraints & Algorithms   \\
\hline
\cite{liu2021reconfigurable} &  AirFL with NOMA  &  Continuous & Rician channel   & Perfect   & Minimizing the gap between converged and optimal  training loss.     & FL device selection, receiver beamforming and RIS phase shifts.  & Device selection, receiver beamform, and phase shifts constraints  & AO, Gibbs sampling, SCA  \\
\hline
\cite{yang2021reconfigurable} & AirFL with RISs  & Continuous  & Rician fading  &   Perfect &   Minimizing global loss  &  Transmit power and RIS phase shifts  & Power and dual constraints  & AO, QCQP, SDP   \\
\hline
\cite{ni2022star} & AirFL with NOMA  & Continuous & Rayleigh fading/  Rician fading     & Perfect   & Minimizing FL training gap  &  BS transmit power and RIS configurations    &  Transmit power, target rate, MSE tolerance, and RIS configuration constraints. & AO, SCA, SDR   \\
\hline
\cite{battiloro2022dynamic} & RIS-enhanced FL   & Discrete  & Generated by simulator   & Perfect   & Minimizing average system power consumption  &  RIS phase shifts and bits, bandwidth, and CPU frequency & Training latency, convergence rate, and learning performance constraints  & Stochastic Lyapunov optimization, greedy algorithm   \\
\hline
\cite{zheng2022balancing} & AirFL with RISs  & Continuous &  Empirical channel fading  & Perfect/ Imperfect   & Minimizing the MSE of the aggregated AirFL model   & Receive and transmit beamformer, RIS phase shifts   &  Total transmit power, phase shifts, and target rate constraints  &  AO, SCA  \\
\hline
\cite{ni2021over} &  AirFL with RISs and NOMA  &  Continuous & Rayleigh fading  &  Perfect & Maximizing the achievable hybrid rate of FL and NOMA   & User transmit power, BS receive scalar, and RIS phase shifts  & Target rate and MSE, phase configuration and total transmit power constraints  &   AO, SCA, SDR   \\
\hline
\cite{ni2021federated} &  AirFL with RISs and NOMA  & Continuous & Rayleigh fading   &  Perfect  &  Minimizing the FL MSE and cardinality  &  Transmit power, receive scalar, reflection coefficients, and learning participants  & Total transmit power, phase configuration, target MSE, and the number of learning devices   &  AO, SDR, SCA  \\
\hline
\cite{liu2021joint} &  AirFL with RISs   & Continuous  & Obtained from \cite{tang2020wireless}  & Perfect   &   Minimizing the effect of device selection and the communication error on the convergence rate  & Device selection, over-the-air transceivers, and RIS phase shifts   & Device selection, receiver beamforming, and phase configurations    &  Gibbs-sampling, SCA   \\
\hline
\cite{yang2022federated} & AirFL with RISs   &  Continuous  & Obtained from testbed   & Perfect     &  Maximizing FL utility  &  RIS phase shifts, user-RIS association, and bandwidth allocation        & Bandwidth allocation, RIS phase configurations and association, target SNR constraints  & Matching game, bisection search      \\
\hline
\cite{zhang2021energy} &  AirFL with RISs     &    Continuous & Rayleigh fading   & Perfect   &  Power minimization   & CPU frequency, power and bandwidth allocation, RIS configurations and accuracy design   &  Task completion time, maximum transmit power, phase configuration, total bandwidth       & AO, SDP, MM     \\
\hline
\cite{li2020enhanced} & FL-aided RIS optimization   &   Continuous  
& Wideband geometric/ Rayleigh fading  &  Predicted    &   Average rate maximization    &\multicolumn{3}{c|}{ \makecell{Local models: local devices train local DNNs to predict channel \\ rate using sampled channel vectors; \\Global model: edge server aggregates local DNN models and average.\tnote{1}\\
}} \\
\hline
\cite{zhong2022mobile} & FL-aided mobile RIS optimization   &   Continuous & Rician fading   &  Predicted    &   Sum-rate maximization    &  \multicolumn{3}{c|}{ \makecell{FL-DDPG is applied. Neural networks are trained at local agents\\ and then aggregated to predict Q-values. Control variables \\ include RIS positions, phase shifts, and AP power allocation.  }} \\
\hline
\end{tabular}}
 \begin{tablenotes}    
        \footnotesize       
        \item[1] The columns are combined because \cite{li2020enhanced} \cite{zhong2022mobile} are different from other studies by using FL as an optimization approach, while FL in other works \\ of Table \ref{tab-fl} is part of the optimization objectives. Therefore, instead of showing control variables and constraints, it is essential to present the local \\ and global models of FL-based optimization algorithms.   
\end{tablenotes} 
      
\end{threeparttable} 
\label{tab-fl}
\vspace{-12pt}
\end{table*}

\subsection{ Federated Learning and RISs}

Different from conventional centralized ML algorithms, FL trains the model across multiple decentralized edge devices or servers that hold local datasets without exchanging data. In FL, each edge device will train a local model using local samples, and then a global model is formed by aggregating local model parameters. Afterwards, edge devices download the global model to update local models. 
Table \ref{tab-fl} summarizes existing works focusing on FL and RIS-aided wireless communications. This subsection first discusses RIS-enhanced over-the-air FL (AirFL), and then introduces how to use FL optimization in RIS-aided environments.

\begin{figure*}[!t]
\centering
\includegraphics[width=1\linewidth]{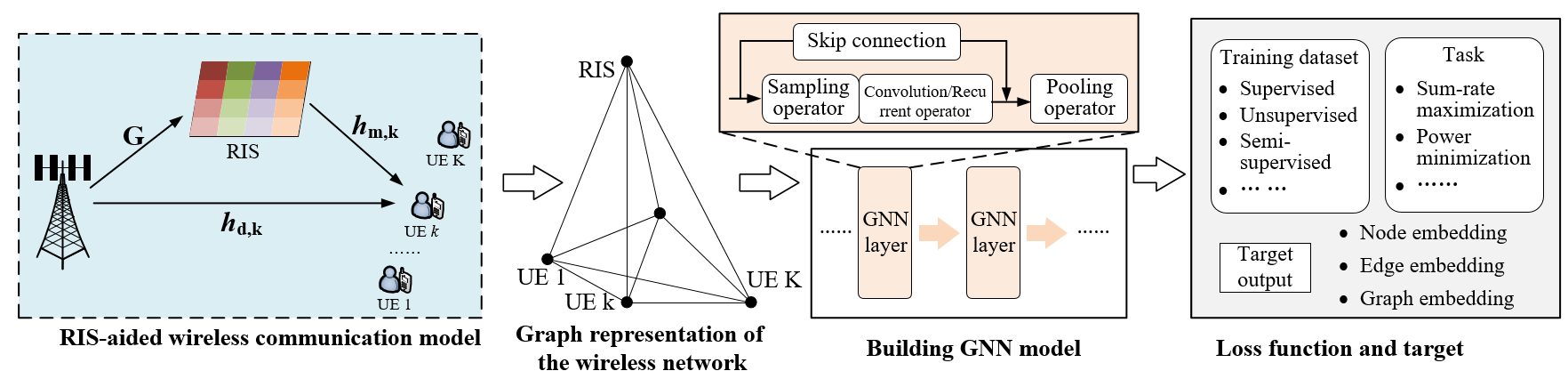}
\caption{Graph learning for RIS-aided wireless networks}
\label{fig-gl}
\setlength{\abovecaptionskip}{-2pt} 
\vspace{-10pt}
\end{figure*}

\subsubsection{RIS-enhanced Over-the-air FL}
The main advantage of FL is that it helps preserve data security and privacy, and the distributed property makes wireless networks an ideal platform for FL training. Therefore AirFL is proposed to combine FL with wireless communications. In particular, AirFL implements FL in wireless networks, using edge devices for local model training and edge servers for model aggregation.

However, the information exchange between local and global servers may be affected by unreliable wireless links, limited bandwidth, signal distortion, dynamic channel conditions, and so on. The uncontrollable signal propagation path can degrade the FL performance, e.g., slow uploading of local models due to low data rate. Therefore, RISs are combined with AirFL to realize the full potential of FL. 

As shown in Fig. \ref{fig-fl}, in conventional Air-FL, obstacles may lead to high penetration loss between edge devices and edge servers, and then the low channel capacity will result in slow model uploading and downloading. Finally, the slow parameter exchange efficiency may degrade the convergence rate and lower the accuracy of Air-FL. By contrast, in Air-FL with RISs, the indirect transmission between UE-RIS-BS provides an alternative transmission path for local model uploading or global model downloading. RISs improve the channel capacity by manipulating the signal propagation path. Therefore, efficient model uploading and downloading will improve the convergence rate and precision of Air-FL.

There are a few works that investigate how to enhance FL performance in RIS-aided wireless networks by minimizing global training loss\cite{liu2021reconfigurable, yang2021reconfigurable},  MSE\cite{ni2022star,ni2021federated}, power consumption\cite{zhang2021energy,battiloro2022dynamic}, maximizing the FL utility\cite{yang2022federated}. In \cite{yang2021reconfigurable}, Yang \textit{et al.} aim to minimize the global training loss of FL by controlling transmit power and RIS phase shifts, and the optimization problem is solved by AO-based QCQP and SDP. 
\cite{yang2022federated} proves that RISs can improve more than 30\% prediction accuracy of AirFL, and a 10 times lower AirFL test error is reported in \cite{ni2021federated} by using multi-RIS. In these works, the FL performance is improved by optimizing the resource allocation and user-RIS association, and then edge users can efficiently upload the local models.
Meanwhile, it is worth noting that these works still rely on model-based optimization algorithms, such as AO, QCQP\cite{yang2021reconfigurable}, SCA\cite{zheng2022balancing,ni2021over,ni2021federated}, and MM\cite{zhang2021energy}.

\subsubsection{ FL for RIS-aided Wireless Communications} 
FL can also be used to optimize the performance of RIS-aided wireless communications. 
For example, deploying a local FL model in RISs may reduce the communication overhead between RISs and the BS, since only local model parameters are shared instead of sharing the whole dataset. In addition, FL can better protect private information such as user CSI, which may be used to infer user locations.
Specifically, FL is used in \cite{li2020enhanced} and \cite{zhong2022mobile} for average rate maximization, in which local models are deployed in user devices and the global model is aggregated by edge servers. In \cite{li2020enhanced}, federated neural networks consider sampled channel vectors as input to predict achievable rates. FL and DDPG are combined in \cite{zhong2022mobile}, and the local neural networks used in DDPG will be aggregated and updated.   

FL is an appealing technique for wireless networks as a distributed ML algorithm. However, the distributed implementation also leads to high communication overhead due to frequent parameter sharing. Meanwhile, the local devices may have different computational capabilities and storage capacities, and such heterogeneity may affect model aggregation and update in FL.

\subsection{Graph Learning }
Graph learning refers to a group of ML techniques in the graph domain, including graph neural networks (GNN), graph attention networks (GAN) and graph convolution networks (GCN). Compared with CNN, which operates on regular Euclidean data like images (2D grid) and text (1D sequence), graph learning is more efficient in describing graphs and structures. Graph learning aims to transform nodes, edges, and their features into low-dimension vector spaces by preserving properties such as graph structure\cite{cui2018survey}. 

Wireless networks are highly dynamic, and wireless data may be collected from non-Euclidean domains, which is represented by graph structure with high dependency on network topology.
The conventional approach of data processing is to convert the data with graph structure into Euclidean domain, but such transformation leads to high complexity and extra overhead. 
By contrast, graph learning enables the graph-structured data to be processed effectively, and transforming the wireless network topology into graphs can better describe the association and interference between network devices\cite{he2021overview}.
Therefore, graph learning has been applied to power control and interference management \cite{naderializadeh2020wireless}, resource allocation \cite{eisen2020optimal,jiang2020dynamic}, network slicing \cite{wang2020graph}, and so on. In the following, GNN is used as an example to introduce graph learning fundamentals, and then we explain how to apply graph learning for RIS control and optimizations.

\subsubsection{GNN Fundamentals} 
The primary motivation for developing GNN is to extend the existing neural network architecture into graph-related data processing capabilities\cite{zhou2020graph}. In a graph, each node is described by its features and related nodes. Suppose that $z_{v}$ is a state vector to describe the features of node $v$, and it is defined by
\begin{equation}\label{eq-gnn1}
 z_v=f(y_v,y_{v}^{ed},y_{v}^{ne},z_{v}^{ne}),    
\end{equation}
where $y_v$ and $y_{v}^{ed}$ are the features of node $v$ and its edge, and $z_{v}^{ne}$ and $y_{v}^{ne}$ are the state and features of neighbour nodes, respectively.
Then, $z_{v}$ and $y_v$ are used to produce an output $o_v$ by
\begin{equation} \label{eq-gnn2}
o_v=g(z_v,y_v),    
\end{equation}
where $g$ is the output function to map the relationship between states, features, and outputs.  

Similarly, by collecting all the states and features, we have
\begin{equation} \label{eq-gnn3}
Z^{l+1}=f(Z^{l+1},Y),    
\end{equation}
\begin{equation}
O=g(Z,Y_N),    
\end{equation}
where $Z^{l+1}$ indicates all the states at $l^{th}$ iteration, $Y$ indicates all the features, $Y_N$ means the node features, and $O$ is the overall output. Equation (\ref{eq-gnn3}) shows that the system state is updated in an iterative manner, which is inspired by Banach’s fixed point theorem \cite{khamsi2011introduction}. Finally, similar to conventional neural networks, GNN aims to minimize the loss function.

\subsubsection{Graph Learning for RIS Control and Optimizations} 
Interference control is an important technique for multi-user environments to maximize the system sum-rate, and the interactions between RISs and UEs are easily described by a graph. The graph in Fig. \ref{fig-gl} includes $K+1$ nodes, in which one node represents the RIS and the rest are $K$ UEs. Given this scheme, GNN is applied to user scheduling and RIS configurations in \cite{zhang2022learning} and \cite{zhang2022user}. In particular, GNN is trained in an unsupervised manner, and the inputs are user weights and pilot sub-frames of the scheduled users, and the outputs are RIS configurations and beamformers. Similarly, unsupervised GNN is applied in \cite{jiang2021learning} for network utility maximization, which takes pilot signals as input to optimize the BS beamforming and RIS configurations.  

In \cite{zhang2022learning,zhang2022user,jiang2021learning}, a useful feature of GNN is used to reduce the interference between users. Specifically, when updating one node in the GNN, all the neighbour nodes will be included in the updating function, which means GNN can better capture the mutual interference between users.  
Meanwhile, RIS node updating is a function of all the user nodes, enabling GNN to configure RIS elements to improve the channel capacity of all users.   
In addition, the authors in \cite{jiang2021learning} note that 
another key advantage of GNN is the generalization capability. 
For instance, when the number of cell users constantly changes, conventional FNN must be re-trained to handle various user numbers. 
In contrast, a GNN can generalize to different numbers of users by simply adding and removing components in its feature extraction and information exchange stages. Such generalization capability can considerably alleviate ML model training efforts.

Graph learning is one of the most state-of-the-art ML techniques. However, the application to wireless networks is still in a very early stage. The real-time wireless environment can produce dynamic and generative changing graphs, which may prevent the application of graph learning.

\subsection{Transfer Learning}
Long training time and slow convergence are common issues of most ML algorithms, and one of the main reasons is that the model must explore the task from scratch. Fast decision-making is critical in wireless communications, but the low sampling efficiency may prevent applying ML to RIS-aided wireless networks. This subsection will introduce transfer learning fundamentals and explain how transfer learning can improve ML-enabled wireless networks with RISs.

\subsubsection{Transfer Learning Fundamentals}
Transfer learning can be combined with many ML algorithms, and here we consider transfer reinforcement learning (TRL) as an example\cite{zhou2022learning}. In conventional RL, the decision-making $\mathcal{D}_{RL}$ of one agent is described by 
\begin{equation} \label{eq-trl}
\mathcal{D}_{RL}:s \times \mathscr{K}\rightarrow a, r ,
\end{equation}
where $\mathscr{K}$ represents the agent's knowledge, $s$, $a$, and $r$ are the current state, selected action, and received reward, respectively. In equation (\ref{eq-trl}), the agent utilizes the collected knowledge $\mathscr{K}$  for decision-making and action selection.  

By contrast, the decision-making in TRL is
\begin{equation} \label{eq-trl2}
\mathcal{D}_{TRL}:s \times \mathcal{M}(\mathscr{K}_{expert}) \times \mathscr{K}_{learner} \rightarrow a,r,
\end{equation}
where $\mathscr{K}_{expert}$ and $\mathscr{K}_{learner}$ are the knowledge of the expert and learner agents, respectively. The learner is designed to solve the target task, and the expert has some existing knowledge of related source tasks.
Considering the similarities between the source and target tasks, the expert's experience may be reused by the learner as prior knowledge.
The $\mathcal{M}$ in equation (\ref{eq-trl2}) defines a mapping function. $\mathcal{M}(\mathscr{K}_{expert})$ indicates that the expert's experience will be transformed into digestible knowledge, boosting the learning process of the learner. With existing prior knowledge, the learner can achieve a jump-start at the exploration phase, achieving a higher exploration efficiency and average reward with faster convergence\cite{zhou2022knowledge}.

\begin{figure}[!t]
\centering
\includegraphics[width=0.95\linewidth]{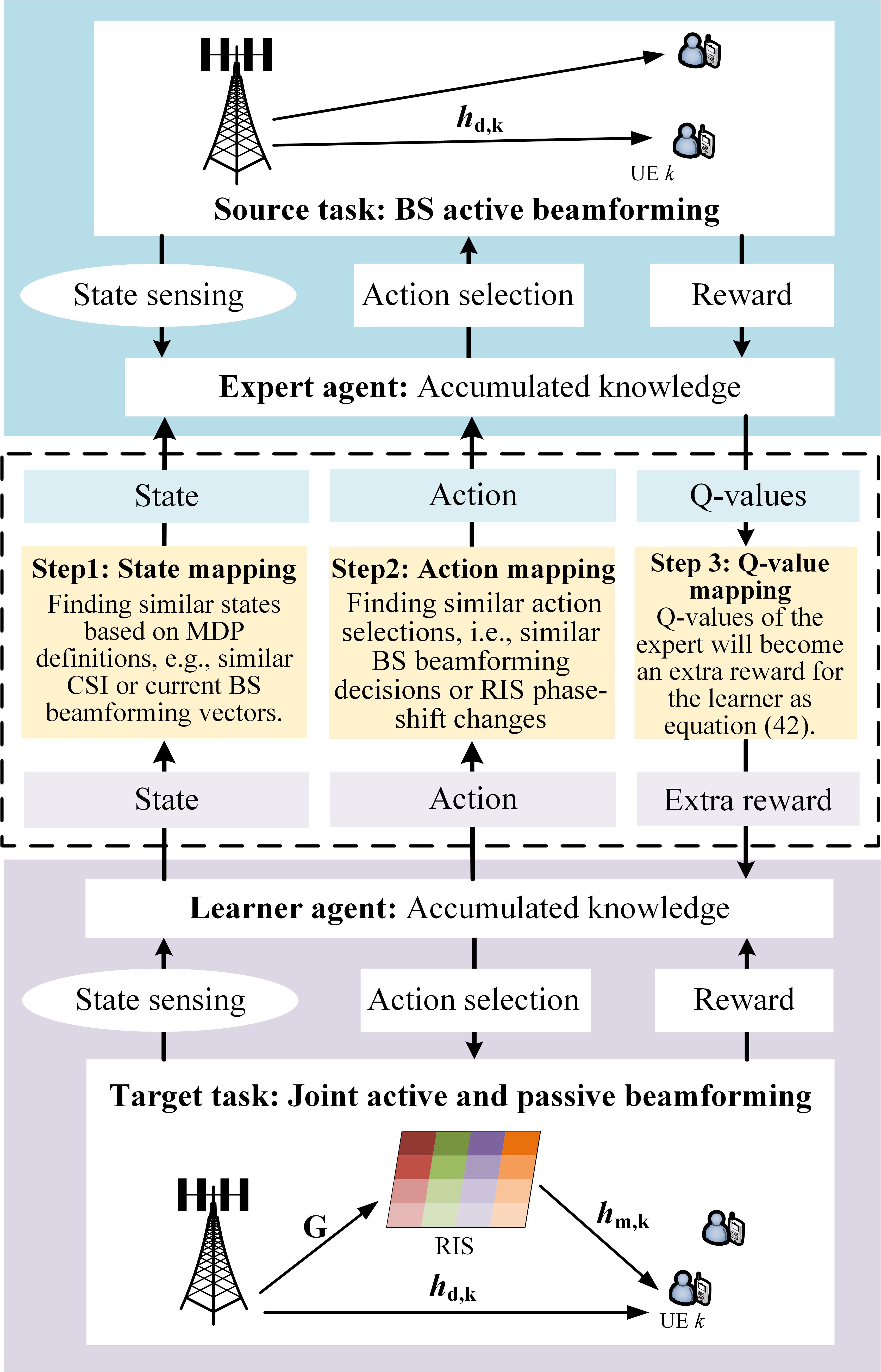}
\caption{\blue{\textbf{Transfer reinforcement learning for RIS-aided wireless networks}}}
\label{fig-trl}
\setlength{\abovecaptionskip}{-2pt} 
\vspace{-10pt}
\end{figure}

\subsubsection{Transfer Learning-boosted Wireless Networks with RISs} 

\blue{Wireless networks can be highly dynamic. For example, user numbers and CSI patterns may change quickly in a short period of time, and then the RIS control policy may need to be retrained to handle these dramatic changes. However, ML algorithms usually require many training iterations, preventing the application to dynamic wireless networks. To this end, TRL may become a promising solution.}
Fig. \ref{fig-trl} illustrates how TRL is used for RIS-aided wireless networks, which includes source and target tasks. We assume that the expert agent has existing knowledge of the source task, BS beamforming, and the learner agent is designed for the target task, joint active and passive beamforming. Due to the potential task similarities, the learner may reuse the expert's experience to better handle target tasks. 
However, note that the expert's knowledge may exist in various ways, e.g., state-action values and action selections, and then the mapping function may be defined in different manners. Fig. \ref{fig-trl} provides an example by finding similar states and actions, and a Q-value-based mapping function can be defined by 
\begin{equation} \label{eq-map}
\resizebox{0.89\hsize}{!}{$\begin{aligned}
Q^{new}(s^{L},a^{L})=  &Q^{E}(\mathcal{M}(s^{L}),\mathcal{M'}(a^{L}))+Q^{old}(s^{L},a^{L})+\\
&\alpha(r+\eta \max\limits_{a} Q(s',a)-Q^{old}(s^{L},a^{L})),
\end{aligned}$}
\end{equation}
where $s^{L}$ and $a^{L}$ are the learner's state and action, $\mathcal{M}$ and $\mathcal{M'}$ are the state and action map functions, respectively, and $Q^{E}$ indicates the state-action value of the expert. 
Compared with conventional RL, the main difference is that $Q^{E}(\mathcal{M}(s^{L}),\mathcal{M'}(a^{L}))$ is involved as an extra reward for selecting $a^{L}$ under $s^{L}$. 
\blue{In particular, Fig. \ref{fig-trl} shows the steps of defining mapping functions for active and passive beamforming tasks. Firstly, the state mapping function $\mathcal{M}$ is defined to find $s^E=\mathcal{M}(s^{L})$, finding similar environment states such as CSI or current BS beamforming vectors between the learner and expert agents. Similarly, the action mapping function $\mathcal{M'}$ aims to find similar beamforming decisions between the learner and expert action spaces.   
Finally, by finding these similar network states and beamforming decisions, as shown in equation (\ref{eq-map}), good actions with high Q-values in the expert can provide extra rewards for the learner. Then the learner is encouraged to select better actions to achieve a higher sum-rate or energy efficiency.}   

With transfer learning, the RL agent can achieve higher exploration efficiency and faster convergence, enhancing the efficiency of RIS-aided wireless networks. Transfer learning has been used in \cite{zhou2022learning} for joint resource allocation of network slicing, and \cite{elsayed2020transfer} for mmWave networks, achieving faster convergence and better network performance. Similarly, transfer learning can be applied to ML-enabled RIS optimization for faster convergence and achieving prompt phase-shift responses. 
Transfer learning is a very useful technique to mitigate ML model training effort. However, note that transfer learning relies on existing experts to reuse prior knowledge, and the mapping function definition may be difficult due to the inherent task difference between experts and learners.

\subsection{Hierarchical Learning }

\begin{figure}[!t]
\centering
\includegraphics[width=1\linewidth]{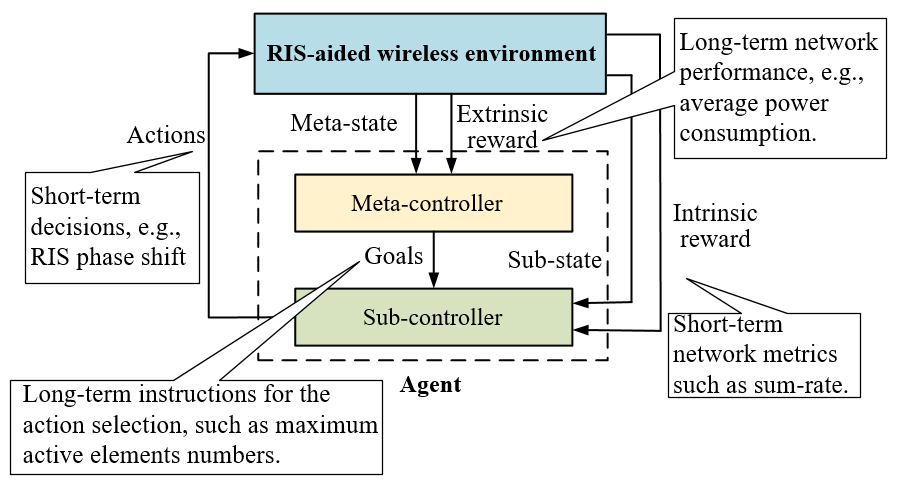}
\caption{Hierarchical reinforcement learning for RIS-aided wireless networks}
\label{fig-hrl}
\setlength{\abovecaptionskip}{-2pt} 
\vspace{-10pt}
\end{figure}

Hierarchical learning is another technique that can be used for optimizing RIS-aided wireless networks. The main idea of hierarchical learning is to decouple the long-term task into multiple achievable goals to increase exploration efficiency\cite{pateria2021hierarchical}. In particular, it defines a meta-controller to select goals and a sub-controller to achieve these goals.
Based on the short-term performance of the sub-controller, the meta-controller can adjust the goal dynamically to guarantee the long-term performance of the whole system. Hierarchical learning can also be applied to optimization problems that include multiple control variables with different time scales\cite{zhou2023hierarchical}.  
For instance, in \cite{zhou2023hierarchical}, Zhou \textit{et al.} consider a meta-controller for sleep control, and sub-controllers for transmission power and RIS control, enabling control variables with different time scales. 

Fig. \ref{fig-hrl} shows how hierarchical reinforcement learning is applied to RIS-aided wireless networks, and the agent consists of a meta-controller and a sub-controller. 
Specifically, the sub-controller can generate long-term policy instructions for the sub-controller, such as the maximum number of active RIS elements that is available. Then, as shown in Fig. \ref{fig-hrl}, given high-level goals, the sub-controllers can select short-term decisions for RIS phase shifts. Meanwhile, the meta-controller focuses on average power consumption in a period as long-term network performance, and the sub-controller accounts for delay or data rate as instant metrics. This scheme can coordinate control variables with different time scales, balancing instant and long-term network metrics. More specifically, the state-action value of the meta-controller is updated by:
\begin{equation} \label{eq-hrl1}
\resizebox{0.89\hsize}{!}{$\begin{aligned}
&Q_{meta}^{new}(s_{meta},g_{meta}) = Q_{meta}^{old}(s_{meta},g_{meta})+\\
&\alpha(r_{ex}+\eta \max\limits_{g} Q_{meta}(s_{meta}',g)-Q_{meta}^{old}(s_{meta},g_{meta})),
\end{aligned}$}
\end{equation}
where $s_{meta}$ and $s_{meta}'$ is the current and next meta-states, $g_{meta}$ is the goal, and $r_{ex}$ is the extrinsic reward, respectively. $Q^{old}_{meta}$ and $Q^{new}_{meta}$ are old and new state-action values for the meta-controller, indicating the accumulated reward by selecting $g_{meta}$ under state $s_{meta}$.  

Similarly, the Q-value of the sub-controller is updated by
\begin{equation} \label{eq-hrl2}
\resizebox{0.89\hsize}{!}{$\begin{aligned}
Q&_{sub}^{new}(s_{sub},g_{meta},a_{sub}) = Q_{sub}^{old}(s_{sub},g_{meta},a_{sub})+\\
&\alpha(r_{in}+\eta \max\limits_{a} Q_{sub}(s_{sub}',g_{meta},a)-Q_{sub}^{old}(s_{sub},g_{meta},a_{sub})),
\end{aligned}$}
\end{equation}
where $s_{sub}$ and $s_{sub}'$ are current and the next sub-states, $a_{sub}$ is the action, and $r_{in}$ is the intrinsic reward. $Q^{new}_{sub}$ and $Q^{old}_{sub}$ are defined similarly as the meta-controller, indicating the expected reward of selecting $a_{sub}$ under state $s_{sub}$ and goal $g_{meta}$. Equation (\ref{eq-hrl2}) shows that the sub-controller is under the policy control of the meta-controller.

Hierarchical learning is a promising technology to enable hierarchical autonomy in RIS-aided wireless networks. However, one key challenge is to define the relationship between different hierarchies, e.g., meta-controller and sub-controllers. In addition, decoupling one task into multiple sub-tasks can be difficult in highly-dynamic wireless networks, which may prevent the application of hierarchical learning.

\begin{figure}[!t]
\centering
\includegraphics[width=1\linewidth]{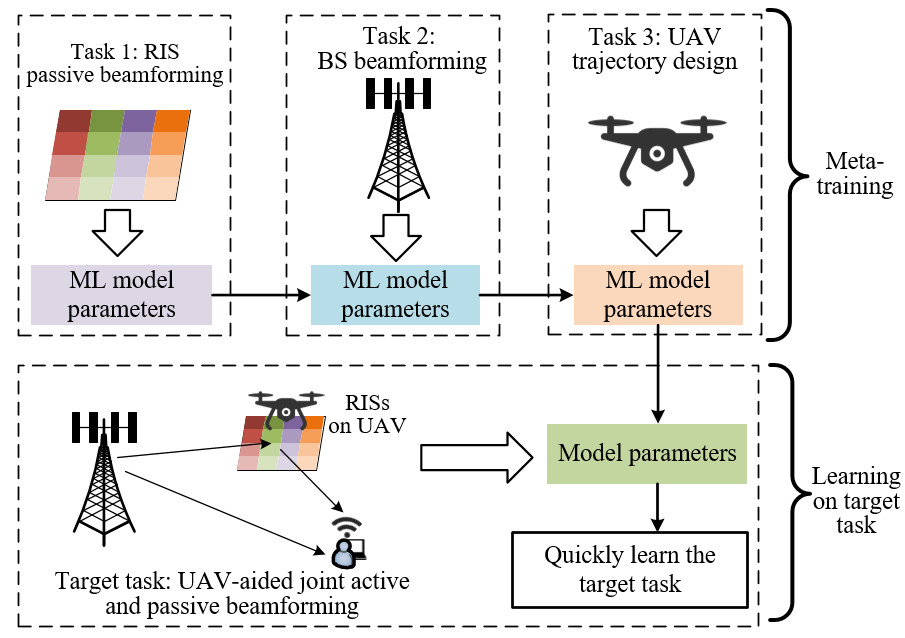}
\caption{Meta-learning for RIS-aided wireless networks}
\label{fig-meta}
\setlength{\abovecaptionskip}{-2pt} 
\vspace{-10pt}
\end{figure}

\begin{table*}[!t]
\caption{Summary of ML-based control and optimization algorithms for RIS-aided wireless networks. }
\centering
\small
\setstretch{1.05}
\resizebox{1\textwidth}{!}{%
\begin{tabular}{|m{2cm}<{\centering}|m{2.5cm}<{\centering}|m{5.5cm}<{\centering}|m{4.2cm}<{\centering}|m{4.5cm}<{\centering}|}
\hline
ML \quad techniques & Typical algorithms & Main features & Difficulties & RIS-related applications\\
\hline
\multirow{1}*{\makecell{Supervised\\ learning}} & Supervised DNN, CNN, decision trees, and support vector machine. & The algorithm is trained to map the relationship between the given input and labeled output for classification and prediction. The input data is fed into the model, and then the model parameters are adjusted until the output is properly fitted.  
& 1) Supervised learning relies on fine-grained datasets to train the algorithm; 2) The algorithm training may be time-consuming; 3) The model is easy to be overfitted.   & Supervised learning is a promising technique if there exist fine-grained datasets, and then various neural networks may be used to predict full channel states \cite{zhang2021deep} or optimal RIS phase shifts\cite{alexandropoulos2020phase}- \cite{hu2021reconfigurable}. \\
\hline
Unsupervised learning  & k-means, DBSCAN, and unsupervised neural networks.  & Unsupervised learning algorithms aim to unveil hidden patterns of unlabeled datasets.  & The result performance is hard to be testified or explain. & Unsupervised DNN can be directly used for optimization problems in RIS-aided networks without involving datasets, which designs RIS phase-shift by defining objectives as loss functions\cite{nguyen2021machine}-\cite{gao2020unsupervised}. \\
\hline
\multirow{20}*{\makecell{Reinforcement \\ learning}}  & Q-learning & The agent interacts with the environment under an MDP framework, recording experience by a Q-table.  & 1)Long convergence time for large state-action problems; 2) Discrete states and actions only.  &\multirow{18}*{\makecell{RL is the most widely \\ applied ML technique for \\ the control and optimization \\  of RIS-aided wireless networks, \\ e.g., power minimization \cite{lin2020deep}, \\ sum-rate\cite{yang2020deep,taha2020deep,samir2021optimizing,yang2021machine},\\ secrecy rate \cite{guo2021learning,yang2020deep2},\\ and energy efficiency\cite{lee2020deep,liu2020ris}.\\ DDPG is especially useful \\ considering the continuous \\ RIS phase-shift control \\ requirements \cite{yang2020deep, guo2021learning}.\\ RL can also be combined with\\ other ML techniques, e.g., \\ transfer reinforcement learning,\\ federated reinforcement learning, \\and meta reinforcement learning. }}\\
\cline{2-4}
& Actor-critic learning & The actor is defined to select actions, while the critic evaluates the actions. & 1) Long convergence iterations; 2) Unstable performance due to the interaction between actor and critic. & \\
\cline{2-4}
& Deep reinforcement learning & DRL applies neural networks to predict state-action values, solving the large state-action issue of tabular Q-learning. 
& \multirow{3}*{\makecell{1) Hyperparameter tuning can \\ be difficult when lacking \\ experience. 2) The sampling \\ efficiency is low. }}  &\\
\cline{2-3}
& Double deep Q-learning & DDQN provides a more accurate Q-value estimation by decoupling the action selection and evaluation.  & &\\
\cline{2-4}
& Multi-agent reinforcement learning & Each agent applies RL or DRL independently to optimize its performance or achieve an overall goal. & The coordination mechanism of multiple agents must be carefully designed.  &\\
\cline{2-4}
& DDPG & DDPG combines actor-critic with policy gradients, optimizing problems with continuous action space.  &  1) Unstable and heavily dependent on appropriate hyperparameters; 2) Overestimation in critic network.   &\\
\hline 
Federated learning & Federated deep learning, federated DRL & Local models are first trained using local datasets, and then the parameters are aggregated to form a global model. Local devices will download the global model to update local models. User privacy is well protected in FL.
& 1) High communication overhead due to parameter exchange; 2) The local device heterogeneity will affect the system performance. & On the one hand, RISs can improve the AirFL performance by improving the channel capacity; on the other hand, FL is used to optimize RIS-aided network performance \cite{li2020enhanced, zhong2022mobile}.    \\
\hline
Graph learning & Graph neural networks, and graph attention networks. & Graph learning refers to ML on graphs. It maps the graph features to vectors with the same dimensions in the embedding space, which is used for link prediction, matching and classification.  &  1) Dynamic and generative changing graph; 2)Interpretability of graph learning.  &  GNN is used for user schedule and RIS configurations in \cite{zhang2022learning, zhang2022user, jiang2021learning} to maximize network utility and sum-rate. The general application of graph learning is still an open issue.   \\
\hline
Transfer learning & Transfer reinforcement learning, transfer supervised learning  & Transfer learning aims to reuse the existing knowledge of experts to accelerate the learning process on target tasks, achieving faster convergence and less training efforts. & 1) The mapping function is hard to design, changing with different algorithms; 2) Transfer learning is vulnerable to adversarial attacks.  & When there are existing experts or source tasks, transfer learning may be used to accelerate ML algorithm training in RIS-aided wireless networks\cite{zhou2022learning}.  \\
\hline
Hierarchical learning & Hierarchical reinforcement learning, hierarchical deep learning  &   Hierarchical learning decouples the task into multiple sub-tasks and goals, increasing the task exploration efficiency.  &  1) The goal and sub-task selection require case-by-case analyses; 2) The relationship between the meta-controller and the sub-controller may be unstable.  & Hierarchical learning is used for optimizing RIS-aided networks with control variables that have different time scales or sparse rewards \cite{zhou2023hierarchical,zhou2023cooperative}.     \\
\hline
 Meta-learning & Meta reinforcement learning, supervised meta-learning & Using experience of former learning tasks to improve the performance on target tasks. The ML model will learn how to learn across tasks.     &  1) Source task distribution must be carefully designed; 2) How to prevent overfitting and underfitting. & Pre-training ML models at the BS for RIS channel estimation\cite{jung2021meta};  Model-agnostic meta-learning for joint RIS phase control and power allocation\cite{zou2021meta}.     \\
\hline
\end{tabular}}
\label{tab-mlsummary}
\vspace{-10pt}
\end{table*}

\subsection{Meta-Learning}

Meta-learning refers to ML algorithms that extract the experience of multiple learning episodes, e.g., a distribution of related tasks, and then use such prior training to improve the performance on target tasks\cite{vanschoren2018meta}. 
In particular, meta-learning is designed to learn how to learn across tasks, 
and this learning-to-learn design can bring several benefits, such as improved training and learning efficiency. In addition, it is better aligned with human learning features, where learning skills are constantly improved on a lifetime timescale and evolutionary policy\cite{hospedales2021meta}. 

RIS-aided networks may include diverse elements, such as RISs, BSs, UAVs, etc, and it can be difficult to train ML models from scratch and meanwhile jointly consider all these network elements. Fig. \ref{fig-meta} shows an example of using meta-learning schemes for UAV-aided joint active and passive beamforming, in which RISs are deployed on the UAV for location flexibility. The ML model is first pre-trained by three existing tasks such as RIS passive beamforming, BS beamforming, and UAV trajectory design. Then, using prior experience, the ML model is expected to learn quickly on the target task, which will jointly consider RISs, BSs, and UAVs. Additionally, such a constant learning scheme can be more useful when other future tasks are expected, and incoming new tasks are always trained based on plenty of former knowledge. There are few works on applying meta-learning to RIS-aided networks. For instance, Jung \textit{et al.} apply meta-learning for RIS channel estimation, and the ML model is pre-trained at the BS by using pilot signals to rapidly estimate RIS channels\cite{jung2021meta}. In \cite{zou2021meta}, model-agnostic meta-learning is used for joint RIS phase-shift control and power allocation, which has a faster convergence rate than baseline ML algorithms.

However, meta-learning must balance the meta-training and self-learning phases. Specifically, meta-training with a wide variety of tasks may lead to underfitting, which means that the agent is unable to specialize to the target task when self-learning. By contrast, if the meta-training tasks are too specific, the knowledge learned on the source tasks may have difficulty in generalizing to target tasks\cite{hospedales2021meta}. Therefore, the source task distribution in the meta-training phase has to be carefully selected.

\subsection{Discussions and Numerical Results}

ML offers promising opportunities for optimizing RIS-aided wireless communications. Table \ref{tab-mlsummary} overviews various ML techniques \footnote{Note that there are many ML algorithms applied to wireless communications. Instead of collecting all the existing ML algorithms, Table \ref{tab-mlsummary} provides a compressed taxonomy to understand the feature of each technique along with RIS control applications.}.  

Supervised learning trains ML models to best map the input to output, e.g., CSI and user position to RIS phase shifts. However, the model training relies on fine-grained labeled datasets, which may be inaccessible in practice. 
By contrast, unsupervised learning has no need for labeled datasets, and it involves the objective function in the loss function for improvement.
Such unsupervised learning approaches can reduce the dependence on labeled datasets, but the generated results are hard to validate due to the absence of labeled data in most circumstances. 

RL is the most widely applied ML technique for optimization problems, and each RL algorithm has its own features and difficulties. For example, DDPG can handle continuous action space of RIS but can be unstable\cite{yang2020deep, guo2021learning, huang2020hybrid,lin2020deep}, and DDQN can prevent overestimation but sampling efficiency is low \cite{liu2020ris}.
FL and graph learning are newly emerging ML techniques. Most existing works consider RIS-enhanced AirFL, demonstrating that RISs can improve the training efficiency and performance of FL \cite{liu2021reconfigurable, yang2021reconfigurable, ni2022star,ni2021federated,zhang2021energy,battiloro2022dynamic}. 
Graph learning has shown great potential in many other fields, and wireless network applications include power control and interference management \cite{naderializadeh2020wireless}, resource allocation \cite{eisen2020optimal,jiang2020dynamic}, and network slicing \cite{wang2020graph}. Despite the advantages, applying graph learning to wireless networks is still an open issue that requires more effort.

\begin{figure}[!t]
\centering
\setlength{\abovecaptionskip}{-2pt} 
\includegraphics[width=7.2cm,height=5.2cm]{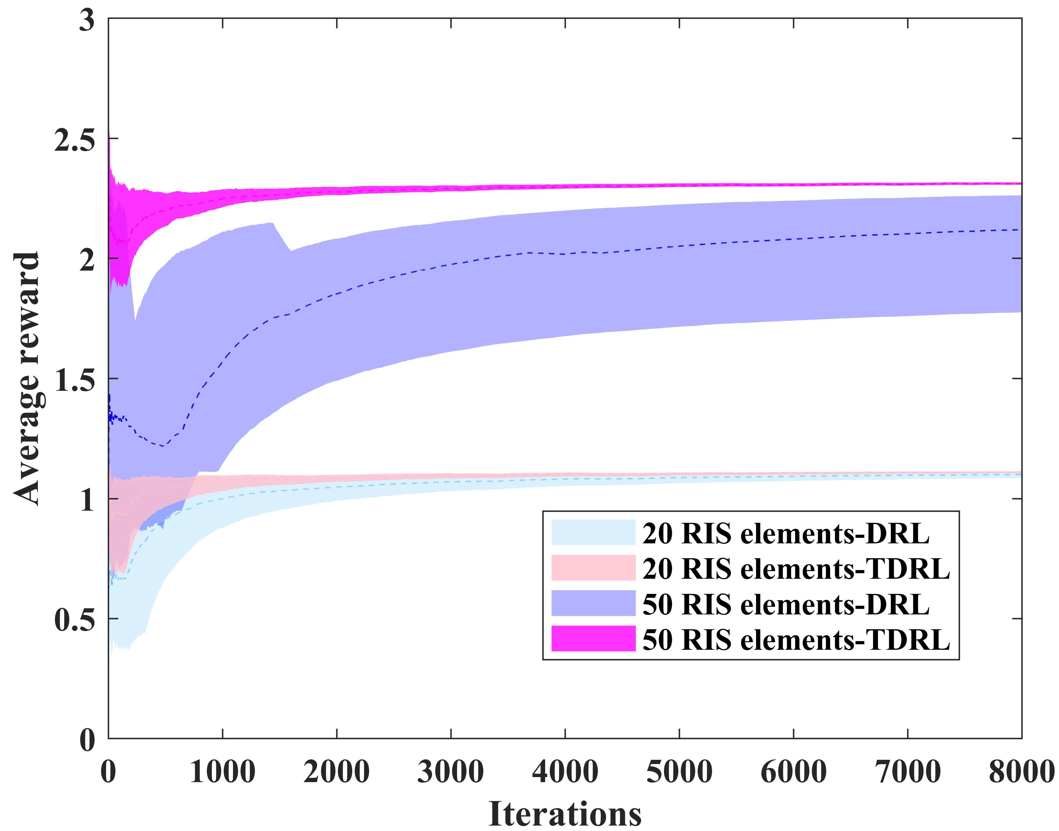}
\caption{Convergence performance of transfer deep reinforcement learning (TDRL) and DRL. 1) TDRL: We assume there is an existing DRL agent that has been trained under a limited number of UEs. Then a TDRL agent will reuse the expert's prior knowledge to adapt to the environment with more diverse UEs. 2) DRL: conventional DQN-based RIS phase-shift control.}
\label{fig-result-tdrl}
%\vspace{-10pt}
\end{figure}

\begin{figure}[!t]
\centering
\setlength{\abovecaptionskip}{-2pt} 
\includegraphics[width=7.2cm,height=5.2cm]{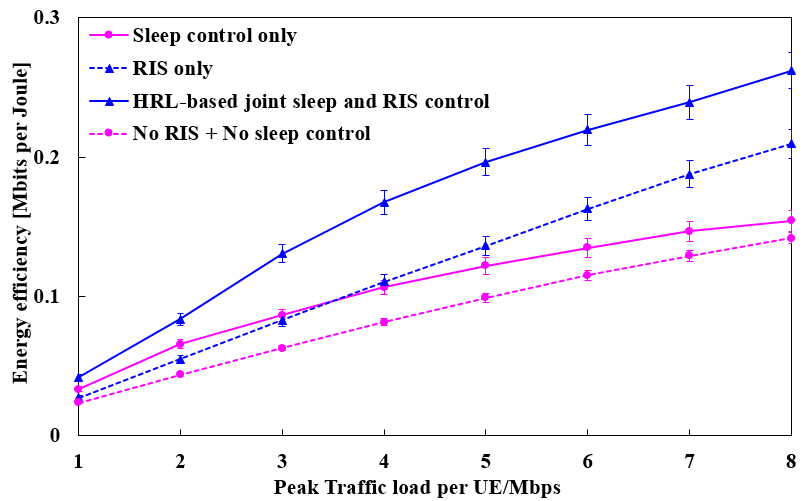}
\caption{Energy efficiency performance by joint sleep control and RIS phase-shift design.
We consider a multi-BS and multi-RIS heterogeneous network. Sleep control is a long-term decision to decide the BS on/off status, while RIS phase-shift control is a short-term optimization based on dynamic channel status. Detailed parameters can be found in \cite{zhou2023cooperative}. }
\label{fig-result-hrl}
\vspace{-10pt}
\end{figure}

Transfer learning and hierarchical learning are both promising ML techniques for RIS-aided wireless networks. Transfer learning can reduce the model training efforts, while hierarchical learning provides a novel architecture for applying ML to wireless communications with hierarchical intelligence, especially when optimization parameters have different timescales. However, more research is needed on these techniques as they are used for RIS-aided wireless networks. 
 Both transfer learning and meta-learning involve source tasks and prior experience.  
The core feature of meta-learning is learn-to-learn, which is an appealing advantage for enabling rapid adaptation to dynamic wireless environments. Compared with transfer learning, meta-learning provides a scheme that can be used to facilitate transfer learning as well as other techniques. In transfer learning, the prior knowledge is usually extracted from the source task without defining a meta-objective. By contrast, the prior experience in meta-learning is usually defined by an outer optimization that evaluates the potential benefit of handling new tasks. Meanwhile, meta-learning involves a wider range of meta-representation problems than transfer learning.

Instead of applying one specific ML algorithm solely, note that these ML algorithms may be jointly used. For instance, federated deep reinforcement learning deploys DRL in each local server for decision-making, and then uses a global server to aggregate the main networks for overall estimation and coordination. Such integration can make the most of each algorithm's advantages, achieving better overall performance.

Finally, Fig. \ref{fig-result-tdrl} and \ref{fig-result-hrl} present examples of using transfer learning and hierarchical learning for RIS-related optimization, respectively. In particular, Fig. \ref{fig-result-tdrl} compares the convergence of transfer deep reinforcement learning (TDRL) and DRL, and TDRL achieves faster convergence with higher average reward. The main reason is that TDRL can reuse the former knowledge of existing experts, which will considerably improve the exploration efficiency of ML algorithms. Such improvement becomes more obvious when the number of RIS elements increases, which indicates higher exploration difficulty for conventional DRL algorithms.
Meanwhile, Fig. \ref{fig-result-hrl} shows the energy efficiency of hierarchical reinforcement learning-enabled joint sleep control and RIS phase-shift optimization\cite{zhou2023cooperative}. It includes a multi-BS and multi-RIS scenario, and sleep control can decide the on/off status of BSs to reduce energy consumption, while RISs can improve the channel capacity. Fig. \ref{fig-result-hrl} demonstrates that combining sleep control with RISs can bring higher energy efficiency than using each technique solely, and hierarchical reinforcement learning can well coordinate different decisions with various time scales.

\begin{table*}[!t]
\caption{Summary of control and optimization techniques for RIS-aided wireless networks }
\centering
\small
\setstretch{1}
\resizebox{1\textwidth}{!}{%
\begin{tabular}{|m{1.5cm}<{\centering}|m{3.5cm}<{\centering}|m{3.3cm}<{\centering}|m{3.5cm}<{\centering}|m{3cm}<{\centering}|m{4cm}<{\centering}|}
\hline 
Optimization approaches  &  Main features   &     Advantage    &  Drawbacks    &  Difficulties   &  Application scenarios for RISs  \\
\hline
Model-based algorithms &  Model-based algorithms aim to find global optimal or at least sub-optimal results for target problems. They usually require full knowledge of the problem to find near-optimal solutions by using transformation, relaxation, and approximation.  &  Model-based algorithms, i.e., SCA, MM, can provide detailed proofs and explanations for the optimality. Target problems are efficiently solved with guaranteed optimality once the closed-form solution is achieved.  & Model-based solutions are usually problem-specific with certain requirements such as convexity and continuity, indicating case-by-case analyses and design. It has difficulty adapting to dynamically changing environments.  &  It has to apply transformations, division, and relaxation to convert the problem to specific forms. These transformations need a dedicated design for each problem.  & Numerous algorithms have been developed to solve RIS-aided optimization problems, e.g., AO to decouple the active and passive beamforming, SDR to relax the rank constraints, and SCA to estimate the sub-optimal results.    \\
\hline
Heuristic algorithms & It applies heuristic rules to find a trade-off between optimality and computational complexity. Heuristic algorithms focus on local optima and low-complexity solutions.  & Heuristic algorithms have much lower computational complexity. It has few requirements for the properties of target problems.  &  It only presents local optima in the current stage, indicating a bad performance in some cases.   & Heuristic rules should be carefully selected and designed, directly affecting the algorithm performance.  &  Considering the high complexity of RIS control problems, heuristic algorithms can provide low-complexity alternatives, i.e., sequential phase shift and on/off control using greedy rule, phase-shift optimization using GA.     \\
\hline
ML techniques & ML techniques are usually data-driven, providing unified control and optimization algorithms for certain types of problems. Most algorithms are easily applied without requiring dedicated design.    &  Data-driven approaches avoid the complexity of building dedicated optimization models. It can better adapt to the dynamic wireless environment given the learning capability.  & It may require many iterations for the algorithm training. ML optimization techniques do not guarantee optimality.   &  Algorithm training is the main difficulty of applying ML, which is data and computation-demanding.  & Various ML techniques have been applied for RIS-related optimizations, e.g., neural networks for CSI prediction and RIS phase control, and DDPG for continuous RIS phase-shift optimization.   \\
\hline
\end{tabular}}
\label{tab-overallcom}
\vspace{-10pt}
\end{table*}

\begin{figure*}[!t]
\centering
\includegraphics[width=0.88\linewidth]{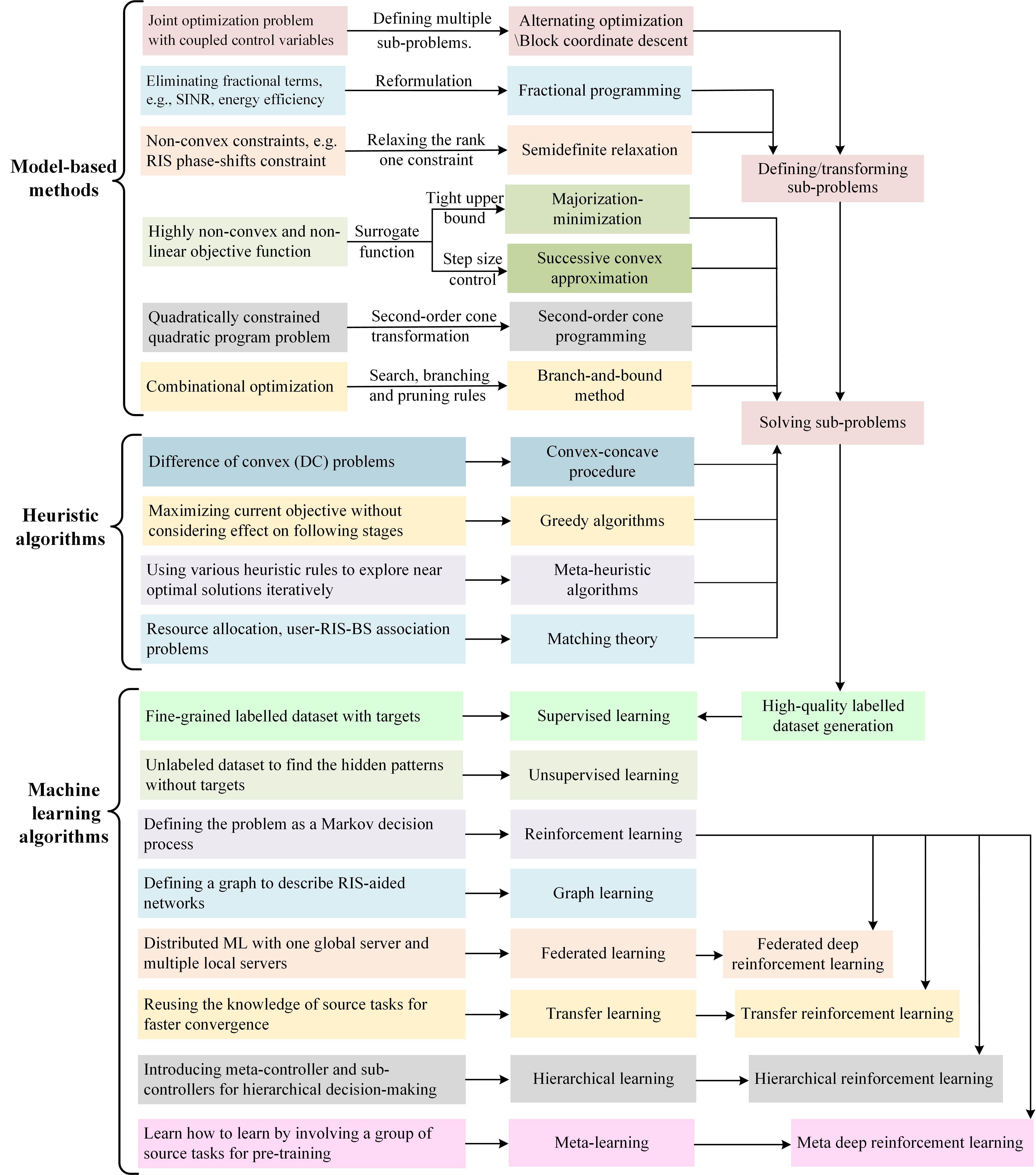}
\caption{Algorithm selection and relationship of model-based, heuristic and ML approaches.}
\label{fig-compar}
\setlength{\abovecaptionskip}{-2pt} 
\vspace{-14pt}
\end{figure*}

\section{Comparison and Relationship between Model-based, Heuristic and ML Approaches}
\label{sec-compa}

This work has introduced three types of optimization techniques: model-based, heuristic, and ML approaches. One intuitive question is how to evaluate the advantages and difficulties of these techniques as well as their relationships. To answer this question, we compare these approaches in Table \ref{tab-overallcom}, including main features, advantages, drawbacks, difficulties, and applications for RISs. In addition, Fig. \ref{fig-compar} summarizes algorithm selection of applying various methods and their relationships. Note that Fig. \ref{fig-compar} provides a general overview for optimizing RIS-aided wireless networks, but the algorithm selection and design should be combined with specific application scenarios.

1) \textbf{Model-based method:} Table \ref{tab-overallcom} shows that model-based approaches can provide efficient and stable solutions once the problem is properly reformulated, especially when closed-form expressions are obtained. 
However, model-based algorithms are usually complicated to design, indicating a series of transformations and relaxations, e.g., decoupling the denominator and numerator in SINR terms and relaxing integer constraints. As a result, the approximation and relaxation can undermine the quality of solutions. 
Additionally, environmental uncertainties can significantly affect the performance of model-based algorithms, since they require full knowledge of the optimization parameters. One possible solution is to assume environment changes follow some specific distributions, but the optimization over distributions will further increase the complexity. Another solution is to use Monte Carlo sampling and repeat the optimization to achieve average results, which is time-consuming.
As illustrated in Fig. \ref{fig-compar}, given a joint optimization problem with coupled control variables, one may use AO or BCD to decouple the joint optimization into multiple sub-problems. Specifically, FP can be used to eliminate fractional terms, e.g., SINR and energy efficiency, and SDR is applied to relax non-convex constraints. Then, various techniques may be applied, such as MM, SCA, SOCP and BnB, to solve each sub-problem under the alternating framework.

2) \textbf{Heuristic algorithms:} The primary benefit of heuristic methods is the low implementation complexity, i.e., optimizing RIS control in an element-by-element manner, achieving a trade-off between optimality and computational complexity. Heuristic methods also show a high generalization capability, e.g., genetic algorithm and PSO apply unified fitness functions to represent the optimization objectives. 
 However, meta-heuristic algorithms are sensitive to key parameters, e.g., population numbers and inertia weight in PSO, which may require find-tuning efforts. But other heuristic methods, especially greedy algorithms and matching theory, can be easily applied with little tuning requirement. 
Meanwhile, Fig. \ref{fig-compar} shows that heuristic algorithms can also be used to solve sub-problems that are defined under an AO scheme, indicating possible combinations between model-based and heuristic algorithms. For instance, to maximize energy efficiency, Yang \textit{et al.} define three sub-problems, and SCA is deployed for active and passive beamforming, while a greedy algorithm is used for RIS on/off control\cite{zhaohui}. Such a combined scheme demonstrates the capability of integrating model-based algorithms with heuristic algorithms.

3) \textbf{ML algorithms:} 
Wireless networks are highly dynamic, and hence optimization techniques must be robust to environmental uncertainties. With the learning capability, ML algorithms can adapt well to dynamic environments.
In particular, ML algorithms present unified optimization schemes, which are applied to diverse problems with few design requirements. For instance, most optimization problems can be converted into unified MDPs that include state, action, transition probability and rewards, and then reinforcement learning is utilized to maximize the reward for a higher sum-rate or energy efficiency. 
Meanwhile, as summarized in Fig. \ref{fig-compar}, reinforcement learning can be integrated with other ML techniques to develop diverse optimization algorithms, such as federated deep reinforcement learning, transfer reinforcement learning, and hierarchical reinforcement learning. For example, transfer reinforcement learning can achieve faster convergence and higher average reward than conventional reinforcement learning algorithms.
However, ML algorithm training is usually computation-demanding, requiring a large number of computational resources, e.g., iterative exploration of RL and back-propagation for neural network training. Finally, datasets are crucial to applying data-driven ML algorithms, especially for supervised learning. Model-based methods provide a useful approach for labeled dataset generation, which indicates the potential to combine model-based and ML algorithms. For example, Hu \textit{et al.} first apply the BCD method for RIS-aided mobile edge computing, and then the produced results serve as datasets for location-based supervised learning algorithms \cite{hu2021reconfigurable}. This reveals the potential benefit of integrating ML techniques with model-based algorithms.

%% file: 7_Applications_and_Challenges.tex
\section{\blue{RIS-assisted 6G Applications: Optimization Analyses and Challenges}}
\label{sec-futu}

%\begin{figure}[!t]
%\centering
%\includegraphics[width=0.6\linewidth]{Image/fig-noma.jpg}
%\caption{ Illustration of cluster-based RIS-aided NOMA.}
%\label{fig-noma}
%\setlength{\abovecaptionskip}{-2pt} 
%\vspace{-10pt}
%\end{figure}

\blue{This section analyzes control and optimization techniques for RIS-assisted 6G applications, e.g., potential optimization difficulties and algorithm selections.} In addition, we identify several research challenges for the optimization of RIS-aided wireless networks.

\begin{table*}[!t]
\caption{\blue{\textbf{Control and optimization analyses for RIS-assisted 6G applications}} }
\centering
\small
\setstretch{1.05}
\resizebox{1\textwidth}{!}{%
\begin{tabular}{|m{1.3cm}<{\centering}|m{3.0cm}<{\centering}|m{4cm}<{\centering}|m{4.3cm}<{\centering}|m{4cm}<{\centering}|}
\hline 
\blue{6G applications}  &  \blue{\quad Key features}   &     \blue{Motivations for integrating RISs}    &  \blue{Potential optimization difficulties}  &   \blue{Analyses of optimization algorithm selections}  \\
\hline
\blue{RIS-NOMA} & \blue{NOMA enables spectrum sharing among users, e.g., multiple users can use the same time and frequency resource blocks, improving user fairness and spectral efficiency. }   & \blue{Due to the resource-sharing nature, the NOMA system is more vulnerable to security issues. Then RISs can be applied to reshape the signal propagation environment for security services against eavesdroppers.}   & \blue{RIS-NOMA integration increases the overall optimization complexity. Specifically, the decoding order may be frequently changed due to the dynamic RIS configuration, increasing the difficulty of applying conventional model-based algorithms.}  & \blue{ML techniques can be applied for intelligent decision-making in RIS-NOMA systems to handle the complexity and meanwhile improve long-term network performance. AO may be used to decouple the RIS control with NOMA optimization.}  \\
\hline
\blue{RIS-SWIPT} & \blue{SWIPT is an attractive solution to transmit electricity without using physical wire links, increasing the mobility, reliability, and safety of electronic devices.}   & \blue{The low energy efficiency at the energy receiver is one of the main issues for practical SWIPT deployment, and RISs become a promising solution, e.g., increasing sum-rate, reducing transmit power, and maximizing the minimum received power.}   & \blue{Existing studies mainly apply model-based methods for joint optimization of RISs and wireless power transfer, requiring dedicated model design. Although some low-complexity algorithms are proposed, i.e., MM \cite{chu2021intelligent} and bi-section search\cite{pan2020intelligent}, they still require full knowledge of the defined problem. }  & \blue{ ML approaches can be promising alternatives to handle the complexity of joint optimizing RISs and SWIPT. For instance, the RL agent can explore the complicated IoT environment without any prior knowledge, and intelligently optimize energy and information transfer and RIS control.  }  \\
\hline
\blue{RIS-mmWave and THz} & \blue{The increasing traffic demand and scarce bandwidth resources make mmWave and THz communications become appealing techniques.}   & \blue{mmWave and THz communications are vulnerable to signal blockages and attenuation, leading to severe path loss and reduced cover range. RISs can be applied to manipulate the signal propagation environment when the direct transmission is blocked.}   & \blue{ The optimization of RIS-mmWave and RIS-THz systems rely on accurate and practical channel estimation to identify the performance limit. Therefore, robust optimization techniques should be developed to handle the uncertainty in channel estimation. 
}  & \blue{ Compared with heuristic or ML algorithms, model-based algorithms are more reliable in guaranteeing the worst-case performance for RIS-mmWave and THz systems, e.g., maximizing worst-case network performance, or considering outage probability constraints. }  \\
\hline
\blue{RIS-NTN} & \blue{The NTN complements the limitations of terrestrial networks, providing flexible and reliable support for remote areas by UAVs, high-altitude platforms (HAPs) and low earth orbit satellites.}   & \blue{Frequent repositioning will increase the UAV power consumption, especially considering that UAVs are powered by a battery\cite{liu2020machine}. In this case, RISs can be applied to overcome this challenge, in which one can configure the RIS phase shifts instead of moving UAVs to save energy.}   & \blue{UAV communications are highly dynamic due to their mobility, and ML algorithms can be used to handle such uncertainty. However, ML model training requires many iterations, which may prevent UAVs from making real-time responses to network dynamics. }  & \blue{To overcome the tedious training iterations of conventional ML algorithms, transfer learning and meta-learning may be used to improve the training efficiency and make rapid responses to UAV dynamics, which is still an open issue.}  \\
\hline
\blue{RIS-V2X} & \blue{V2X is a key paradigm for envisioned 6G networks, enabling intelligent transportation with higher road safety and traffic efficiency.}   & \blue{Latency and reliability are the most critical requirements of V2X networks for road safety. However, the V2X transmission can be unstable due to fast-moving vehicles and the dynamic nature of wireless communications. Therefore, RISs can be exploited to improve channel capacity, coverage, signal strength and reliability.}   & \blue{A critical feature of V2X communications is the stringent requirement for reliability and safety, which means the proposed algorithm should guarantee the worst-case network performance. In addition, this indicates that the control and optimization algorithms should be efficient, robust and reliable.}  & \blue{ML algorithms can quickly adapt to dynamic environments, but the model training is time-consuming. Meanwhile, model-based algorithms can produce stable results, but they require full knowledge of the environment. Therefore, developing efficient optimization algorithms for V2X networks is still challenging for current studies.}  \\
\hline
\blue{RIS-ISAC} & \blue{ISAC is recently emerging as a key technology to support ubiquitous wireless connectivity and accurate sensing\cite{liu2022survey22}.}   & \blue{Target detection and parameter estimation are two primary tasks in radar sensing\cite{yao2022joint}, and RISs can be deployed to provide virtual LoS signal transmission, enabling the radar to sense targets in blocked areas\cite{10050406}. }   & \blue{ISAC offers significant potential by combining sensing and communication, but it also leads to extra complexity for network management, and such complexity further increases by involving RISs.}  & \blue{It is crucial to develop efficient optimization techniques to realize the full potential of the RIS-ISAC system, e.g., decoupling joint optimization into multiple sub-problems using AO, and applying ML algorithms for joint control.}  \\
\hline
\end{tabular}}
\label{tab-6gapp}
\vspace{-10pt}
\end{table*}

\subsection{ \blue{ Control and Optimization Analyses of RIS-assisted 6G applications}}

%\orange{{[Note: Here we have removed 6 separated subsections in the previous manuscript, e.g., RIS-NOMA, RIS-SWIPT, and RIS-V2X, and use Table.\ref{tab-6gapp} to summarize various RIS-aided 6G application scenarios.]}} 

\blue{Table \ref{tab-6gapp} summarizes RIS-assisted 6G applications, including NOMA, SWIPT, mmWave and THz communications, NTNs, V2X communications, and ISAC.}
\blue{For example, due to the resource-sharing nature, the NOMA system is more vulnerable to security issues. Then RISs can be applied to reshape the signal propagation environment for security services against eavesdroppers. The low energy efficiency at the energy receiver is one of the main issues for practical SWIPT deployment, and RISs become a promising solution to increase sum-rate, reduce transmit power, and maximize the minimum received power. In addition, Table \ref{tab-6gapp} also summarizes the motivations for integrating RISs with other 6G applications such as RIS-NTN, RIS-V2X, and RIS-ISAC.}

\blue{However, integrating RISs with 6G techniques also increases the difficulties for network management. In RIS-NOMA systems, the decoding order may be frequently changed due to the dynamic RIS configuration, increasing the difficulty of applying conventional model-based algorithms. For V2X networks, a critical feature is the stringent requirement for reliability and safety, which means the proposed algorithm should guarantee the worst-case network performance. Such a requirement means that the control and optimization algorithms should be efficient, robust and reliable. The potential optimization difficulties of other RIS-assisted 6G applications are also reviewed in Table \ref{tab-6gapp}. }

\blue{Finally, we analyze optimization algorithm selections for various RIS-assisted 6G applications. Integrating RISs will substantially increase the network management complexity, since RIS phase-shift control is highly coupled with other control variables such as decoding order in NOMA, beam selection in mmWave networks, and UAV altitude control. ML algorithms become promising solutions to handle such complexity, such as DDQN\cite{liu2020ris} and DDPG\cite{yang2020deep,yang2021machine}. In particular, these studies apply unified schemes to optimize network performance, overcoming the difficulties of reformulation and transformation for convexity. However, conventional ML algorithms require many iterations for model training, which may prevent the application to highly dynamic environments such as RIS-UAV. To this end, transfer learning and meta-learning may be used to improve training efficiency and make rapid responses.   
On the other hand, other applications such as V2X have more stringent service requirements to guarantee worst-case performance. In this case, model-based methods can usually provide more stable performance than ML or heuristic approaches, providing detailed proofs and explanations for the algorithm output.   }

\subsection{Challenges and Future Directions}
This subsection identifies research challenges and possible future directions.

1) \textbf{Practical RIS Phase-shift Design}:  
Most existing RIS optimization studies rely on perfect CSI acquisition and static user conditions, which are impractical assumptions in the real world. Specifically, the wireless environment is highly dynamic due to various channel conditions and diverse user demands. Therefore, developing robust and practical algorithms for RIS control is of great importance for the real-world deployment of RISs, e.g., imperfect CSI acquisition and UEs with high mobility, which requires more research efforts.   

2) \textbf{Low-overhead Control}:
Many advanced control and optimization techniques have been proposed for RISs, but the communication and control overhead is neglected in most works. For example, frequent parameter exchange between the BS and RISs may lead to high overhead, and the model training overhead of ML algorithms can hamper the system efficiency. These issues are still open challenges, and agile optimization algorithms with low complexity and overhead are yet to be developed.    

3) \textbf{ML-enabled Intelligent RIS Beamforming}:
ML is one of the most promising techniques to facilitate future 6G networks, and integrating ML with RISs can bring intelligent prediction, clustering, and decision-making for RIS-aided wireless networks.   
Despite the significant potential, some critical questions, e.g., algorithm deployment, offline or online training, and training cost, are neglected in many existing studies.
Addressing these problems can further enable an intelligent future wireless network.

4) \textbf{Practical RIS Location Optimization}: In many existing studies, RIS location is considered as a predefined parameter for simulation. However, RIS location can considerably affect the system performance and therefore should be very carefully handled. Moreover, the real-world environment is more complicated when considering dense buildings and other obstacles. RIS location and scale should be jointly optimized by considering the wireless environment, user distribution, and service requirements, which still require research effort.

5) \textbf{Flexible Control Framework}:
The former analyses have shown that each optimization approach has its advantages and difficulties.  Model-based methods have higher stability and optimality, and heuristic methods have lower complexity, while ML techniques are more robust. 
One intuitive direction is to combine these methods to form a flexible optimization framework that can make the most of each approach's advantages and complement the difficulties. However, many existing studies stick with one type of optimization technique, and flexible control schemes are considered future challenges.

%% file: 8_Conclusion.tex
\section{Conclusion}
\label{sec-con}
RIS technology is a key enabler for 6G networks, and control and optimization techniques are critical to exploiting the full potential of RISs. In this work, we have surveyed various approaches for optimizing RIS-aided wireless networks, including model-based, heuristic, and ML approaches. We have provided in-depth analyses of the algorithms' features, difficulties, and applications towards RIS, and we have further compared the advantages and disadvantages of nearly 20 techniques. Our analyses reveal that model-based methods exhibit satisfying performance and stability, but the algorithm design is complicated with low generalization capability. Heuristic algorithms can obtain low-complexity sub-optimal solutions, which are usually considered as baselines or supplements for other techniques. ML techniques have high generalization capability and optimality, but ML model training is computationally demanding and requires experience. Finally, the algorithm selection depends on specific optimization requirements, which should be jointly considered based on application scenarios.
It is hoped that this survey will serve as a roadmap for researchers to investigate advanced optimization techniques for RIS-aided wireless networks.

%% file: 0_Main.bbl
\begin{thebibliography}{100}
\providecommand{\url}[1]{#1}
\csname url@samestyle\endcsname
\providecommand{\newblock}{\relax}
\providecommand{\bibinfo}[2]{#2}
\providecommand{\BIBentrySTDinterwordspacing}{\spaceskip=0pt\relax}
\providecommand{\BIBentryALTinterwordstretchfactor}{4}
\providecommand{\BIBentryALTinterwordspacing}{\spaceskip=\fontdimen2\font plus
\BIBentryALTinterwordstretchfactor\fontdimen3\font minus \fontdimen4\font\relax}
\providecommand{\BIBforeignlanguage}[2]{{%
\expandafter\ifx\csname l@#1\endcsname\relax
\typeout{** WARNING: IEEEtran.bst: No hyphenation pattern has been}%
\typeout{** loaded for the language `#1'. Using the pattern for}%
\typeout{** the default language instead.}%
\else
\language=\csname l@#1\endcsname
\fi
#2}}
\providecommand{\BIBdecl}{\relax}
\BIBdecl

\bibitem{zhang20196g}
Z.~Zhang, Y.~Xiao, Z.~Ma, M.~Xiao, Z.~Ding, X.~Lei, G.~K. Karagiannidis, and P.~Fan, ``{6G} wireless networks: {Vision}, requirements, architecture, and key technologies,'' \emph{IEEE Veh. Technol. Mag.}, vol.~14, no.~3, pp. 28--41, Sep. 2019.

\bibitem{ertu}
E.~{Basar}, M.~D. {Renzo}, J.~D. {Rosny}, M.~{Debbah}, M.-S. {Alouini}, and R.~{Zhang}, ``Wireless communications through reconfigurable intelligent surfaces,'' \emph{IEEE Access}, vol.~7, pp. 116\,753--116\,773, Aug. 2019.

\bibitem{emil}
E.~{Björnson}, H.~{Wymeersch}, B.~{Matthiesen}, P.~{Popovski}, L.~{Sanguinetti}, and E.~{de Carvalho}, ``Reconfigurable intelligent surfaces: {A} signal processing perspective with wireless applications,'' \emph{IEEE Signal Process. Mag.}, vol.~39, no.~2, pp. 135--158, Feb. 2022.

\bibitem{mu2021simultaneously}
X.~Mu, Y.~Liu, L.~Guo, J.~Lin, and R.~Schober, ``Simultaneously transmitting and reflecting {(STAR) {RIS}} aided wireless communications,'' \emph{IEEE Trans. Wireless Commun.}, vol.~21, no.~5, pp. 3083--3098, May 2021.

\bibitem{mu2021intelligent}
X.~Mu, Y.~Liu, L.~Guo, J.~Lin, and H.~V. Poor, ``Intelligent reflecting surface enhanced multi-{UAV NOMA} networks,'' \emph{IEEE J. Sel. Areas Commun.}, vol.~39, no.~10, pp. 3051--3066, Oct. 2021.

\bibitem{marco}
M.~D. {Renzo}, A.~{Zappone}, M.~{Debbah}, M.-S. {Alouini}, C.~{Yuen}, J.~de~{Rosny}, and S.~{Tretyakov}, ``Smart radio environments empowered by reconfigurable intelligent surfaces: {How} it works, state of research, and the road ahead,'' \emph{IEEE J. Sel. Areas Commun.}, vol.~38, no.~11, pp. 2450--2525, Jul. 2020.

\bibitem{Almo}
A.~{Almohamad}, A.~M. {Tahir}, A.~{Al-Kababji}, H.~{M. Furqan}, T.~{Khattab}, M.~{O. Hasna}, and H.~{Arslan}, ``Smart and secure wireless communications via reflecting intelligent surfaces: {A} short survey,'' \emph{IEEE Open J. Commun. Soc.}, vol.~1, pp. 1442--1456, Sep. 2020.

\bibitem{gong}
S.~{Gong}, X.~{Lu}, D.~{T. Hoang}, D.~{Niyato}, L.~{Shu}, D.~I. { Kim}, and Y.-C. {Liang}, ``Toward smart wireless communications via intelligent reflecting surfaces: {A} contemporary survey,'' \emph{IEEE Commun. Surveys Tuts.}, vol.~22, no.~4, pp. 2283--2314, 4th quarter 2020.

\bibitem{moha}
M.~A. {ElMossallamy}, H.~{Zhang}, L.~{Song}, K.~G. {Seddik}, Z.~{Han}, and G.~{Ye Li}, ``Reconfigurable intelligent surfaces for wireless communications: {Principles,} challenges, and opportunities,'' \emph{IEEE Trans. Cogn. Commun. Netw.}, vol.~3, no.~6, pp. 990--1002, Sep. 2020.

\bibitem{mohadz}
M.~Z. {Siddiqi} and T.~{Mir}, ``Reconfigurable intelligent surface-aided wireless communications: {An} overview,'' \emph{Intell. and Converged Netw.}, vol.~3, no.~1, pp. 33--63, Mar. 2022.

\bibitem{cunh}
C.~{Pan}, G.~{Zhou}, K.~{Zhi}, S.~{Hong}, T.~{Wu}, Y.~{Pan}, H.~{Ren}, M.~D. {Renzo}, A.~L. {Swindlehurst}, R.~{Zhang}, and A.~Y. {Zhang}, ``An overview of signal processing techniques for {RIS/{RIS}}-aided wireless systems,'' \emph{IEEE J. Sel. Top. Signal Process.}, vol.~5, no.~16, pp. 883--917, Aug. 2022.

\bibitem{rawa}
R.~{Alghamdi}, R.~{Alhadrami}, D.~{Alhothali}, H.~{Almorad}, A.~{Faisal}, S.~{Helal}, R.~{Shalabi}, R.~{Asfour}, N.~{Hammad}, A.~{Shams}, N.~{Saeed}, H.~{Dahrouj}, T.~Y. {Al-Naffouri}, and M.-S. {Alouini}, ``Intelligent surfaces for {6G} wireless networks: {A} survey of optimization and performance analysis techniques,'' \emph{IEEE Access}, vol.~8, pp. 202\,795--202\,818, Oct. 2020.

\bibitem{kfai}
K.~{Faisal} and W.~{Choi}, ``Machine learning approaches for reconfigurable intelligent surfaces: {A} survey,'' \emph{IEEE Access}, vol.~10, pp. 27\,343--27\,367, Jul. 2022.

\bibitem{yliu}
Y.~{Liu}, X.~{Liu}, X.~{Mu}, T.~{Hou}, J.~{Xu}, M.~{D. Renzo}, and N.~{Al-Dhahir}, ``Reconfigurable intelligent surfaces: {Principles} and opportunities,'' \emph{IEEE Commun. Surveys Tuts.}, vol.~23, no.~3, pp. 1546--1577, 3rd quarter 2021.

\bibitem{zheng2022survey}
B.~Zheng, C.~You, W.~Mei, and R.~Zhang, ``A survey on channel estimation and practical passive beamforming design for intelligent reflecting surface aided wireless communications,'' \emph{IEEE Commun. Surveys Tuts.}, vol.~24, no.~2, pp. 1035--1071, 2nd quarter 2022.

\bibitem{eldar2022machine}
Y.~C. Eldar, A.~Goldsmith, D.~G{\"u}nd{\"u}z, and H.~V. Poor, \emph{{Machine Learning and Wireless Communications}}.\hskip 1em plus 0.5em minus 0.4em\relax Cambridge, UK: Cambridge University Press, 2022.

\bibitem{huang2019indoor}
C.~Huang, G.~C. Alexandropoulos, C.~Yuen, and M.~Debbah, ``Indoor signal focusing with deep learning designed reconfigurable intelligent surfaces,'' in \emph{Proc. Int. Workshop Signal Process. Adv. Wireless commun. (SPAWC)}, Aug. 2019, pp. 1--5.

\bibitem{lu2021aerial}
H.~Lu, Y.~Zeng, S.~Jin, and R.~Zhang, ``Aerial intelligent reflecting surface: {Joint} placement and passive beamforming design with {3D} beam flattening,'' \emph{IEEE Trans. on Wireless Commun.}, vol.~20, no.~7, pp. 4128--4143, Jul 2021.

\bibitem{tekbiyik2022reconfigurable}
K.~Tekb{\i}y{\i}k, G.~K. Kurt, A.~R. Ekti, and H.~Yanikomeroglu, ``Reconfigurable intelligent surfaces in action for nonterrestrial networks,'' \emph{IEEE Veh. Technol. Mag.}, vol.~17, no.~3, pp. 45--53, Sep. 2022.

\bibitem{ruoc}
R.~{Su}, L.~{Dai}, J.~{Tan}, M.~{Hao}, and R.~{MacKenzie}, ``Capacity enhancement for irregular reconfigurable intelligent surface-aided wireless communications,'' in \emph{Proc. IEEE Global Commun. Conf.}, Dec. 2020, pp. 1--6.

\bibitem{cunhua}
C.~{Pan}, H.~{Ren}, K.~{Wang}, W.~{Xu}, M.~{Elkashlan}, A.~{Nallanathan}, and L.~{Hanzo}, ``Multicell \text{{MIMO}} communications relying on intelligent reflecting surfaces,'' \emph{IEEE Trans. Wireless Commun.}, vol.~19, pp. 5218--5233, Oct. 2020.

\bibitem{shuowen}
S.~{Zhang} and R.~{Zhang}, ``Capacity characterization for intelligent reflecting surface aided \text{{MIMO}} communication,'' \emph{IEEE J. Sel. Areas Commun.}, vol.~38, pp. 1823--1838, Jun. 2020.

\bibitem{yu}
Y.~{Zhang}, C.~{Zhong}, Z.~{Zhang}, and W.~{Lu}, ``Sum rate optimization for two way communications with intelligent reflecting surface,'' \emph{IEEE Commun. Lett.}, vol.~24, pp. 1090--1094, Mar. 2020.

\bibitem{ming}
M.-M. {Zhao}, Q.~{Wu}, M.-J. {Zhao}, and R.~{Zhang}, ``Intelligent reflecting surface enhanced wireless networks: {Two}-timescale beamforming optimization,'' \emph{IEEE Trans. Wireless Commun.}, vol.~20, pp. 2--17, Jan. 2021.

\bibitem{jiey}
J.~{Yuan}, Y.-C. {Liang}, J.~{Joung}, G.~{Feng}, and E.~G. {Larsson}, ``Intelligent reflecting surface-assisted cognitive radio system,'' \emph{IEEE Trans. Commun.}, vol.~69, pp. 675 -- 687, Jan. 2021.

\bibitem{boya}
B.~{Di}, H.~{Zhang}, L.~{Li}, L.~{Song}, Y.~{Li}, and Z.~{Han}, ``Practical hybrid beamforming with finite-resolution phase shifters for reconfigurable intelligent surface based multi-user communications,'' \emph{IEEE Trans. Vehi. Technol.}, vol.~69, pp. 4565 -- 4570, Apr. 2020.

\bibitem{huayan}
H.~{Guo}, Y.-C. {Liang}, J.~{Chen}, and E.~G. {Larsson}, ``Weighted sum-rate maximization for reconfigurable intelligent surface aided wireless networks,'' \emph{IEEE Trans. Wireless Commun.}, vol.~19, pp. 3064--3076, May 2020.

\bibitem{chang}
C.~{You}, B.~{Zheng}, and R.~{Zhang}, ``Intelligent reflecting surface with discrete phase shifts: {Channel} estimation and passive beamforming,'' in \emph{Proc. IEEE Int. Conf. Commun. (ICC)}, Jun. 2020, pp. 1--6.

\bibitem{xidong}
X.~{Mu}, Y.~{Liu}, L.~{Guo}, J.~{Lin}, and N.~{Al-Dhahir}, ``Exploiting intelligent reflecting surfaces in {NOMA} networks: {Joint} beamforming optimization,'' \emph{IEEE Trans. Wireless Commun.}, vol.~19, pp. 6884--6898, Oct. 2020.

\bibitem{yuanbin}
Y.~{Chen}, Y.~{Wang}, and L.~{Jiao}, ``Robust transmission for reconfigurable intelligent surface aided millimeter wave vehicular communications with statistical {CSI},'' \emph{IEEE Trans. Wireless Commun.}, vol.~21, pp. 928--944, Aug. 2021.

\bibitem{dai2021reconfigurable}
J.~Dai, Y.~Wang, C.~Pan, K.~Zhi, H.~Ren, and K.~Wang, ``Reconfigurable intelligent surface aided massive {MIMO} systems with low-resolution {DACs},'' \emph{IEEE Commun. Lett.}, vol.~25, no.~9, pp. 3124--3128, Sep. 2021.

\bibitem{zhi2021statistical}
K.~Zhi, C.~Pan, H.~Ren, and K.~Wang, ``Statistical {CSI}-based design for reconfigurable intelligent surface-aided massive {MIMO} systems with direct links,'' \emph{IEEE Wireless Commun. Lett.}, vol.~10, no.~5, pp. 1128--1132, May 2021.

\bibitem{rivera}
M.~Rivera, M.~Chegini, W.~Jaafar, S.~Alfattani, and H.~Yanikomeroglu, ``Optimization of quantized phase shifts for reconfigurable smart surfaces assisted communications,'' in \emph{Proc. Consum. Commun. Netw. Conf. (CCNC)}, Feb. 2022, pp. 901--904.

\bibitem{ni2021resource}
W.~Ni, X.~Liu, Y.~Liu, H.~Tian, and Y.~Chen, ``Resource allocation for multi-cell {RIS}-aided {NOMA} networks,'' \emph{IEEE Trans. Wireless Commun.}, vol.~20, no.~7, pp. 4253--4268, Jul. 2021.

\bibitem{chen2021qos}
Y.~Chen, Y.~Wang, J.~Zhang, and M.~Di~Renzo, ``{QoS}-driven spectrum sharing for reconfigurable intelligent surfaces {(RISs)} aided vehicular networks,'' \emph{IEEE Trans. Wireless Commun.}, vol.~20, no.~9, pp. 5969--5985, Sep. 2021.

\bibitem{yu2007transmitter}
W.~Yu and T.~Lan, ``Transmitter optimization for the multi-antenna downlink with per-antenna power constraints,'' \emph{IEEE Transactions on signal processing}, vol.~55, no.~6, pp. 2646--2660, 2007.

\bibitem{gang}
G.~{Yang}, X.~{Xu}, and Y.-C. {Liang}, ``Intelligent reflecting surface assisted non-orthogonal multiple access,'' in \emph{Proc. IEEE Wireless Commun. Netw. Conf (WCNC)}, May. 2020, pp. 1--6.

\bibitem{qingqing}
Q.~{Wu} and R.~{Zhang}, ``Intelligent reflecting surface enhanced wireless network via joint active and passive beamforming,'' \emph{IEEE Trans. Wireless Commun.}, vol.~18, pp. 5394--5409, Aug. 2019.

\bibitem{qingqing2}
Q.~{Wu} and R.~{Zhang}, ``Beamforming optimization for wireless network aided by intelligent reflecting surface with discrete phase shifts,'' \emph{IEEE Trans. Commun.}, vol.~68, pp. 1838--1851, Mar. 2020.

\bibitem{guizhou2}
G.~{Zhou}, C.~{Pan}, H.~{Ren}, K.~{Wang}, M.~D. {Renzo}, and A.~{Nallanathan}, ``Robust beamforming design for intelligent reflecting surface aided {MISO} communication systems,'' \emph{IEEE Wireless Commun. Lett.}, vol.~9, pp. 1658--1662, Oct. 2020.

\bibitem{huimei}
H.~{Han}, J.~{Zhao}, W.~{Zhai}, Z.~{Xiong}, D.~{Niyato}, M.~D. {Renzo}, Q.-V. {Pham}, W.~{Lu}, and K.-Y. {Lam}, ``Reconfigurable intelligent surface aided power control for physical-layer broadcasting,'' \emph{IEEE Trans. Commun.}, vol.~69, pp. 7821--7836, Nov. 2021.

\bibitem{guiz3}
G.~{Zhou}, C.~{Pan}, H.~{Ren}, K.~{Wang}, and A.~{Nallanathan}, ``A framework of robust transmission design for {RIS}-aided {MISO} communications with imperfect cascaded channels,'' \emph{IEEE Trans. Signal Process.}, vol.~68, pp. 5092--5106, Aug. 2020.

\bibitem{beixiong}
B.~{Zheng}, Q.~{Wu}, and R.~{Zhang}, ``Intelligent reflecting surface-assisted multiple access with user pairing: {NOMA} or {OMA}?'' \emph{IEEE Wireless Commun. Lett.}, vol.~24, pp. 753--757, Jan. 2020.

\bibitem{yiqing}
Y.~{Li}, M.~{Jiang}, Q.~{Zhang}, and J.~{Qin}, ``Joint beamforming design in multi-cluster {MISO} {NOMA} reconfigurable intelligent surface-aided downlink communication networks,'' \emph{IEEE Trans. Commun.}, vol.~69, pp. 664--674, Oct. 2020.

\bibitem{minf}
M.~{Fu}, Y.~{Zhou}, Y.~{Shi}, and K.~B. {Letaief}, ``Reconfigurable intelligent surface empowered downlink non-orthogonal multiple access,'' \emph{IEEE Trans. Commun.}, vol.~69, pp. 3802--3817, Jun. 2021.

\bibitem{jianyue}
J.~{Zhu}, Y.~{Huang}, J.~{Wang}, K.~{Navaie}, and Z.~{Ding}, ``Power efficient {IRS}-assisted {NOMA},'' \emph{IEEE Trans. Commun.}, vol.~69, pp. 900--913, Oct. 2020.

\bibitem{yang2021optimal}
Z.~Yang, C.~Huang, J.~Shi, Y.~Chau, W.~Xu, Z.~Zhang, and M.~Shikh-Bahaei, ``Optimal control for full-duplex communications with reconfigurable intelligent surface,'' in \emph{Proc. IEEE Int. Conf. Commun.(ICC)}, Jun. 2021, pp. 1--6.

\bibitem{zhiyang}
Z.~{Li}, M.~{Chen}, Z.~{Yang}, J.~{Zhao}, Y.~{Wang}, J.~{Shi}, and C.~{Huang}, ``Energy efficient reconfigurable intelligent surface enabled mobile edge computing networks with {NOMA},'' \emph{IEEE Trans. Cogn. Commun. Netw.}, vol.~7, pp. 427--440, Mar. 2021.

\bibitem{cao2022joint}
X.~Cao, X.~Hu, and M.~Peng, ``Joint mode selection and beamforming for {RIS}-aided maritime cooperative communication systems,'' \emph{IEEE Trans. Green Commun. Netw. (Early Access)}, Mar. 2022.

\bibitem{chongwen}
C.~{Huang}, A.~{Zappone}, G.~C. {Alexandropoulos}, M.~{Debbah}, and C.~{Yuen}, ``Reconfigurable intelligent surfaces for energy efficiency in wireless communication,'' \emph{IEEE Trans. Wireless Commun.}, vol.~18, pp. 4157--4170, Aug. 2019.

\bibitem{linsong}
L.~{Du}, W.~{Zhang}, J.~{Ma}, and Y.~{Tang}, ``Reconfigurable intelligent surfaces for energy efficiency in multicast transmissions,'' \emph{IEEE Trans. Vehi. Technol.}, vol.~70, pp. 6266--6271, Jun. 2021.

\bibitem{shuaiqi}
S.~{Jia}, X.~{Yuan}, and Y.-C. {Liang}, ``Reconfigurable intelligent surfaces for energy efficiency in {D2D} communication network,'' \emph{IEEE Wireless Commun. Lett.}, vol.~10, pp. 683--687, Mar. 2021.

\bibitem{zhaohui2}
Z.~{Yang}, J.~{Shi}, Z.~{Li}, M.~{Chen}, W.~{Xu}, and M.~{Shikh-Bahaei}, ``Energy efficient rate splitting multiple access {(RSMA)} with reconfigurable intelligent surface,'' in \emph{Proc. IEEE Int. Conf. Commun. Workshops (ICC Workshops)}, Jun 2020, pp. 1--6.

\bibitem{zhaohui}
Z.~{Yang}, M.~{Chen}, W.~{Saad}, W.~{Xu}, M.~{Shikh-Bahaei}, H.~V. {Poor}, and S.~{Cui}, ``Energy-efficient wireless communications with distributed reconfigurable intelligent surfaces,'' \emph{IEEE Trans. Wireless Commun.}, vol.~21, pp. 665--679, Jul. 2021.

\bibitem{yutong}
Y.~{Zhang}, B.~{Di}, H.~{Zhang}, J.~{Lin}, C.~{Xu}, D.~{Zhang}, Y.~{Li}, and L.~{Song}, ``Beyond cell-free {MIMO}: {Energy} efficient reconfigurable intelligent surface aided cell-free {MIMO} communications,'' \emph{IEEE Trans. Cogn. Commun. Netw.}, vol.~7, pp. 412--426, Feb. 2021.

\bibitem{zhang2020joint}
M.~Zhang, M.~Chen, Z.~Yang, H.~Asgari, and M.~Shikh-Bahaei, ``Joint user clustering and passive beamforming for downlink {NOMA} system with reconfigurable intelligent surface,'' in \emph{Proc. IEEE Int. Symp. Personal, Indoor and Mobile Radio Commun.(PIMRC)}, Oct. 2020, pp. 1--6.

\bibitem{zhou2023hierarchical}
H.~Zhou, L.~Kong, M.~Elsayed, M.~Bavand, R.~Gaigalas, S.~Furr, and M.~Erol-Kantarci, ``Hierarchical reinforcement learning for {RIS}-assisted energy-efficient {RAN},'' in \emph{Proc. IEEE Global Commun. Conf. (GLOBECOM)}, Dec. 2022, pp. 3326--3331.

\bibitem{guiz}
G.~{Zhou}, C.~{Pan}, H.~{Ren}, K.~{Wang}, and A.~{Nallanathan}, ``Intelligent reflecting surface aided multigroup multicast {MISO} communication systems,'' \emph{IEEE Trans. Signal Process.}, vol.~68, pp. 3236--3251, Apr. 2020.

\bibitem{qurrat}
Q.-U.-A. {Nadeem}, A.~{Kammoun}, A.~{Chaaban}, M.~{Debbah}, and M.-S. {Alouini}, ``Asymptotic max-min {SINR} analysis of reconfigurable intelligent surface assisted {MISO} systems,'' \emph{IEEE Trans. Wireless Commun.}, vol.~19, pp. 7748--7764, Dec. 2020.

\bibitem{hailiang}
H.~{Xie}, J.~{Xu}, and Y.-F. {Liu}, ``Max-min fairness in {RIS}-aided multi-cell {MISO} systems with joint transmit and reflective beamforming,'' \emph{IEEE Trans. Wireless Commun.}, vol.~20, pp. 1379--1393, Feb. 2021.

\bibitem{manij}
M.~{Bashar}, K.~{Cumanan}, A.~G. {Burr}, P.~{Xiao}, and M.~D. {Renzo}, ``On the performance of reconfigurable intelligent surface-aided cell-free massive \text{{MIMO}} uplink,'' in \emph{Proc. IEEE Int. Conf. Commun. (ICC)}, Dec. 2020, pp. 1--6.

\bibitem{shiqi}
S.~{Gong}, Z.~{Yang}, C.~{Xing}, J.~{An}, and L.~{Hanzo}, ``Beamforming optimization for intelligent reflecting surface-aided {SWIPT IoT} networks relying on discrete phase shifts,'' \emph{IEEE Internet Things J.}, vol.~8, pp. 8585--8602, Dec. 2020.

\bibitem{zaid}
Z.~{Abdullah}, G.~{Chen}, S.~{Lambotharan}, and J.~A. {Chambers}, ``Optimization of intelligent reflecting surface assisted full-duplex relay networks,'' \emph{IEEE Wireless Commun. Lett.}, vol.~10, pp. 363--367, Oct. 2020.

\bibitem{menghua}
M.~{Hua}, Q.~{Wu}, D.~W.~K. {Ng}, J.~{Zhao}, and L.~{Yang}, ``Intelligent reflecting surface-aided joint processing coordinated multipoint transmission,'' \emph{IEEE Trans. Commun.}, vol.~69, pp. 1650--1665, Dec. 2020.

\bibitem{zhi2022power}
K.~Zhi, C.~Pan, H.~Ren, and K.~Wang, ``Power scaling law analysis and phase shift optimization of {RIS}-aided massive {MIMO} systems with statistical {CSI},'' \emph{IEEE Trans. Commun.}, vol.~70, no.~5, pp. 3558--3574, May 2022.

\bibitem{Atapattu}
S.~Atapattu, R.~Fan, P.~Dharmawansa, G.~Wang, J.~Evans, and T.~A. Tsiftsis, ``Reconfigurable intelligent surface assisted two-way communications: {Performance} analysis and optimization,'' \emph{IEEE Trans. Commun.}, vol.~68, no.~10, pp. 6552--6567, Oct. 2020.

\bibitem{zheng}
Z.~{Chu}, W.~{Hao}, P.~{Xiao}, and J.~{Shi}, ``Intelligent reflecting surface aided multi-antenna secure transmission,'' \emph{IEEE Wireless Commun. Lett.}, vol.~9, pp. 108--112, Jan. 2020.

\bibitem{hong}
H.~{Shen}, W.~{Xu}, S.~{Gong}, Z.~{He}, and C.~{Zhao}, ``Secrecy rate maximization for intelligent reflecting surface assisted multi-antenna communications,'' \emph{IEEE Wireless Commun. Lett.}, vol.~23, pp. 1488--1492, Jun. 2019.

\bibitem{xianghao}
X.~{Yu}, D.~{Xu}, Y.~{Sun}, D.~W.~K. {Ng}, and R.~{Schober}, ``Robust and secure wireless communications via intelligent reflecting surfaces,'' \emph{IEEE J. Sel. Areas Commun.}, vol.~38, pp. 2637--2652, Jul. 2020.

\bibitem{biqian}
B.~{Feng}, Y.~{Wu}, and M.~{Zheng}, ``Secure transmission strategy for intelligent reflecting surface enhanced wireless system,'' in \emph{Proc. Int. Conf. Wireless Commun. Signal Process. (WCSP)}, Oct. 2019, pp. 1--6.

\bibitem{miao}
M.~{Cui}, G.~{Zhang}, and R.~{Zhang}, ``Secure wireless communication via intelligent reflecting surface,'' \emph{IEEE Wireless Commun. Lett.}, vol.~8, pp. 1410--1414, May 2019.

\bibitem{zijie}
Z.~{Ji}, P.~L. {Yeoh}, D.~{Zhang}, G.~{Chen}, Y.~{Zhang}, Z.~{He}, H.~{Yin}, and Y.~{Li}, ``Secret key generation for intelligent reflecting surface assisted wireless communication networks,'' \emph{IEEE Trans. Vehi. Technol.}, vol.~70, pp. 1030--1034, Dec. 2020.

\bibitem{huiming2}
H.-M. {Wang}, J.~{Bai}, and L.~{Dong}, ``Intelligent reflecting surfaces assisted secure transmission without eavesdropper's {CSI},'' \emph{IEEE Signal Process. Lett.}, vol.~27, pp. 1300--1304, Jul. 2020.

\bibitem{wei}
W.~{Sun}, Q.~{Song}, L.~{Guo}, and J.~{Zhao}, ``Secrecy rate maximization for intelligent reflecting surface aided {SWIPT} systems,'' in \emph{Proc. IEEE Int. Conf. Commun. in China (ICCC)}, Aug. 2020, pp. 1--6.

\bibitem{limeng}
L.~{Dong} and H.-M. {Wang}, ``Secure \text{{MIMO}} transmission via intelligent reflecting surface,'' \emph{IEEE Wireless Commun. Lett.}, vol.~9, pp. 787--790, Jan. 2020.

\bibitem{limeng2}
L.~{Dong} and H.-M. {Wang}, ``Enhancing secure \text{{MIMO}} transmission via intelligent reflecting surface,'' \emph{IEEE Trans. Wireless Commun.}, vol.~19, pp. 7543--7556, Nov. 2020.

\bibitem{jie}
J.~{Chen}, Y.-C. {Liang}, Y.~{Pei}, and H.~{Guo}, ``Intelligent reflecting surface: {A} programmable wireless environment for physical layer security,'' \emph{IEEE Access}, vol.~7, pp. 82\,599--82\,612, Jun. 2019.

\bibitem{niu2021simultaneous}
H.~Niu, Z.~Chu, F.~Zhou, and Z.~Zhu, ``Simultaneous transmission and reflection reconfigurable intelligent surface assisted secrecy {MISO} networks,'' \emph{IEEE Commun. Lett.}, vol.~25, no.~11, pp. 3498--3502, Nov. 2021.

\bibitem{liu2020machine}
X.~Liu, Y.~Liu, and Y.~Chen, ``Machine learning empowered trajectory and passive beamforming design in {UAV-RIS} wireless networks,'' \emph{IEEE J. Sel. Areas Commun.}, vol.~39, no.~7, pp. 2042--2055, Jul. 2021.

\bibitem{wu2022resource}
C.~Wu, X.~Mu, Y.~Liu, X.~Gu, and X.~Wang, ``Resource allocation in {STAR-RIS}-aided networks: {OMA and NOMA},'' \emph{IEEE Transactions on Wireless Commun.}, vol.~21, no.~9, pp. 7653--7667, Sep 2022.

\bibitem{zuo2020resource}
J.~Zuo, Y.~Liu, Z.~Qin, and N.~Al-Dhahir, ``Resource allocation in intelligent reflecting surface assisted {NOMA} systems,'' \emph{IEEE Trans. Commun.}, vol.~68, no.~11, pp. 7170--7183, Nov. 2020.

\bibitem{cao2021sum}
Y.~Cao, T.~Lv, W.~Ni, and Z.~Lin, ``Sum-rate maximization for multi-reconfigurable intelligent surface-assisted device-to-device communications,'' \emph{IEEE Trans. Commun.}, vol.~69, no.~11, pp. 7283--7296, Nov. 2021.

\bibitem{yang2021reconfigurable2}
G.~Yang, Y.~Liao, Y.-C. Liang, O.~Tirkkonen, G.~Wang, and X.~Zhu, ``Reconfigurable intelligent surface empowered device-to-device communication underlaying cellular networks,'' \emph{IEEE Trans. Commun.}, vol.~69, no.~11, pp. 7790--7805, Nov. 2021.

\bibitem{peng2022active}
Z.~Peng, R.~Weng, Z.~Zhang, C.~Pan, and J.~Wang, ``Active reconfigurable intelligent surface for mobile edge computing,'' \emph{IEEE Wireless Commun. Lett.}, vol.~11, no.~12, pp. 2482--2486, Dec. 2022.

\bibitem{li2021joint}
A.~Li, Y.~Liu, M.~Li, Q.~Wu, and J.~Zhao, ``Joint scheduling design in wireless powered {MEC IoT} networks aided by reconfigurable intelligent surface,'' in \emph{Proc. Int. Conf. Commun. in China (ICCC Workshops)}, Sep. 2021, pp. 159--164.

\bibitem{lu2020robust}
X.~Lu, W.~Yang, X.~Guan, Q.~Wu, and Y.~Cai, ``Robust and secure beamforming for intelligent reflecting surface aided {mmWave} {MISO} systems,'' \emph{IEEE Wireless Commun. Lett.}, vol.~9, no.~12, pp. 2068--2072, Dec. 2020.

\bibitem{hao2021robust}
W.~Hao, G.~Sun, M.~Zeng, Z.~Chu, Z.~Zhu, O.~A. Dobre, and P.~Xiao, ``Robust design for intelligent reflecting surface-assisted {MIMO}-{OFDMA} terahertz {IoT} networks,'' \emph{IEEE Internet Things J.}, vol.~8, no.~16, pp. 13\,052--13\,064, Aug. 2021.

\bibitem{zhao2021outage}
M.-M. Zhao, A.~Liu, and R.~Zhang, ``Outage-constrained robust beamforming for intelligent reflecting surface aided wireless communication,'' \emph{IEEE Trans. Signal Process.}, vol.~69, pp. 1301--1316, Feb. 2021.

\bibitem{guo2021learning}
X.~Guo, Y.~Chen, and Y.~Wang, ``Learning-based robust and secure transmission for reconfigurable intelligent surface aided millimeter wave {UAV} communications,'' \emph{IEEE Wireless Commun. Lett.}, vol.~10, no.~8, pp. 1795--1799, Aug. 2021.

\bibitem{jcbe}
{J. C. {Bezdek} and R. J. {Hathaway}}, ``Some notes on alternating optimization,'' in \emph{Proc. IEEE Int. Conf. Fuzzy Syst.}, Jan. 2002, pp. 288--300.

\bibitem{Julie}
M.~S. Julie~Nutini, Issam~Laradji, ``Let's make block coordinate descent converge faster: {Faster} greedy rules, message-passing, active-set complexity, and superlinear convergence,'' \emph{arXiv:1712.08859}, p. 1–50, Jun. 2022.

\bibitem{ying}
Y.~{Sun}, P.~{Babu}, and D.~P. {Palomar}, ``{Majorization-Minimization} algorithms in signal processing, communications, and machine learning,'' \emph{IEEE Trans. Signal Process.}, vol.~65, pp. 794 -- 816, Feb. 2017.

\bibitem{palomar}
G.~{Scutari}, F.~{Facchinei}, P.~{Song}, D.~P. {Palomar}, , and J.-S. {Pang}, ``Decomposition by partial linearization: {Parallel} optimization of multi-agent systems,'' \emph{IEEE Trans. Signal Process.}, vol.~62, pp. 641-- 656, Nov. 2014.

\bibitem{shuzhong}
S.~{Zhang}, ``Quadratic maximization and semidefinite relaxation,'' \emph{Math. Program.}, vol.~87, pp. 453--465, May 2000.

\bibitem{zhiquan}
Z.-Q. Luo, W.-K. Ma, A.~M.-C. So, Y.~Ye, and S.~Zhang, ``Semidefinite relaxation of quadratic optimization problems,'' \emph{IEEE Signal Processing Magazine}, vol.~27, no.~3, pp. 20--34, May 2010.

\bibitem{peilan}
P.~{Wang}, J.~{Fang}, X.~{Yuan}, Z.~{Chen}, and H.~{Li}, ``Intelligent reflecting surface-assisted millimeter wave communications: {Joint} active and passive precoding design,'' \emph{IEEE Trans. Vehi. Technol.}, vol.~69, pp. 14\,960--14\,973, Oct. 2020.

\bibitem{miguso}
M.~S. {Lobo}, L.~{Vandenberghe}, S.~{Boyd}, and H.~{Lebret}, ``Applications of second-order cone programming,'' \emph{Linear Algebra Appl.}, vol. 284, pp. 193--228, Nov. 1998.

\bibitem{nest}
Y.~Nesterov and A.~Nemirovskii, \emph{Interior-point polynomial algorithms in convex programming}.\hskip 1em plus 0.5em minus 0.4em\relax Philadelphia, Pennsylvania, US: SIAM, 1994.

\bibitem{zappone}
A.~Zappone, E.~Jorswieck \emph{et~al.}, ``Energy efficiency in wireless networks via fractional programming theory,'' \emph{Found. Trends Commun. Inf. Theory}, vol.~11, no. 3-4, pp. 185--396, Mar. 2015.

\bibitem{Dinke}
W.~Dinkelbach, ``On nonlinear fractional programming,'' \emph{Manage Sci.}, vol.~13, no.~7, pp. 492--498, Mar. 1967.

\bibitem{shenk}
K.~Shen and W.~Yu, ``Fractional programming for communication systems—{Part I: Power} control and beamforming,'' \emph{IEEE Trans. Signal Process.}, vol.~66, no.~10, pp. 2616--2630, May 2018.

\bibitem{clau}
J.~Clausen, ``Branch and bound algorithms-principles and examples,'' \emph{Department of Computer Science, University of Copenhagen}, pp. 1--30, Mar 1999.

\bibitem{zhou2023cooperative}
\BIBentryALTinterwordspacing
H.~Zhou, M.~Elsayed, M.~Bavand, R.~Gaigalas, S.~Furr, and M.~Erol-Kantarci, ``Cooperative hierarchical deep reinforcement learning based joint sleep, power, and ris control for energy-efficient hetnet,'' Apr 2023. [Online]. Available: \url{https://arxiv.org/abs/2304.13226}
\BIBentrySTDinterwordspacing

\bibitem{lipp}
T.~Lipp and S.~Boyd, ``Variations and extension of the convex--concave procedure,'' \emph{Optim. Eng.}, vol.~17, no.~2, pp. 263--287, Nov. 2016.

\bibitem{yulie}
A.~L. Yuille and A.~Rangarajan, ``The concave-convex procedure {(CCCP)},'' in \emph{Adv. Neural Inf. Process. Syst.}, Dec. 2001, pp. 1--8.

\bibitem{xu2022reconfigurable}
Y.~Xu, T.~Zhang, Y.~Zou, and Y.~Liu, ``Reconfigurable intelligence surface aided {UAV-MEC} systems with {NOMA},'' \emph{IEEE Commun. Lett.}, vol.~26, no.~9, pp. 2121--2125, Sep. 2022.

\bibitem{bozorg2017meta}
O.~Bozorg-Haddad, M.~Solgi, and H.~A. Lo{\'a}iciga, \emph{Meta-heuristic and evolutionary algorithms for engineering optimization}.\hskip 1em plus 0.5em minus 0.4em\relax Hoboken, NJ, US: John Wiley \& Sons, 2017.

\bibitem{beheshti2013review}
Z.~Beheshti and S.~M.~H. Shamsuddin, ``A review of population-based meta-heuristic algorithms,'' \emph{Int. J. Adv. Soft Comput. Appl}, vol.~5, no.~1, pp. 1--35, Mar. 2013.

\bibitem{martins2006metaheuristics}
S.~L. Martins and C.~C. Ribeiro, ``Metaheuristics and applications to optimization problems in telecommunications,'' in \emph{Handbook of optimization in telecommunications}.\hskip 1em plus 0.5em minus 0.4em\relax Springer, 2006, pp. 103--128.

\bibitem{tewes}
S.~Tewes, M.~Heinrichs, P.~Staat, R.~Kronberger, and A.~Sezgin, ``Full-duplex meets reconfigurable surfaces: {RIS-assisted SIC} for full-duplex radios,'' in \emph{Proc. IEEE Int. Conf. Commun. (ICC)}, Aug. 2022, pp. 1106--1111.

\bibitem{angli}
A.~Li, L.~Song, B.~Vucetic, and Y.~Li, ``Interference exploitation precoding for reconfigurable intelligent surface aided multi-user communications with direct links,'' \emph{IEEE Wireless Commun. Lett.}, vol.~9, no.~11, pp. 1937--1941, Nov. 2020.

\bibitem{muham}
A.~Muhammad, M.~Elhattab, M.~Shokry, and C.~Assi, ``Leveraging reconfigurable intelligent surface to minimize age of information in wireless networks,'' in \emph{Proc. IEEE Int. Conf. Commun. (ICC)}, Aug. 2022, pp. 2525--2530.

\bibitem{zhang2021joint}
Z.~Zhang and L.~Dai, ``A joint precoding framework for wideband reconfigurable intelligent surface-aided cell-free network,'' \emph{IEEE Trans. Signal Process.}, vol.~69, pp. 4085--4101, Jun. 2021.

\bibitem{el2022latency}
E.~El~Haber, M.~Elhattab, C.~Assi, S.~Sharafeddine, and K.~K. Nguyen, ``Latency and reliability aware edge computation offloading in {RIS}-aided networks,'' in \emph{Proc. IEEE Int. Conf. Commun. (ICC)}, Aug. 2022, pp. 5035--5040.

\bibitem{taghavi2021user}
E.~M. Taghavi, A.~Alizadeh, N.~Rajatheva, M.~Vu, and M.~Latva-aho, ``User association in millimeter wave cellular networks with intelligent reflecting surfaces,'' in \emph{Proc. IEEE 93rd Vehi. Technol. Conf. (VTC2021-Spring)}, Jun. 2021, pp. 1--6.

\bibitem{zhou2023heuristic}
\BIBentryALTinterwordspacing
H.~Zhou, M.~Erol-Kantarci, Y.~Liu, and H.~V. Poor, ``Heuristic algorithms for {RIS-assisted} wireless networks: {Exploring} heuristic-aided machine learning,'' Jun 2023. [Online]. Available: \url{https://arxiv.org/abs/2307.01205}
\BIBentrySTDinterwordspacing

\bibitem{zhang2021deep}
S.~Zhang, S.~Zhang, F.~Gao, J.~Ma, and O.~A. Dobre, ``Deep learning optimized sparse antenna activation for reconfigurable intelligent surface assisted communication,'' \emph{IEEE Trans. Commun.}, vol.~69, no.~10, pp. 6691--6705, Oct. 2021.

\bibitem{stylianopoulos2022deep}
K.~Stylianopoulos, N.~Shlezinger, P.~Del~Hougne, and G.~C. Alexandropoulos, ``Deep-learning-assisted configuration of reconfigurable intelligent surfaces in dynamic rich-scattering environments,'' in \emph{Proc. IEEE Int. Conf. Acoust. Speech Signal Process. (ICASSP)}, Apr. 2022, pp. 8822--8826.

\bibitem{taha2021enabling}
A.~Taha, M.~Alrabeiah, and A.~Alkhateeb, ``Enabling large intelligent surfaces with compressive sensing and deep learning,'' \emph{IEEE Access}, vol.~9, pp. 44\,304--44\,321, Mar. 2021.

\bibitem{aygul2021deep}
M.~A. Ayg{\"u}l, M.~Nazzal, and H.~Arslan, ``Deep learning-based optimal {RIS} interaction exploiting previously sampled channel correlations,'' in \emph{Proc. IEEE Wireless Commun. Netw. Conf (WCNC)}, May 2021, pp. 1--6.

\bibitem{alexandropoulos2020phase}
G.~C. Alexandropoulos, S.~Samarakoon, M.~Bennis, and M.~Debbah, ``Phase configuration learning in wireless networks with multiple reconfigurable intelligent surfaces,'' in \emph{Proc. IEEE Global Commun. Conf. (GLOBECOM Workshops)}, Dec. 2020, pp. 1--6.

\bibitem{ozdougan2020deep}
{\"O}.~{\"O}zdo{\u{g}}an and E.~Bj{\"o}rnson, ``Deep learning-based phase reconfiguration for intelligent reflecting surfaces,'' in \emph{Proc. Asilomar Conf. Signals Syst. Comput.}, Nov. 2020, pp. 707--711.

\bibitem{yang2021intelligent}
B.~Yang, X.~Cao, C.~Huang, C.~Yuen, L.~Qian, and M.~Di~Renzo, ``Intelligent spectrum learning for wireless networks with reconfigurable intelligent surfaces,'' \emph{IEEE Trans. Vehi. Technol.}, vol.~70, no.~4, pp. 3920--3925, Apr. 2021.

\bibitem{song2021truly}
Y.~Song, M.~R. Khandaker, F.~Tariq, K.-K. Wong, and A.~Toding, ``Truly intelligent reflecting surface-aided secure communication using deep learning,'' in \emph{Proc. IEEE Vehi. Technol. Conf. (VTC2021-Spring)}, Jun. 2021, pp. 1--6.

\bibitem{hu2021reconfigurable}
X.~Hu, C.~Masouros, and K.-K. Wong, ``Reconfigurable intelligent surface aided mobile edge computing: {From} optimization-based to location-only learning-based solutions,'' \emph{IEEE Trans. Commun.}, vol.~69, no.~6, pp. 3709--3725, Mar. 2021.

\bibitem{song2020unsupervised}
H.~Song, M.~Zhang, J.~Gao, and C.~Zhong, ``Unsupervised learning-based joint active and passive beamforming design for reconfigurable intelligent surfaces aided wireless networks,'' \emph{IEEE Commun. Lett.}, vol.~25, no.~3, pp. 892--896, Mar. 2020.

\bibitem{nguyen2021machine}
N.~T. Nguyen, L.~V. Nguyen, T.~Huynh-The, D.~H. Nguyen, A.~L. Swindlehurst, and M.~Juntti, ``Machine learning-based reconfigurable intelligent surface-aided {MIMO} systems,'' in \emph{Proc. Int. Workshop Signal Process. Adv. Wireless commun. (SPAWC)}, Nov. 2021, pp. 101--105.

\bibitem{dinh2022unsupervised}
S.~Dinh-Van, T.~M. Hoang, R.~Trestian, and H.~X. Nguyen, ``Unsupervised deep-learning-based reconfigurable intelligent surface-aided broadcasting communications in industrial {IoTs},'' \emph{IEEE Internet of Things J.}, vol.~9, no.~19, pp. 19\,515--19\,528, Oct. 2022.

\bibitem{gao2020unsupervised}
J.~Gao, C.~Zhong, X.~Chen, H.~Lin, and Z.~Zhang, ``Unsupervised learning for passive beamforming,'' \emph{IEEE Commun. Lett.}, vol.~24, no.~5, pp. 1052--1056, May 2020.

\bibitem{lopez2022deep}
G.~L{\'o}pez-Lanuza, K.~Chen-Hu, and A.~G. Armada, ``Deep learning-based optimization for reconfigurable intelligent surface-assisted communications,'' in \emph{Proc. IEEE Wireless Commun. Netw. Conf (WCNC)}, May 2022, pp. 764--769.

\bibitem{bjornson2021configuring}
E.~Bj{\"o}rnson and L.~Marcenaro, ``Configuring an intelligent reflecting surface for wireless communications: {Highlights} from the 2021 ieee signal processing cup student competition [sp competitions],'' \emph{IEEE Signal Process. Mag.}, vol.~39, no.~1, pp. 126--131, Jan. 2021.

\bibitem{yang2020deep}
Z.~Yang, Y.~Liu, Y.~Chen, and J.~T. Zhou, ``Deep reinforcement learning for {RIS}-aided non-orthogonal multiple access downlink networks,'' in \emph{Proc. IEEE Global Commun. Conf. (GLOBECOM)}, Dec. 2020, pp. 1--6.

\bibitem{taha2020deep}
A.~Taha, Y.~Zhang, F.~B. Mismar, and A.~Alkhateeb, ``Deep reinforcement learning for intelligent reflecting surfaces: {Towards} standalone operation,'' in \emph{Proc. Int. Workshop Signal Process. Adv. Wireless commun. (SPAWC)}, Aug. 2020, pp. 1--5.

\bibitem{yang2020intelligent}
H.~Yang, Z.~Xiong, J.~Zhao, D.~Niyato, Q.~Wu, H.~V. Poor, and M.~Tornatore, ``Intelligent reflecting surface assisted anti-jamming communications: {A} fast reinforcement learning approach,'' \emph{IEEE Trans. Wireless Commun.}, vol.~20, no.~3, pp. 1963--1974, Mar. 2020.

\bibitem{yang2020deep2}
H.~Yang, Z.~Xiong, J.~Zhao, D.~Niyato, L.~Xiao, and Q.~Wu, ``Deep reinforcement learning-based intelligent reflecting surface for secure wireless communications,'' \emph{IEEE Trans. Wireless Commun.}, vol.~20, no.~1, pp. 375--388, Jan. 2021.

\bibitem{feng2020deep}
K.~Feng, Q.~Wang, X.~Li, and C.-K. Wen, ``Deep reinforcement learning based intelligent reflecting surface optimization for {MISO} communication systems,'' \emph{IEEE Wireless Commun. Lett.}, vol.~9, no.~5, pp. 745--749, May 2020.

\bibitem{huang2020hybrid}
C.~Huang, Z.~Yang, G.~C. Alexandropoulos, K.~Xiong, L.~Wei, C.~Yuen, and Z.~Zhang, ``Hybrid beamforming for {RIS}-empowered multi-hop terahertz communications: {A DRL}-based method,'' in \emph{Proc. IEEE Global Commun. Conf. (GLOBECOM Workshops)}, Dec. 2020, pp. 1--6.

\bibitem{lee2020deep}
G.~Lee, M.~Jung, A.~T.~Z. Kasgari, W.~Saad, and M.~Bennis, ``Deep reinforcement learning for energy-efficient networking with reconfigurable intelligent surfaces,'' in \emph{Proc. IEEE Int. Conf. Commun. (ICC)}, Jul. 2020, pp. 1--6.

\bibitem{lin2020deep}
J.~Lin, Y.~Zout, X.~Dong, S.~Gong, D.~T. Hoang, and D.~Niyato, ``Deep reinforcement learning for robust beamforming in {RIS}-assisted wireless communications,'' in \emph{Proc. IEEE Global Commun. Conf. (GLOBECOM)}, Dec. 2020, pp. 1--6.

\bibitem{huang2020reconfigurable}
C.~Huang, R.~Mo, and C.~Yuen, ``Reconfigurable intelligent surface assisted multiuser {MISO} systems exploiting deep reinforcement learning,'' \emph{IEEE J. Sel. Areas Commun.}, vol.~38, no.~8, pp. 1839--1850, Aug. 2020.

\bibitem{zhang2021millimeter}
Q.~Zhang, W.~Saad, and M.~Bennis, ``Millimeter wave communications with an intelligent reflector: {Performance} optimization and distributional reinforcement learning,'' \emph{IEEE Trans. Wireless Commun.}, vol.~21, no.~3, pp. 1836--1850, Mar. 2021.

\bibitem{liu2020ris}
X.~Liu, Y.~Liu, Y.~Chen, and H.~V. Poor, ``{RIS} enhanced massive non-orthogonal multiple access networks: {Deployment} and passive beamforming design,'' \emph{IEEE J. Sel. Areas Commun.}, vol.~39, no.~4, pp. 1057--1071, Apr. 2020.

\bibitem{kim2021multi}
J.~Kim, S.~Hosseinalipour, T.~Kim, D.~J. Love, and C.~G. Brinton, ``Multi-{{RIS}}-assisted multi-cell uplink {MIMO} communications under imperfect {{CSI}}: {A} deep reinforcement learning approach,'' in \emph{Proc. IEEE Int. Conf. Commun. Workshops (ICC Workshops)}, Jul. 2021, pp. 1--7.

\bibitem{samir2021optimizing}
M.~Samir, M.~Elhattab, C.~Assi, S.~Sharafeddine, and A.~Ghrayeb, ``Optimizing age of information through aerial reconfigurable intelligent surfaces: {A} deep reinforcement learning approach,'' \emph{IEEE Trans. Vehi. Technol.}, vol.~70, no.~4, pp. 3978--3983, Apr. 2021.

\bibitem{yang2021machine}
Z.~Yang, Y.~Liu, Y.~Chen, and N.~Al-Dhahir, ``Machine learning for user partitioning and phase shifters design in {RIS}-aided {NOMA} networks,'' \emph{IEEE Trans. Commun.}, vol.~69, no.~11, pp. 7414--7428, Nov. 2021.

\bibitem{moerland2023model}
T.~M. Moerland, J.~Broekens, A.~Plaat, C.~M. Jonker \emph{et~al.}, ``Model-based reinforcement learning: A survey,'' \emph{Foundations and Trends{\textregistered} in Machine Learning}, vol.~16, no.~1, pp. 1--118, 2023.

\bibitem{sutton2018reinforcement}
R.~S. Sutton and A.~G. Barto, \emph{Reinforcement {Learning}: An {Introduction}}.\hskip 1em plus 0.5em minus 0.4em\relax MIT Press, 2018.

\bibitem{liu2021reconfigurable}
H.~Liu, X.~Yuan, and Y.-J.~A. Zhang, ``Reconfigurable intelligent surface enabled federated learning: {A} unified communication-learning design approach,'' \emph{IEEE Trans. Wireless Commun.}, vol.~20, no.~11, pp. 7595--7609, Nov. 2021.

\bibitem{yang2021reconfigurable}
Y.~Yang, Y.~Zhou, T.~Wang, and Y.~Shi, ``Reconfigurable intelligent surface assisted federated learning with privacy guarantee,'' in \emph{Proc. IEEE Int. Conf. Commun. Workshops (ICC Workshops)}, Jul. 2021, pp. 1--6.

\bibitem{ni2022star}
W.~Ni, Y.~Liu, Y.~C. Eldar, Z.~Yang, and H.~Tian, ``{STAR-RIS} integrated nonorthogonal multiple access and {Over-the-Air} federated learning: {Framework}, analysis, and optimization,'' \emph{IEEE Internet Things J.}, vol.~9, no.~18, pp. 17\,136--17\,156, Sep. 2022.

\bibitem{battiloro2022dynamic}
C.~Battiloro, M.~Merluzzi, P.~Di~Lorenzo, and S.~Barbarossa, ``Dynamic resource optimization for adaptive federated learning empowered by reconfigurable intelligent surfaces,'' in \emph{Proc. IEEE Int. Conf. Acoust. Speech Signal Process. (ICASSP)}, Apr. 2022, pp. 4083--4087.

\bibitem{zheng2022balancing}
J.~Zheng, H.~Tian, W.~Ni, W.~Ni, and P.~Zhang, ``Balancing accuracy and integrity for reconfigurable intelligent surface-aided {Over-the-Air} federated learning,'' \emph{IEEE Trans. Wireless Commun.}, vol.~21, no.~12, pp. 10\,964--10\,980, Dec. 2022.

\bibitem{ni2021over}
W.~Ni, Y.~Liu, Z.~Yang, and H.~Tian, ``Over-the-air federated learning and non-orthogonal multiple access unified by reconfigurable intelligent surface,'' in \emph{Proc. IEEE Conf. on Comput. Commun. (INFOCOM Workshops)}, Jul. 2021, pp. 1--6.

\bibitem{ni2021federated}
W.~Ni, Y.~Liu, Z.~Yang, H.~Tian, and X.~Shen, ``Federated learning in multi-ris-aided systems,'' \emph{IEEE Internet of Things Journal}, vol.~9, no.~12, pp. 9608--9624, Jun. 2021.

\bibitem{liu2021joint}
H.~Liu, X.~Yuan, and Y.-J.~A. Zhang, ``Joint communication-learning design for {RIS}-assisted federated learning,'' in \emph{Proc. IEEE Int. Conf. Commun. Workshops (ICC Workshops)}, Jul. 2021, pp. 1--6.

\bibitem{tang2020wireless}
W.~Tang, M.~Z. Chen, X.~Chen, J.~Y. Dai, Y.~Han, M.~Di~Renzo, Y.~Zeng, S.~Jin, Q.~Cheng, and T.~J. Cui, ``Wireless communications with reconfigurable intelligent surface: {Path} loss modeling and experimental measurement,'' \emph{IEEE Trans. Wireless Commun.}, vol.~20, no.~1, pp. 421--439, Jan. 2021.

\bibitem{yang2022federated}
B.~Yang, X.~Cao, C.~Huang, C.~Yuen, M.~Di~Renzo, Y.~L. Guan, D.~Niyato, L.~Qian, and M.~Debbah, ``Federated spectrum learning for reconfigurable intelligent surfaces-aided wireless edge networks,'' \emph{IEEE Transactions on Wireless Communications}, vol.~21, no.~11, pp. 9610--9626, Nov. 2022.

\bibitem{zhang2021energy}
T.~Zhang and S.~Mao, ``Energy-efficient federated learning with intelligent reflecting surface,'' \emph{IEEE Trans. Green Commun. Netw.}, vol.~6, no.~2, pp. 845--858, Jun. 2021.

\bibitem{li2020enhanced}
L.~Li, D.~Ma, H.~Ren, D.~Wang, X.~Tang, W.~Liang, and T.~Bai, ``Enhanced reconfigurable intelligent surface assisted {mmWave} communication: {A} federated learning approach,'' \emph{China Commun.}, vol.~17, no.~10, pp. 115--128, Oct. 2020.

\bibitem{zhong2022mobile}
R.~Zhong, X.~Liu, Y.~Liu, Y.~Chen, and Z.~Han, ``Mobile reconfigurable intelligent surfaces for {NOMA} networks: {Federated} learning approaches,'' \emph{IEEE Trans. Wireless Commun.}, vol.~21, no.~11, pp. 10\,020--10\,034, Nov. 2022.

\bibitem{cui2018survey}
P.~Cui, X.~Wang, J.~Pei, and W.~Zhu, ``A survey on network embedding,'' \emph{IEEE Trans. Knowl. Data Eng.}, vol.~31, no.~5, pp. 833--852, May 2018.

\bibitem{he2021overview}
S.~He, S.~Xiong, Y.~Ou, J.~Zhang, J.~Wang, Y.~Huang, and Y.~Zhang, ``An overview on the application of graph neural networks in wireless networks,'' \emph{IEEE Open J. Commu. Soc.}, vol.~2, pp. 2547--2565, Nov. 2021.

\bibitem{naderializadeh2020wireless}
N.~Naderializadeh, M.~Eisen, and A.~Ribeiro, ``Wireless power control via counterfactual optimization of graph neural networks,'' in \emph{Proc. Int. Workshop Signal Process. Adv. Wireless commun. (SPAWC)}, Aug. 2020, pp. 1--5.

\bibitem{eisen2020optimal}
M.~Eisen and A.~Ribeiro, ``Optimal wireless resource allocation with random edge graph neural networks,'' \emph{IEEE Trans. Signal Process.}, vol.~68, pp. 2977--2991, Apr. 2020.

\bibitem{jiang2020dynamic}
H.~Jiang, H.~He, and L.~Liu, ``Dynamic spectrum access for femtocell networks: {A} graph neural network based learning approach,'' in \emph{Proc. Int. Conf. Comput., Netw. and Commun. (ICNC)}, Mar. 2020, pp. 927--931.

\bibitem{wang2020graph}
H.~Wang, Y.~Wu, G.~Min, and W.~Miao, ``A graph neural network-based digital twin for network slicing management,'' \emph{IEEE Trans. Industr. Inform.}, vol.~18, no.~2, pp. 1367--1376, Feb. 2020.

\bibitem{zhou2020graph}
J.~Zhou, G.~Cui, S.~Hu, Z.~Zhang, C.~Yang, Z.~Liu, L.~Wang, C.~Li, and M.~Sun, ``Graph neural networks: {A} review of methods and applications,'' \emph{AI Open}, vol.~1, pp. 57--81, Apr. 2020.

\bibitem{khamsi2011introduction}
M.~A. Khamsi and W.~A. Kirk, \emph{An introduction to metric spaces and fixed point theory}.\hskip 1em plus 0.5em minus 0.4em\relax Hoboken, NJ, US: John Wiley \& Sons, 2011.

\bibitem{zhang2022learning}
Z.~Zhang, T.~Jiang, and W.~Yu, ``Learning based user scheduling in reconfigurable intelligent surface assisted multiuser downlink,'' \emph{IEEE J. Sel. Topics Signal Process.}, vol.~16, no.~5, pp. 1026--1039, Feb. 2022.

\bibitem{zhang2022user}
Z.~Zhang, T.~Jiang, and W.~Yu, ``User scheduling using graph neural networks for reconfigurable intelligent surface assisted multiuser downlink communications,'' in \emph{Proc. IEEE Int. Conf. Acoust. Speech Signal Process. (ICASSP)}, Apr. 2022, pp. 8892--8896.

\bibitem{jiang2021learning}
T.~Jiang, H.~V. Cheng, and W.~Yu, ``Learning to reflect and to beamform for intelligent reflecting surface with implicit channel estimation,'' \emph{IEEE J. Sel. Areas Commun.}, vol.~39, no.~7, pp. 1931--1945, Jul. 2021.

\bibitem{zhou2022learning}
H.~Zhou, M.~Erol-Kantarci, and H.~V. Poor, ``Learning from peers: {Deep} transfer reinforcement learning for joint radio and cache resource allocation in {5G} {RAN} slicing,'' \emph{IEEE Trans. Cogn. Commun. Netw.}, vol.~8, no.~4, pp. 1925--1941, Sep. 2022.

\bibitem{zhou2022knowledge}
H.~Zhou, M.~Erol-Kantarci, and V.~Poor, ``Knowledge transfer and reuse: {A} case study of {AI}-enabled resource management in {RAN} slicing,'' \emph{IEEE Wireless Commun.}, pp. 1--10, Dec. 2022.

\bibitem{elsayed2020transfer}
M.~Elsayed, M.~Erol-Kantarci, and H.~Yanikomeroglu, ``Transfer reinforcement learning for {5G} new radio {mmWave} networks,'' \emph{IEEE Trans. Wireless Commun.}, vol.~20, no.~5, pp. 2838--2849, May 2021.

\bibitem{pateria2021hierarchical}
S.~Pateria, B.~Subagdja, A.-h. Tan, and C.~Quek, ``Hierarchical reinforcement learning: {A} comprehensive survey,'' \emph{ACM Comput. Surveys (CSUR)}, vol.~54, no.~5, pp. 1--35, Jun. 2021.

\bibitem{jung2021meta}
M.~Jung and W.~Saad, ``Meta-learning for {6G} communication networks with reconfigurable intelligent surfaces,'' in \emph{Proc. IEEE Int. Conf. Acoust. Speech Signal Process. (ICASSP)}, Jun 2021, pp. 8082--8086.

\bibitem{zou2021meta}
Y.~Zou, Y.~Liu, K.~Han, X.~Liu, and K.~K. Chai, ``Meta-learning for {RIS-assisted} {NOMA} networks,'' in \emph{Proc. IEEE Global Commu. Conf.(GLOBECOM)}, Dec. 2021, pp. 1--6.

\bibitem{vanschoren2018meta}
\BIBentryALTinterwordspacing
J.~Vanschoren, ``Meta-learning: {A survey},'' pp. 1--29, Oct. 2018. [Online]. Available: \url{arXiv preprint arXiv:1810.03548}
\BIBentrySTDinterwordspacing

\bibitem{hospedales2021meta}
T.~Hospedales, A.~Antoniou, P.~Micaelli, and A.~Storkey, ``Meta-learning in neural networks: {A} survey,'' \emph{IEEE Trans. pattern analysis and machine intelligence}, vol.~44, no.~9, pp. 5149--5169, Sep 2021.

\bibitem{chu2021intelligent}
Z.~Chu, Z.~Zhu, F.~Zhou, M.~Zhang, and N.~Al-Dhahir, ``Intelligent reflecting surface assisted wireless powered sensor networks for internet of things,'' \emph{IEEE Trans. Commun.}, vol.~69, no.~7, pp. 4877--4889, Jul. 2021.

\bibitem{pan2020intelligent}
C.~Pan, H.~Ren, K.~Wang, M.~Elkashlan, A.~Nallanathan, J.~Wang, and L.~Hanzo, ``Intelligent reflecting surface aided {MIMO} broadcasting for simultaneous wireless information and power transfer,'' \emph{IEEE J. Sel. Areas Commun.}, vol.~38, no.~8, pp. 1719--1734, Aug. 2020.

\bibitem{liu2022survey22}
A.~Liu, Z.~Huang, M.~Li, Y.~Wan, W.~Li, T.~X. Han, C.~Liu, R.~Du, D.~K.~P. Tan, J.~Lu \emph{et~al.}, ``A survey on fundamental limits of integrated sensing and communication,'' \emph{IEEE Commun. Surveys Tuts.}, vol.~24, no.~2, pp. 994--1034, 2nd quarter 2022.

\bibitem{yao2022joint}
Y.~Yao, H.~Zhou, and M.~Erol-Kantarci, ``Joint sensing and communications for deep reinforcement learning-based beam management in 6g,'' in \emph{Proc. IEEE Global Commu. Conf.(GLOBECOM)}.\hskip 1em plus 0.5em minus 0.4em\relax IEEE, Dec. 2022, pp. 5019--5024.

\bibitem{10050406}
Z.~Wang, X.~Mu, and Y.~Liu, ``{STARS} enabled integrated sensing and communications,'' \emph{IEEE Trans. on Wireless Commu. (Early access)}, pp. 1--16, Feb. 2023.

\end{thebibliography}
